\documentclass[superscriptaddress,showkeys]{revtex4-2}
\usepackage{float}
\usepackage{multirow}
\usepackage{amsmath}
\usepackage[normalem]{ulem}
\usepackage{amssymb}
\usepackage{mathtools}
\usepackage{graphicx}
\usepackage{aas_macros}
\usepackage{xcolor}
\usepackage{stackrel}
\usepackage{natbib}
\definecolor{tl}{RGB}{0,180,120}
\usepackage[colorlinks,citecolor=tl,linkcolor=blue,urlcolor=magenta,unicode=true]{hyperref}

\begin{document}
\title{Optical depth to reionization in a Universe with multiple inhomogeneous domains}
\author{Shashank Shekhar Pandey}
\email{shashankpandey7347@gmail.com}
\affiliation{Department of Astrophysics and High Energy Physics, S. N. Bose National Centre for Basic Sciences, JD Block, Sector III, Salt Lake City, Kolkata-700106, India}
\author{Ruchika} \email{ruchika.science@usal.es}
\affiliation{Departamento de Física Fundamental and IUFFyM, Universidad de Salamanca, E-37008 Salamanca, Spain}
\affiliation{INFN, Sezione di Roma, Piazzale Aldo Moro 2, I-00185 Roma, Italy}
\affiliation{Dipartimento di Fisica, Università di Roma ``La Sapienza'', Piazzale Aldo Moro 2, I-00185 Roma, Italy}
\author{Subhadeep Mukherjee}   \email{subhadeep.avg@gmail.com}
\affiliation{Department of Astrophysics and High Energy Physics, S. N. Bose National Centre for Basic Sciences, JD Block, Sector III, Salt Lake City, Kolkata-700106, India}
\author{A. S. Majumdar }   \email{archan@bose.res.in}
\affiliation{Department of Astrophysics and High Energy Physics, S. N. Bose National Centre for Basic Sciences, JD Block, Sector III, Salt Lake City, Kolkata-700106, India}
\date{\today}
\begin{abstract}
\begin{center}
\textbf{Abstract}
\end{center}

We study the optical depth to reionization in a cosmological setting that includes backreaction from matter inhomogeneities, using the Buchert averaging formalism. We construct a spacetime model consisting of multiple inhomogeneous domains, hereafter referred to as the backreaction model, characterized by a set of parameters. We first examine how these parameters influence the computation of the optical depth to reionization, $\tau_{reion}$. Next, we carry out a Markov Chain Monte Carlo (MCMC) analysis based on the PantheonPlus+SH0ES Type Ia supernova sample to infer the best-fit values of the model parameters, and then use these to evaluate $\tau_{reion}$. We obtain $\tau_{reion} = 0.0581^{+0.0105}_{-0.0096}$ (68$\%$ confidence limits). This result indicates that, when PantheonPlus+SH0ES data are used to constrain the model parameters, our backreaction model yields a value of $\tau_{reion}$ that aligns more closely with observational estimates than the value predicted by the standard cosmological model. We further demonstrate that the backreaction model leads to a modest reduction of the Hubble tension, while avoiding the need for exotic or non-standard physics.

\end{abstract}
\keywords{cosmology: inhomogeneous universe, backreaction formalism, cosmology: Hubble tension, optical depth to reionization}
\pacs{}
\maketitle
\section{Introduction }\label{sec:intro}

Since its formulation, the $\Lambda$CDM (lambda cold dark matter) model has successfully explained numerous cosmological observations. However, in recent years, improved precision and consistency of observations of various cosmic phenomena \cite{Planck, Panth_full_dataset_Scolnic_2022, union} have revealed growing tensions between the theoretical predictions of the $\Lambda$CDM model and the observations \cite{LCDM_challenges, LCDM_small_scale_problems, cosmology_intertwined, eleonora1, CosmoVerse, Leandros1}. The most statistically significant of these discrepancies is the Hubble tension - a disagreement of 4.7$\sigma$ to 6.5$\sigma$ \cite{Hubble_tension_review1, Hubble_tension_review2, Hubble_tension3, Nils_review} between two independent measurements of the Hubble constant, $H_0$. On the one hand, the Planck collaboration derives $H_0$ from high-redshift cosmic microwave background (CMB) observations combined with the assumptions of the $\Lambda$CDM model \cite{Planck}. On the other hand, measurements from low-redshift late-time observations using Type Ia supernovae calibrated with classical Cepheid variables \cite{Pantheon_likelihood, Panth_full_dataset_Scolnic_2022, Riess_2022_fidu}
leads to persistent disagreement with the early-universe determination of $H_0$, thereby constituting the Hubble tension.

\par The $\Lambda$CDM model relies on the Friedmann-Lemaître-Robertson-Walker (FLRW) metric, which is fundamentally grounded in the cosmological principle, the assumption that the Universe exhibits statistical homogeneity and isotropy on large scales. However, recent large-scale structure (LSS) observations suggest the need to incorporate additional complexities beyond the standard cosmological framework. Although the Universe may indeed be uniform and isotropic at the largest scales, numerous astrophysical surveys have unveiled significant matter distribution inhomogeneities extending to substantial scales \cite{Labini_2009, Buchert_obs_challenges, Labini_obs_theo_2010, Labini_obs_challenges, Verevkin2011, Aluri_obs_2023}. The deviations from homogeneity are statistically significant, with luminous red galaxy samples showing departures exceeding 3$\sigma$ from $\Lambda$CDM mock catalogs on scales as large as 500 $h^{-1}$ Mpc \cite{wiegand_scale}. The discovery of a giant arc of galaxies spanning approximately 1 Gpc (proper size at the present epoch) presents a notable challenge to the $\Lambda$CDM framework \cite{lopez}. However, subsequent studies have questioned its cosmological significance \cite{lopez2025,sawala2025,sawala2025_1}.

\par The observed departures from the smooth, homogeneous Universe assumed in the $\Lambda$CDM paradigm suggest that incorporating the effects of inhomogeneities may be essential for accurate analyses of cosmological phenomena. Models of backreaction that analyze the effect of matter distribution inhomogeneities on cosmological dynamics \cite{Weigand_et_al, Rasanen_2006_accelerated} systematically incorporate inhomogeneity effects into cosmological analyses through various averaging schemes  \cite{Ellis1984, Futamase, Zalaletdinov1992, Zalaletdinov1993, Gasperini_2011}. Buchert introduced a particularly effective averaging procedure \cite{Buchert, Buchert2001} focusing on scalar quantities defined on space-like hypersurfaces. The Buchert averaging framework has subsequently been employed in numerous studies investigating how the backreaction of inhomogeneities affects cosmological dynamics, with particular attention to explaining the Universe's observed accelerated expansion \cite{Coley, Buchert, Buchert2001, Korzynski_2010, Clifton, Skarke, Buchert_2015, Buchert4, Weigand_et_al, Rasanen_2004, wiltshire, Kolb_2006, Koksbang_2019, Koksbang2, Koksbang_PRL, Rasanen_2008, bose, Bose2013, Ali_2017, Pandey_2022, Pandey2023, Koksbang5, Koksbang6, Koksbang7, Koksbang8, Koksbang2023}. Backreaction effects have also been applied to address other observational puzzles. For example, \cite{pandey_analyzing_2024} argues that the backreaction from the inhomogeneous matter distribution could explain the Experiment to Detect Global Epoch of reionization Signal (EDGES) 21-cm signal observations without requiring exotic physics. \cite{Mukherjee_2025} investigates how 21-cm brightness temperature can be utilized to constrain the Hubble parameter within the context of an inhomogeneous cosmological model.

\par In this context, it may be noted that a recent work related to Dark Energy Spectroscopic Instrument (DESI) baryon acoustic oscillation (BAO) measurements \cite{lodha2025extendeddarkenergyanalysis} have reported a past phantom-like dark energy ($\omega(z)<-1$, where $\omega(z)$ is the equation of state of the dark energy) transitioning to non-phantom-like ($\omega(z)>-1$) today. This is accompanied by a decrease in the universe's acceleration, possibly leading to deceleration. This analysis is further supported by a recent study of supernova cosmology \cite{supernova_deceleration}, which reports a significant tension ($>9\sigma$) with the $\Lambda$CDM model. Such observations of a decelerating Universe can be accounted for by backreaction models, without resorting to exotic or non-standard physics \cite{Halder_future_2023, Ali_2017}. Moreover, a two-parameter extension of the flat $\Lambda$CDM model that incorporates the effects of matter inhomogeneities on cosmological interpretations has been introduced recently \cite{backreaction_PRL}, suggesting the possibility of addressing the Hubble tension through backreaction from the cosmic web structure, which could reconcile early- and late-universe cosmological measurements.

\par Buchert's averaging procedure for evaluating backreaction effects from matter distribution inhomogeneities provides a promising theoretical framework for connecting spatially averaged quantities to observationally measurable phenomena, particularly redshift-distance relations \cite{rasanen1, rasanen2, Koksbang_2019, Koksbang2, Koksbang3, Koksbang4}. Studies of electromagnetic wave propagation through inhomogeneous spacetime have revealed distinctive modifications to the redshift-distance relation arising from backreaction effects \cite{Koksbang_2019, Koksbang2, Koksbang3}. Analogous phenomena have been demonstrated for gravitational waves from compact binary systems propagating through inhomogeneous matter distributions \cite{Pandey_2022, Pandey2023}. Previous studies have examined the effects of inhomogeneous matter distribution on the Hubble tension with varying degrees of success \cite{Hubble_tension_inhomogeneous1, Hubble_tension_inhomogeneous2, Hubble_tension_inhomogeneous3, Hubble_tension_inhomogeneous4, Hubble_tension_inhomogeneous5, Heinesen_Hubble_tension, Hubble_tension_PRL_schwarz}. Given these theoretical developments and their potential observational implications, we apply Buchert's averaging procedure to analyze another significant cosmological parametrization, {\it viz.},  the optical depth to reionization in the Universe with an inhomogeneous matter distribution.

\par The reionization of the Universe is closely linked to the formation of the first stars and galaxies, which are the primary sources of ionizing photons. Reionization affects our ability to measure the CMB radiation as it propagates through the intergalactic medium. The optical depth to reionization, $\tau_{reion}$, is a dimensionless quantity which provides a measure of the line-of-sight free-electron opacity to CMB radiation \cite{griffiths_tau}. Cosmic reionization is a topic of significant importance in both astrophysics and cosmology, and the parameter $\tau_{reion}$ plays a crucial role in its analysis. Determining $\tau_{reion}$ from observational data is therefore of strong interest \cite{tau_intro, tau_intro2, tau_intro3}, and may have a direct bearing on the tension between CMB and BAO measurements \cite{jhaveri_PRD_optical_depth, huang_BAO}. The Planck public-release PR3 results \cite{Planck} provide the optical depth and reionization redshift as $\tau_{reion} = 0.0544^{+0.0070}_{-0.0081}$, and $z_{reion} = 7.68\pm0.79$. This value of $\tau_{reion}$ is for the base-$\Lambda$CDM model and is obtained from the TT,TE,EE+lowE spectra and is weakly dependent on the cosmological model. In \cite{Planck}, the value of $z_{reion}$ is obtained by incorporating the tanh model of reionization with the $\tau_{reion}$ value (see \autoref{subsec:tau_re}). In the present study, we investigate $\tau_{reion}$ within a backreaction model.

\par In the present work, our approach considers a multidomain spacetime model that provides a more realistic representation of the actual Universe compared to the simplified two-domain toy models predominantly used in earlier backreaction studies \cite{bose, Bose2013, Ali_2017, Pandey_2022, Koksbang_2019}. The backreaction model considered here incorporates six parameters: four associated with matter distribution inhomogeneities, the dimensionless Hubble parameter $h$ (where $H_0 = 100\hspace{0.1 cm} h\hspace{0.1 cm} \text{km}\hspace{0.1 cm} \text{s}^{-1} \text{Mpc}^{-1}$), and $n$, representing the total number of subregions of each type. Following previous optimizations \cite{Halder_future_2023, pandey_analyzing_2024}, we adopt $n = 100$. We explore the backreaction model as an alternative to the standard $\Lambda$CDM cosmological framework. We perform a Markov Chain Monte Carlo (MCMC) analysis to constrain the parameters of the backreaction model, using a low-redshift observational dataset which includes PantheonPlus+SH0ES (Supernova H0 for the Equation of State) Type Ia supernova data. We constrain the parameter $h$, thereby placing constraints on $H_0$ and enabling us to address the Hubble tension. Based on these constrained model parameters, we evaluate the $\tau_{reion}$ value and compare it with values obtained for the $\Lambda$CDM model for various observational datasets.

\par The paper is organized as follows. We briefly introduce the cosmological models that have been employed in this analysis in \autoref{subsec:cosmo_model}. We also briefly outline Buchert's averaging procedure and our multi-domained backreaction model here. Next, we introduce a formalism to evaluate the optical depth to reionization in \autoref{subsec:tau_re}. We then present a multidomain analysis of the optical depth to reionization in the context of the multidomain backreaction model in \autoref{sec:param_space}. In \autoref{sec:obs_analysis}, we use the PantheonPlus+SH0ES type Ia supernova dataset to constrain the backreaction model parameters and calculate the optical depth to reionization using the optimal parameter values. Finally, we summarize our results in \autoref{sec:conclusions}.

\section{Model and Methodology}\label{sec:formalism}

\subsection{Background Cosmological Models}\label{subsec:cosmo_model}

\begin{enumerate}
\item \textit{$\Lambda$CDM model}:\\
\par The $\Lambda$CDM model is the current standard framework of cosmology. It assumes a homogeneous and isotropic Universe described by the FLRW metric, with dynamics governed by General Relativity. The energy budget consists of radiation, baryonic matter, cold dark matter, and a cosmological constant $\Lambda$ that drives late-time acceleration. The Hubble parameter in this model evolves with redshift as

\begin{equation}
    H_{\Lambda{\rm CDM}}(z) = H_0 \sqrt{\Omega_m (1+z)^3 + \Omega_r (1+z)^4 + \Omega_\Lambda},
\end{equation}

where $\Omega_m$, $\Omega_r$, and $\Omega_\Lambda$ are the present-day density parameters for matter, radiation, and dark energy, respectively.  Because of its simplicity and its success in reproducing a broad spectrum of observations, $\Lambda$CDM serves as the standard reference model against which alternative approaches, such as the backreaction framework, are evaluated.

\item \textit{ Backreaction model}:
\par In this analysis, we utilize Buchert's averaging method tailored to a pressure-less model (Buchert's backreaction formalism), explicitly referring to the dust universe model \cite{Buchert, buchert_rasanen}. Buchert's backreaction framework streamlines the averaging process by focusing solely on scalar quantities. The division of spacetime is accomplished through flow-orthogonal hypersurfaces characterized by the line element \cite{Buchert, Weigand_et_al}

\begin{equation}\label{eq:line_element}
    ds^2 = -dt^2+g_{ij}dX^idX^j,
\end{equation}

where $t$ is the proper time, $X^i$ are Gaussian normal coordinates in the hypersurfaces and $g_{ij}$ is the spatial three-metric of the hypersurfaces of constant $t$. The volume of a compact spatial domain $\mathcal{D}$ on these hypersurfaces is defined as,

\begin{equation}\label{eq:volD}
    |\mathcal{D}|_g := \int_{\mathcal{D}} d\mu_g
\end{equation}

where $d\mu_g:= \sqrt{\prescript{(3)}{}{g(t,X^1,X^2,X^3)}}dX^1dX^2dX^3$. Now, we define a dimensionless (`effective') scale factor

\begin{equation}\label{eq:scale_factor}
    a_{\mathcal{D}}(t) := \left(\frac{|\mathcal{D}|_g}{|\mathcal{D}_i|_g}\right)^{1/3},
\end{equation}

normalized by the volume of the initial domain $|\mathcal{D}_i|_g$, which can be considered the domain's current volume, at the present time $t_0$. The average over a scalar quantity $f$ is defined as,

\begin{equation}\label{eq:average_scalar}
    \langle f\rangle_{\mathcal{D}}(t) := \frac{\int_{\mathcal{D}}f(t,X^1,X^2,X^3)d\mu_g}{\int_{\mathcal{D}}d\mu_g}
\end{equation}

Using this averaging procedure and scalar parts of the Einstein equations, that is, the Hamiltonian constraint and the Raychaudhuri evolution equation for the expansion scalar, together with the continuity equation, gives us evolution equations \cite{Buchert},

\begin{eqnarray}\label{eq:aD_ddot}
    3\frac{\ddot{a_{\mathcal{D}}}}{a_{\mathcal{D}}} = -4\pi G\langle\rho\rangle_{\mathcal{D}} + \mathcal{Q}_{\mathcal{D}} \\
    3H^2_{\mathcal{D}} = 8\pi G\langle\rho\rangle_{\mathcal{D}} - \frac{1}{2}\langle\mathcal{R}\rangle_{\mathcal{D}} - \frac{1}{2}\mathcal{Q}_{\mathcal{D}} \label{eq:aD_dot}\\
    0 = \partial_t\langle\rho\rangle_{\mathcal{D}} + 3H_D\langle\rho\rangle_{\mathcal{D}}\label{eq:continuity}
\end{eqnarray}

where $\langle\rho\rangle_{\mathcal{D}}$, $\langle\mathcal{R}\rangle_{\mathcal{D}}$ and $H_{\mathcal{D}}$ are the averaged matter density, averaged spatial Ricci scalar and the Hubble parameter ($H_{\mathcal{D}}:= \dot{a_{\mathcal{D}}}/a_{\mathcal{D}}$) of the domain $\mathcal{D}$, respectively. $\mathcal{Q}_{\mathcal{D}}$ is called the kinematical backreaction and is defined as 

\begin{equation}\label{eq:Q_D}
    \mathcal{Q}_{\mathcal{D}}:= \frac{2}{3}(\langle\theta^2\rangle_{\mathcal{D}} - \langle\theta\rangle^2_{\mathcal{D}}) - 2\langle\sigma^2\rangle_{\mathcal{D}},
\end{equation}

where $\theta$ is the local expansion rate and $\sigma^2 := 1/2 \sigma_{ij}\sigma^{ij}$ is the squared rate of shear. The Hubble parameter $H_{\mathcal{D}}$ and $\langle\theta\rangle_{\mathcal{D}}$ are related by the relation $H_{\mathcal{D}} = 1/3\langle\theta\rangle_{\mathcal{D}}$. $\mathcal{Q}_{\mathcal{D}}$ is zero for an FLRW-like domain. The necessary condition of integrability connecting \autoref{eq:aD_ddot} and \autoref{eq:aD_dot} is given by,

\begin{equation}\label{eq:integrability}
    \frac{1}{a^2_{\mathcal{D}}}\partial_t(a^2_{\mathcal{D}}\langle\mathcal{R}\rangle_{\mathcal{D}}) + \frac{1}{a^6_{\mathcal{D}}}\partial_t(a^6_{\mathcal{D}}\mathcal{Q}_{\mathcal{D}}) = 0.
\end{equation}

\par In \autoref{eq:integrability}, a key aspect of the averaged equations is illustrated; it links the evolution of the averaged intrinsic curvature $(\langle\mathcal{R}\rangle_{\mathcal{D}})$ to the kinematic backreaction term $(\mathcal{Q}_{\mathcal{D}})$. This term represents the incorporation of matter inhomogeneities into the analysis. The interdependence of $\langle\mathcal{R}\rangle_{\mathcal{D}}$ and $\mathcal{Q}_{\mathcal{D}}$, alongside $\mathcal{Q}_{\mathcal{D}}$, signifies the deviation from homogeneity.

\par In the framework of the Buchert formalism, we adopt a particular method where groups of disjoint regions collectively define the global domain \cite{Buchert_2015, Buchert4, Weigand_et_al, Rasanen_2004, wiltshire, Kolb_2006, Koksbang_2019, Koksbang2, Koksbang_PRL, Rasanen_2008, bose, Bose2013, Ali_2017, Pandey_2022, Pandey2023, Koksbang5, Koksbang6, Koksbang7, Koksbang8}. In this context, the global domain $\mathcal{D}$ is envisioned as being divided into subregions $\mathcal{F}_l$, each of which is composed of more fundamental spatial units $\mathcal{F}_l^{(\alpha)}$. Thus, in mathematical terms, we express this as $\mathcal{D} = \cup_l\mathcal{F}_l$, where $\mathcal{F}_l:=\cup_{\alpha}\mathcal{F}^{(\alpha)}_l$ and $\mathcal{F}_l^{(\alpha)}\cap \mathcal{F}_m^{(\beta)} = \emptyset$ for all $\alpha\neq\beta$ and $l\neq m$.

\par Average of a scalar-valued function $f$ on the domain $\mathcal{D}$ is given by,

\begin{equation}\label{eq:averaging}
\begin{split}
    \langle f\rangle :&= |\mathcal{D}|^{-1}_g \int_\mathcal{D} fd\mu_g
    = \sum_l |\mathcal{D}|_g^{-1}\sum_{\alpha}\int_{\mathcal{F}_l^{(\alpha)}} f d\mu_g\\
    &= \sum_l\frac{|\mathcal{F}_l|_g}{|\mathcal{D}|_g}\langle f\rangle_{\mathcal{F}_l} = \sum_l \lambda_l\langle f\rangle_{\mathcal{F}_l}
\end{split}
\end{equation}

where 

\begin{equation}\label{eq:lambda}
    \lambda_l := \frac{|\mathcal{F}_l|_g}{|\mathcal{D}|_g}
\end{equation}

is the volume fraction of the subregion $\mathcal{F}_l$ such that $\sum_l\lambda_l = 1$ and $\langle f\rangle_{\mathcal{F}_l}$ is the average of $f$ on subregion $\mathcal{F}_l$. The scalar quantities $\rho$, $\mathcal{R}$, and $\mathcal{H}_{\mathcal{D}}$ are governed by \autoref{eq:averaging}, but $\mathcal{Q}_{\mathcal{D}}$, due to the presence of $\langle\theta\rangle_{\mathcal{D}}^2$ - term, do not adhere to the above equation and instead follow,

\begin{equation}\label{eq:QDsum}
    \mathcal{Q}_{\mathcal{D}} = \sum_l\lambda_l\mathcal{Q}_l + 3\sum_{l\neq m}\lambda_l\lambda_m(H_l-H_m)^2
\end{equation}

where $\mathcal{Q}_l$ and $\mathcal{H}_l$ are defined in subregion $\mathcal{F}_l$ in the same way as $\mathcal{Q}_{\mathcal{D}}$ and $\mathcal{H}_{\mathcal{D}}$ are defined in the domain $\mathcal{D}$. We can also define the scale factor $a_l$ for a subregion $\mathcal{F}_l$. By definition, the different subregions are disjoint; therefore it follows that $|\mathcal{D}|_g = \sum_l|\mathcal{F}_l|_g$ and hence, using \autoref{eq:scale_factor}, we have 

\begin{equation}\label{eq:aD3_sum}
    a^3_{\mathcal{D}} = \sum_l\lambda_{l_i}a_l^3
\end{equation}

where $\lambda_{l_i} = \frac{|\mathcal{F}_{l_i}|_g}{|\mathcal{D}_i|_g}$ is the initial volume fraction which can also be taken as the volume fraction at present and can be represented as $\lambda_{l_0}$, where the subscript $0$ stands for quantities calculated at the present time. Differentiating this relation twice with respect to the foliation time results in,

\begin{equation}\label{eq:aD_sum}
    \frac{\ddot{a}_D}{a_D} = \sum_l\lambda_l\frac{\ddot{a}_l(t)}{a_l(t)}+\sum_{l\neq m}\lambda_l\lambda_m(H_l -H_m)^2 \, .
\end{equation}

\textit{\textbf{A model of multiple subregions}}:

\par Here, using Buchert's backreaction framework, we consider a model of the Universe in which domain $\mathcal{D}$ comprises multiple underdense and overdense subregions. Similar models have been used in \cite{Halder_future_2023} to study the future evolution of the accelerating universe with multiple inhomogeneous domains and in \cite{pandey_analyzing_2024} to analyze the 21-cm signal in the Universe with inhomogeneities. The underdense subregions have lower densities than the overdense subregions. The underdense subregions are modeled to mimic almost-empty matter-dominated FLRW regions with very little matter (dust). The overdense subregions are modeled to mimic matter-dominated FLRW models with matter (dust) content. The underdense subregions are taken to have Friedmann-like $1/a^2$ negative curvature, while the overdense subregions have Friedmann-like $1/a^2$ positive curvature. The time evolution of the scale factor of $i^{th}$ overdense subregions, $a_{o_i}$ is given in terms of a development angle $\phi_{o_i}$ of the $i^{th}$ overdense subregion \cite{weinberg},

\begin{eqnarray}
    a_{o_i} &=& \frac{q_{o_{i,0}}}{2q_{o_{i,0}} - 1}(1 - \cos{\phi_{o_i}})\label{eq:ao}\\
    t &=& t_0\frac{(\phi_{o_i} - \sin{\phi_{o_i}})}{(\phi_{o_{i,0}} - \sin{\phi_{o_{i,0}}})} \label{eq:ao_t}
\end{eqnarray}

where $q_{o_{i,0}}$ and $\phi_{o_{i,0}}$ are respectively the deceleration parameter of the $i^{th}$ overdense subregion and the value of $\phi_{o_i}$ at time $t_0$,  which is the present time. For an overdense region such as the one considered here, $q_{o_{i,0}}$ must be greater than $1/2$ \cite{weinberg}. Here, we have set $q_{o_{i,0}}$ to take values between $1/2$ and $1$. The time $t$ in \autoref{eq:ao_t} is the cosmic time, although for each overdense subregion, this $t$ is a function of $(\phi_{o_{i,0}}, \phi_{o_i})$ ($\phi_{o_{i,0}}$ itself is a function of $q_{o_{i,0}}$). Therefore, each underdense subregion has a different evolution of $t$ as a function of $(q_{o_{i,0}}, \phi_{o_i})$ but value of $t_0$ is the same across all overdense subregions as well as for the global domain, which is ensured by the specific form of \autoref{eq:ao_t}. The time evolution of the scale factor of $i^{th}$ underdense subregions, $a_{u_i}$ is given in terms of a development angle $\phi_{u_i}$ of the $i^{th}$ underdense subregion \cite{weinberg},

\begin{eqnarray}
    a_{u_i} &=& \frac{q_{u_{i,0}}}{1 - 2q_{u_{i,0}}}(\cosh{\phi_{u_i}} - 1)\label{eq:au}\\
    t &=& t_0\frac{(\sinh{\phi_{u_i}} - \phi_{u_i})}{(\sinh{\phi_{u_{i,0}}} - \phi_{u_{i,0}})} \label{eq:au_t}
\end{eqnarray}

where $q_{u_{i,0}}$ and $\phi_{u_{i,0}}$ are, respectively, the deceleration parameter of the $i^{th}$ underdense subregion and the value of $\phi_{u_i}$ at time $t_0$,  which is the present time. For an underdense region such as the one considered here, $q_{u_{i,0}}$ ranges from 0 to 1/2 \cite{weinberg}. The time $t$ in \autoref{eq:au_t} is the cosmic time, although for each underdense subregion, this $t$ is a function of $(\phi_{u_{i,0}}, \phi_{u_i})$ ($\phi_{u_{i,0}}$ itself is a function of $q_{u_{i,0}}$). Therefore, each underdense subregion has a different evolution of $t$ as a function of $(q_{u_{i,0}}, \phi_{u_i})$, but the value of $t_0$ is the same across all underdense and overdense subregions as well as for the global domain which is ensured by the specific form of \autoref{eq:au_t} and \autoref{eq:ao_t}. Since the values of $t_0$ and $H_{\mathcal{D}_0}$ are interrelated, one needs to fix either of them \cite{Koksbang2023}. The value of $t_0$ is calculated using the procedure used in \cite{pandey_analyzing_2024}.


\par Note that $a_{o_i}$ and $a_{u_i}$ can be expressed in terms of the volume of the respective subregions using \autoref{eq:scale_factor}, which gives us

\begin{equation}\label{eq:scale_fact_subreg}
    a_{o_i}(t) := \left(\frac{|\mathcal{D}|_{o_i}}{|\mathcal{D}_0|_{o_i}}\right)^{1/3}; \hspace{0.5 cm} a_{u_i}(t) := \left(\frac{|\mathcal{D}|_{u_i}}{|\mathcal{D}_0|_{u_i}}\right)^{1/3}
\end{equation}

where, $|\mathcal{D}|_{o_i}$ is the volume of the $i^{th}$ overdense subregions at time $t$, $|\mathcal{D}_0|_{o_i}$ is the volume of the $i^{th}$ overdense subregion at time $t_0$ as was done in \autoref{eq:scale_factor}, and similarly for the case of the underdense subregions.
\autoref{eq:scale_factor} and \autoref{eq:scale_fact_subreg} require that at $t = t_0$, $a_\mathcal{D}$ = $a_{o_i}$ = $a_{u_i} = 1$, leading to,

\begin{equation}\label{eq:c_values}
    \cos{\phi_{o_{i,0}}} = \left(\frac{1}{q_{o_{i,0}}} - 1\right); \hspace{0.5 cm} \cosh{\phi_{u_{i,0}}} = \left(\frac{1}{q_{u_{i,0}}} - 1\right)
\end{equation}

\par For a given value of $q_{o_{i,0}}$ and $q_{o_{i,0}}$; $a_{o_i}(t)$ and $a_{u_i}(t)$ can be calculated using \autoref{eq:ao}, \autoref{eq:ao_t}, \autoref{eq:au}, \autoref{eq:au_t} and \autoref{eq:c_values}. Then using \autoref{eq:aD3_sum}, $a_\mathcal{D}(t)$ can be obtained provided $\lambda_{l_0}$ which is the set of all $\lambda_{{u_i},0}$ and $\lambda_{{o_i},0}$, is known.

\par It may be noted that $a_\mathcal{D}$ can also be obtained from solving the second-order differential equation \autoref{eq:aD_sum}. Using \autoref{eq:ao} and \autoref{eq:au} in \autoref{eq:aD_sum}, we get,

\begin{equation}\label{eq:aD_model}
\begin{split}
\frac{\ddot{a}_{\mathcal{D}}}{a_{\mathcal{D}}}=\left(\sum_i{\lambda_{o_i}\dfrac{\ddot{a}_{o_i}}{a_{o_i}}}\right)+\left(
\sum_j{\lambda_{u_j}\dfrac{\ddot{a}_{u_i}}{a_{u_i}}}\right)
+\left(\sum_{k}\sum_{l}{ \lambda_k \lambda_l \left(H_l-H_k\right)^2}\right).
\end{split}
\end{equation}

where, $\lambda_{o_i}$ is the volume fraction of the $i^{th}$ overdense subregion, $\lambda_{u_j}$ is the volume fraction of the $j^{th}$ underdense subregion, $\lambda$ is the set of all $\lambda_{o_i}$ and $\lambda_{u_i}$ and $H$ is respectively, the set of all $H_{o_i}$ and $H_{u_i}$. The combined volume fraction of all the underdense subregions is given by $\lambda_u$, i.e., $\sum_i\lambda_{u_i} = \lambda_u$. Similarly, the total volume fraction of all the overdense subregions is given by $\sum_i\lambda_{o_i} = \lambda_o$. Clearly, $\lambda_o +\lambda_u = 1$. The evaluation of $a_\mathcal{D}$ obtained from these two methods is identical, as confirmed through our analysis.

\par The volume fraction of the $i^{th}$ overdense subregion can be written as,

\begin{equation}\label{eq:lambda_o_relation}
\begin{split}
    \lambda_{o_i} = \frac{|\mathcal{F}_{o_i}|_g}{|\mathcal{D}|_g}= \frac{a^3_{o_i}|\mathcal{F}_{o_i,0}|_g}{a^3_\mathcal{D}|\mathcal{D}_0|_g} = \lambda_{{o_i},0}\frac{a^3_{o_i}}{a^3_\mathcal{D}}
\end{split}
\end{equation} 

where $t_0$ is a reference time which can be taken as the present time, $|\mathcal{F}_{o_i}|_g$ is the volume of the $i^{th}$ overdense subregion, $|\mathcal{F}_{o_i,0}|_g$ is the volume of the $i^{th}$ overdense subregion at time $t_0$, $|\mathcal{D}_0|_g$ is the volume of the domain $\mathcal{D}$ at time $t_0$ and $\lambda_{{o_i},0}$ is the volume fraction of the $i^{th}$ overdense subregion at time $t_0$. The present time ($t_0$) value of $(\lambda_o,\lambda_u)$ is given by $(\lambda_{o,0},\lambda_{u,0})$ which we have taken to be (0.09,0.91) \cite{Weigand_et_al}.

\par In this backreaction model, we consider the present time volume fraction of $i^{th}$ underdense subregion, $\lambda_{{u_i},0}$ to have a Gaussian distribution within the allowed range of $q_{u_{i,0}}$ from $0$ to $1/2$, given by, 

\begin{equation}
    \lambda_{{u_i},0} = \dfrac{N_u}{\sigma_u \sqrt{2 \pi}} e^{-(q_{u_{i,0}}-\mu_u)^2/2\sigma_u^2},\label{eq:gauss_u}
\end{equation}

where $N_u$ is a normalization constant which ensures that $\sum_i\lambda_{{u_i},0} = \lambda_{u,0} = 0.91$, $\mu_u$ is the mean value of $q_{u_{i,0}}$ and $\sigma_u$ is the standard deviation of $q_{u_{i,0}}$. Therefore, each $i^{th}$ underdense subregion is associated with a particular value of $q_{u_{i,0}}$ and $\lambda_{{u_i},0}$ such that $q_{u_{i,0}}$ varies from $0$ to $1/2$ across the $i$ number of underdense subregions and $\sum_i\lambda_{{u_i},0}  = 0.91$. Hence, this model's allowed range for $\mu_u$ is also from $0$ to $1/2$. And we are imposing the allowed range for $\sigma_u$ from $0.01$ to $0.09$.

\par The present-time volume fraction of $i^{th}$ overdense subregion, $\lambda_{{o_i},0}$ is considered to have a Gaussian profile within the allowed range of $q_{o_{i,0}}$ from $1/2$ to $1$ given by, 

\begin{equation}
    \lambda_{{o_i},0} = \dfrac{N_o}{\sigma_o \sqrt{2 \pi}} e^{-(q_{o_{i,0}}-\mu_o)^2/2\sigma_o^2}\label{eq:gauss_o},
\end{equation}

where $N_o$ is a normalization constant which ensures that $\sum_i\lambda_{{o_i},0} = \lambda_{u,0} = 0.09$, $\mu_o$ is the mean value of $q_{o_{i,0}}$ and $\sigma_o$ is the standard deviation of $q_{o_{i,0}}$. In this case, each $i^{th}$ overdense subregion is associated with a particular value of $q_{o_{i,0}}$ and $\lambda_{{o_i},0}$, where $q_{o_{i,0}}$ lies within the range $1/2$ to $1$ across the $i$ number of overdense subregions and $\sum_i\lambda_{{o_i},0}  = 0.09$. Hence, this model's allowed range for $\mu_o$ is also from $1/2$ to $1$. And we are imposing the allowed range for $\sigma_o$ from $0.01$ to $0.09$. The volume fraction of the $i^{th}$ underdense subregion at a time $t$, $\lambda_{u_i}$ is related to the volume fraction at present time $t_0$ by,

\begin{equation}\label{eq:lambda_u_i}
    \lambda_{u_i} = \lambda_{{u_i},0}\left(\dfrac{1-\sum_i \lambda_{o_i}}{1-\sum_i \lambda_{{o_i},0}}\right),
\end{equation}

\par We employed the Gaussian distribution to characterize the current volume fraction for different subregions. To gain a genuine understanding of the physical distribution, extensive galactic surveys are required to map how matter is spread throughout the Universe. While surveys have examined the local Universe, no studies have been conducted for the redshifts pertinent to our analysis. In the absence of these surveys, we assume a normal distribution, as it is a standard choice known to model unbiased physical conditions effectively. The Gaussian distribution is a familiar and widely utilized tool in various physical studies.

\par Using \autoref{eq:QDsum}, the kinematical backreaction term for the domain $\mathcal{D}$ for the backreaction model effectively becomes 

\begin{equation}\label{eq:QDsum2}
    \mathcal{Q}_{\mathcal{D}} = \sum_i \lambda_{o_i}\mathcal{Q}_{o_i} + \sum_j \lambda_{u_j}\mathcal{Q}_{u_j} + 3\sum_{l\neq m}\lambda_l\lambda_m(H_l-H_m)^2   \, , 
\end{equation}

where $\mathcal{Q}_{o_i}$ is the kinematical backreaction term for the $i^{th}$ overdense subregion, $\mathcal{Q}_{u_i}$ is for the $i^{th}$ underdense subregion. The summation in the last term runs over the sets of all  $\lambda_{o_i}$, $\lambda_{u_i}$,  $H_{o_i}$, and $H_{u_i}$. The subregions are also governed by this coupling of the kinematical backreaction term to the Ricci scalar (\autoref{eq:integrability}). Therefore, by selectively choosing the curvatures of the subregions (FLRW-like), we can make the respective kinematical backreaction terms for these subregions equal to zero \cite{Weigand_et_al, Rasanen_2006_accelerated}. Hence, in this case, the global kinematical backreaction is governed by only the interplay of the sub-domain Hubble evolutions and volume fractions (third term of \autoref{eq:QDsum2}). Note that the above assumptions are made in the context of the present model. On the other hand, if the subdomains are endowed with dynamical curvature, other intricate effects could arise through kinematical backreaction, as may also happen in a more general case where the subregions may not necessarily be FLRW. Obtaining the values of $\lambda_{o_i,0}$ and $\lambda_{u_i,0}$ from \autoref{eq:gauss_o} and \autoref{eq:gauss_u} respectively, and using these in \autoref{eq:lambda_o_relation} and \autoref{eq:lambda_u_i} gives us $\lambda_{o_i}$ and $\lambda_{u_i}$. Hubble parameters for the subregions can be obtained from \autoref{eq:ao}, \autoref{eq:ao_t}, and \autoref{eq:au}, \autoref{eq:au_t}. We can then use \autoref{eq:aD_model} to get $a_\mathcal{D}(t)$ and $H_{\mathcal{D}}(t)$. 

\par We next relate these quantities  calculated theoretically from the backreaction model with observational quantities (redshift and angular diameter distance by using the covariant scheme \cite{rasanen1, rasanen2}, given by,

\begin{align}
    1+z &= \frac{1}{a_\mathcal{D}}\label{eq:covariant_sch_1}\\
    H_\mathcal{D}\frac{d}{dz}\left((1+z)^2H_\mathcal{D}\frac{dD_A}{dz}\right) &= -4\pi G\langle\rho\rangle_\mathcal{D} D_A.\label{eq:covariant_sch_2}
\end{align}

\autoref{eq:covariant_sch_1} relates $a_\mathcal{D}(t)$ with the cosmological redshift $z(t)$ and \autoref{eq:covariant_sch_2} relates the angular diameter distance $D_A$ with $\langle\rho\rangle_\mathcal{D}$ and $H_{\mathcal{D}}$. Here, we use \autoref{eq:covariant_sch_1} to obtain $z(t)$ from $a_\mathcal{D}(t)$. We can thus evaluate $H_{\mathcal{D}}(z)$ using  $H_{\mathcal{D}}(t)$ (from \autoref{eq:aD_model}) and $z(t)$ (from \autoref{eq:covariant_sch_1}). This particular model has been characterized thoroughly in \cite{pandey_analyzing_2024}.
\end{enumerate}

\subsection{Formulation of Optical Depth}\label{subsec:tau_re}

The optical depth $\tau$ is defined as the integral of the electron density times the Thomson cross section over the geometrical path length between the present redshift and the redshift of reionization. It can be expressed as \cite{griffiths_tau}

\begin{equation}
    \tau(t) = \sigma_T \int_t^{t_0} n_e(t) c \, dt, \label{eq:tau_t}
\end{equation}

where $n_e$ is the electron number density, $\sigma_T$ is the Thomson scattering cross-section, and $t_0$ is the present time. 

\par To express this in terms of the redshift, we define the ionization fraction as $\chi(z) = n_e(z)/n_H(z)$ \cite{Planck}, where $n_H(z)$ is the number density of hydrogen nuclei. The number density of hydrogen nuclei evolves as

\begin{equation}
    n_H(z) = n_{H,0}(1+z)^3, \label{eq:np}
\end{equation}

where $n_{H,0}$ is the present-day value of $n_H(z)$. The time-redshift relation is

\begin{equation}
    \frac{dz}{dt} = -(1+z)H(z). \label{eq:dzdt}
\end{equation}

Substituting into \autoref{eq:tau_t}, the optical depth becomes

\begin{equation}\label{eq:tau_z}
    \tau(z) = n_{H,0} c \sigma_T \int_0^z (1+z')^2 \frac{\chi(z')}{H(z')} \, dz'.
\end{equation}

\par Assuming a primordial helium fraction of $24\%$ \cite{He_abundance}, we have $n_H(z) = 0.76\hspace{0.05cm} n_B(z)$, where $n_B(z)$ is the baryon number density. Now, $n_{H,0} = 0.76\hspace{0.05cm}n_{B,0}$, where $n_{B,0}$ is the present-day value of $n_B(z)$ and is related to the baryon density parameter $\Omega_B$ via

\begin{equation}
    \Omega_B = \frac{8\pi G}{3H_0^2} m_p n_{B,0}, \label{eq:omegaB}
\end{equation}

where $m_p$ is the proton mass and $H_0 = 100 h$ km s$^{-1}$ Mpc$^{-1}$ is the Hubble constant. For simplicity, both helium and hydrogen are assumed to be fully ionized. 

\par We denote the total integrated optical depth to reionization as $\tau_{reion}$, and it is given by \cite{Planck}

\begin{equation}\label{eq:final_tau}
    \tau_{reion} = n_{H,0} c \sigma_T \int_0^{z_{max}} (1+z')^2 \frac{\chi(z')}{H(z')} \, dz'.
\end{equation}

where $z_{max}=50$, which is early enough to capture the full expected contribution from reionization.

\par The computation of the $\tau_{reion}$ (the integral in \autoref{eq:final_tau}) relies on two fundamental ingredients: (i) the evolution of the Hubble parameter \(H(z)\), determined by the underlying cosmological model, and (ii) the ionization fraction \(\chi(z)\), specified by the reionization model. In this work, these two components are consistently combined within a unified pipeline: the cosmological sector provides the background expansion history as an input to the integral, while the reionization module computes the corresponding ionization fraction. The two parts are interfaced at the background evolution level, ensuring a coherent evaluation of the optical depth.\\  We consider two cosmological scenarios:

\begin{itemize}
    \item the standard $\Lambda$CDM model, and
    \item the multi-domained backreaction model.
\end{itemize}

In both cases, the background expansion history is combined with the $\tanh$ reionization parameterization \cite{Planck,Lewis2008} to model the ionization fraction and compute the corresponding optical depth.

\par Each of these cosmological models is combined with the $\tanh$ reionization model \cite{Planck,Lewis2008}:

\begin{equation}\label{eq:tanh_model}
\chi(z) = \frac{1+n_{He}/n_H}{2} \left[ 1 + \tanh\left( \frac{y(z_{reion}) - y(z)}{\Delta y} \right) \right],
\end{equation}
 where $y(z)=(1+z)^{3/2}$, $\Delta y=\tfrac{3}{2}(1+z_{reion})^{1/2}\Delta z$, and $\Delta z=0.5$. Throughout this study, we have used $z_{reion} = 7.68$, unless otherwise stated (see \autoref{subsec:tau_LCDM}).
\vspace{0.5 cm}

\par We then analyze two distinct scenarios:
\[
(\Lambda{\rm CDM} + \tanh), \quad
({\rm backreaction} + \tanh).
\]

Using $H_\mathcal{D}(z)$ derived from the backreaction formalism and $\chi(z)$ from the reionization prescription, we compute $\tau_{reion}$ via \autoref{eq:final_tau}. This calculation is performed for both cosmological models ($\Lambda$CDM and backreaction), each paired with the $\tanh$ reionization prescription. Hence, two distinct values of $\tau_{reion}$ are obtained and compared with the observational results.

\subsection{Observational values of Optical Depth to Reionization}\label{subsec:tau_LCDM}

In the Planck collaboration PR2 release \cite{Planck_2015}, 68$\%$ confidence limits for the base $\Lambda$CDM model from Planck CMB power spectra, in combination with lensing reconstruction ``lensing") and external data ``ext", BAO+JLA+$H_0$), are given as (here, tanh prescription of ionization fraction is not used)

\begin{equation}
    \tau_{reion} = 0.066\pm0.012, \hspace{0.25 cm} \text{and} \hspace{0.25 cm} z_{reion} = 8.8^{+1.2}_{-1.1} \hspace{0.5 cm} (\text{68$\%$ TT,TE,EE+lowP+lensing+ext}),
\end{equation}

\par The Planck 2018 legacy release (Planck Collaboration V 2020) \cite{Planck_V_2018} cites $\tau = 0.063\pm 0.020$ as inferred from Low Frequency Instrument (LFI) data and 0.0506$\pm$0.0086 from High Frequency Instrument (HFI) data.

\par In the Planck collaboration PR3 release (Planck Collaboration VI 2020) \cite{Planck}, the optical depth is well constrained by the large-scale polarization measurements from the Planck HFI, with the joint constraint

\begin{equation}
    \tau_{reion} = 0.0544^{+0.0070}_{-0.0081} \hspace{0.5 cm} (\text{68$\%$ TT,TE,EE+lowE}),\label{eq:Planck_6_tau}
\end{equation}

and with the tanh reionization prescription, this implies a mid-point redshift of reionization

\begin{equation}
    z_{reion} = 7.68\pm0.79 \hspace{0.5 cm} (\text{68$\%$ TT,TE,EE+lowE})\label{eq:Planck_6_z}
\end{equation}

In the same release, the base-$\Lambda$CDM model from Planck CMB power spectra, in combination with CMB lensing reconstruction and Baryon Acoustic Oscillation (BAO) \cite{Planck}, gives

\begin{equation}
    \tau_{reion} = 0.0561\pm0.0071, \hspace{0.25 cm} \text{and} \hspace{0.25 cm} z_{reion} = 7.82\pm0.71 \hspace{0.5 cm} (\text{68$\%$ TT,TE,EE+lowE+lensing+BAO}).
\end{equation}

\par Assuming the $\Lambda$CDM cosmological model and combining the Planck large-scale temperature likelihood with the high-\textit{l} temperature and polarization likelihood, \cite{tau_intro2} derives a value of $\tau_{reion}=0.059\pm0.006$ (68$\%$ confidence level), which implies a reionization mid-point (assuming tanh parametrization) at $z_{reion}=8.14\pm0.61$ (68$\%$ confidence level).

\par Through further reprocessing of the Planck large-scale polarization measurements, \cite{Planck_Int_LVII} obtains a revised value for the reionization optical depth of $\tau_{reion}=0.051\pm0.006$.

\par Relative to the previous Planck cosmological analysis in \cite{Planck}, the constraints reported in \cite{tristram_planck} are more stringent, without any significant shifts in the central values. The reionization optical depth is now measured to a precision of a few percent, yielding $\tau_{reion} = 0.058\pm0.006$.

\par In \cite{Li_2025}, estimates based on the Cosmology Large Angular Scale Surveyor (CLASS) 90 GHz data correlated with Planck data (Planck PR4, \cite{Planck_Int_LVII}) give $\tau_{reion} = 0.058^{+0.018}_{-0.019}$. In \cite{Louis_2025}, Atacama Cosmology Telescope (ACT) DR6 data in combination with Planck, lensing, and BAO data give $\tau_{reion} = 0.0632^{+0.0055}_{-0.0066}$.

\par These observational constraints are shown in \autoref{fig:optical_depth_final} for comparison. For the purposes of our analysis, we adopt as fiducial reference values those reported in \autoref{eq:Planck_6_tau} and \autoref{eq:Planck_6_z}, which serve as the baseline constraints throughout this work.

\section{Exploration of Parameter Space}\label{sec:param_space}

\par Before performing a complete statistical comparison with observational data, we first explore the sensitivity of the optical depth of reionization $\tau_{reion}$ to variations in the parameters of the backreaction model. This preliminary analysis allows us to identify the approximate ranges of parameters that yield values of $\tau_{reion}$ compatible with Planck Collaboration VI 2020 constraints ($\tau_{reion} = 0.0544^{+0.0070}_{-0.0081}$) \cite{Planck}. 

\par
\begin{figure*}
    \centering
    \begin{tabular}{cc}
	\includegraphics[width=0.5\textwidth]{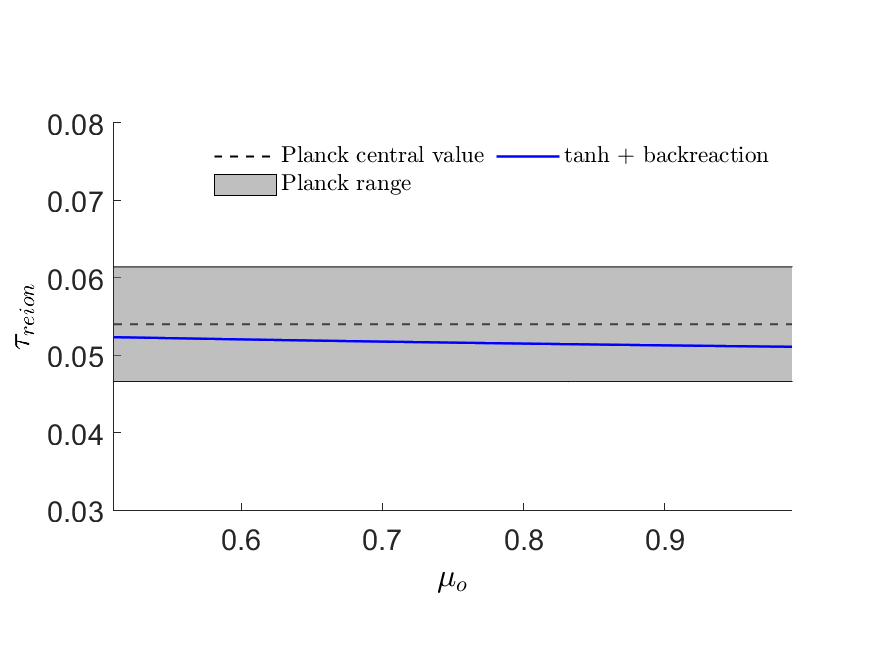}&
	\includegraphics[width=0.5\textwidth]{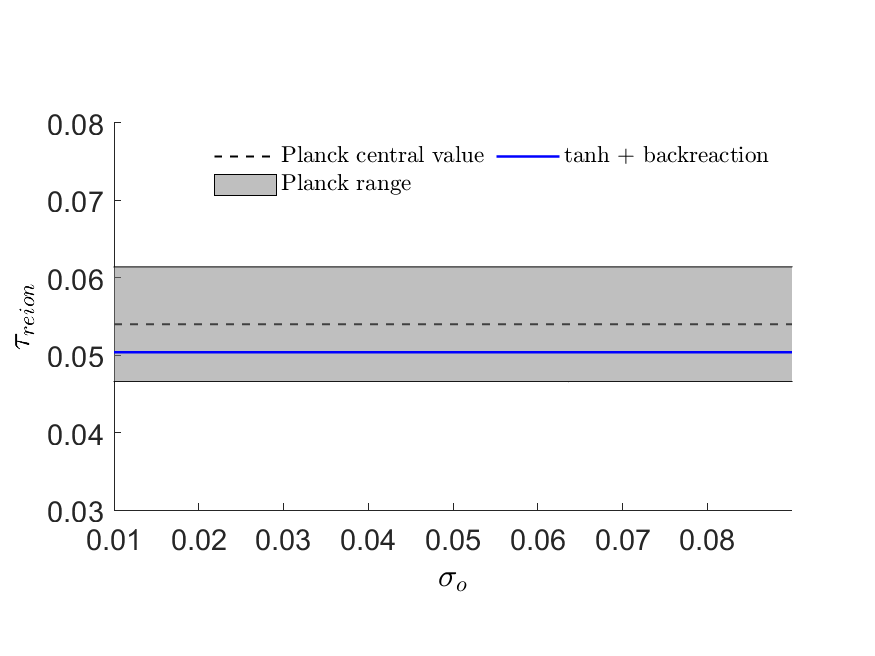}\\
	(a)&(b)\\
	\includegraphics[width=0.5\textwidth]{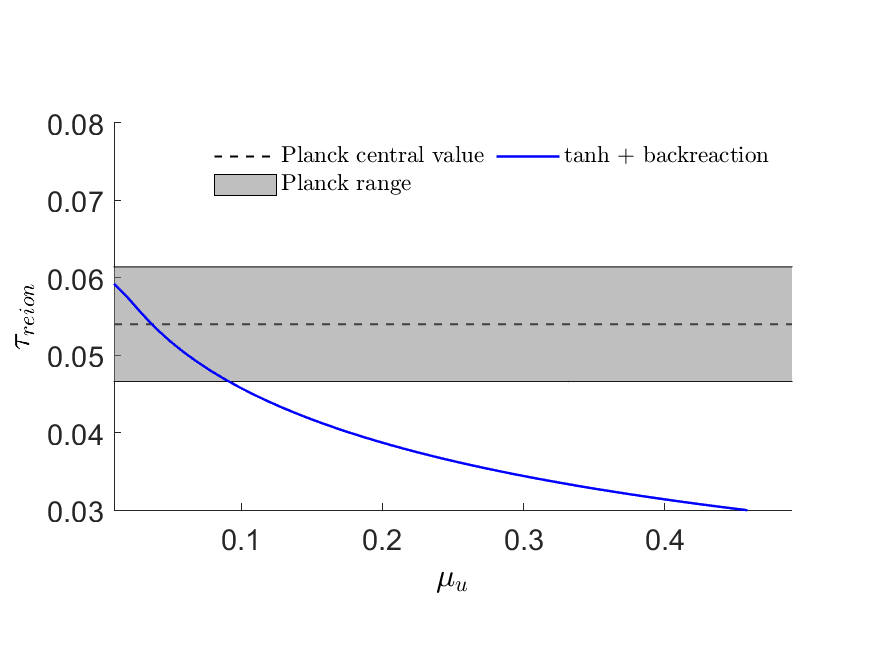}&
	\includegraphics[width=0.5\textwidth]{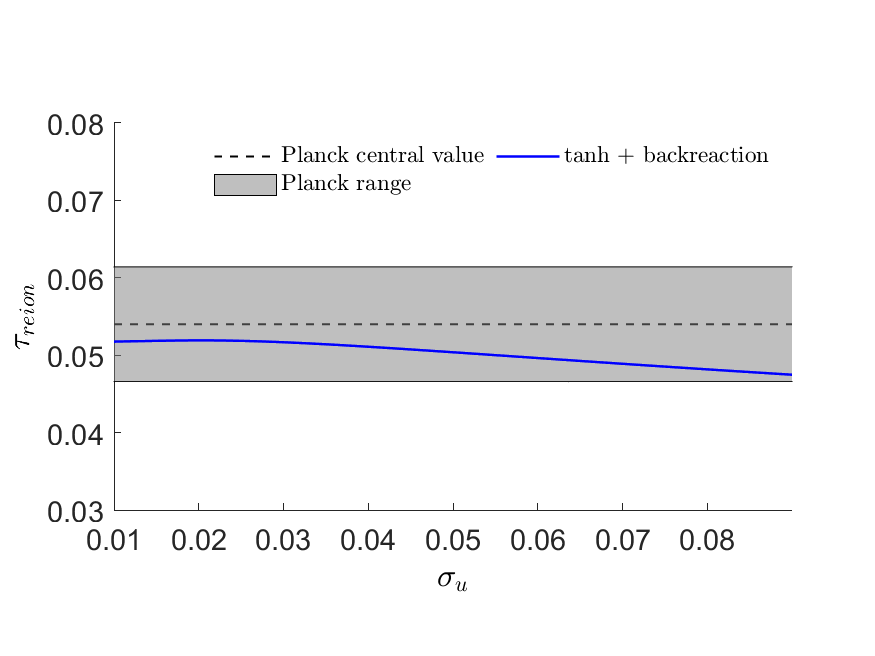}\\
	(c)&(d)\\
    \end{tabular}
\caption{\label{fig:tau_mu_sigma_vary} Plots of $\tau_{reion}$ for the backreaction model as a function of $z$. The backreaction model has six parameters. We have fixed two, $n = 100$ and $h = 0.7$. The other four parameters that can be varied are: $\mu_u, \sigma_u, \mu_o, \sigma_o$. In (a) $\mu_o$ is varied with $\sigma_o = \sigma_u = 0.01$ and $\mu_u = 0.09$ fixed. In (b) $\sigma_o$ is  varied with $\mu_o = 0.99, \sigma_u = 0.01$ and $\mu_u = 0.08$ fixed. In (c), $\mu_u$ is varied with $\sigma_o = \sigma_u = 0.01$ and $\mu_o = 0.95$ fixed. In (d), $\sigma_u$ is varied with $\sigma_o = 0.01, \mu_o = 0.99$ and $\mu_u = 0.07$ fixed. The value of $H_0$ (Hubble parameter at present) used is 100 $h$ km $\rm{s^{-1} Mpc^{-1}}$ where $h = 0.7$. The dashed black line and the grey band show the Planck reported value of $0.054$ and the 68$\%$ confidence interval. We have used $z_{reion} = 7.68$.}
\end{figure*}

\begin{figure*}
    \centering
    \begin{tabular}{cc}
	\includegraphics[width=0.5\textwidth]{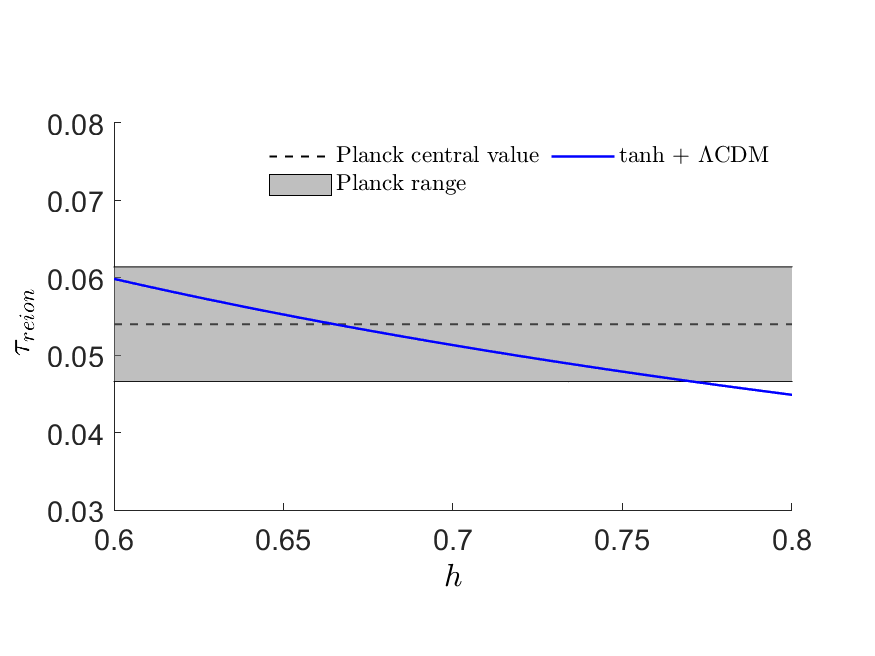}&
	\includegraphics[width=0.5\textwidth]{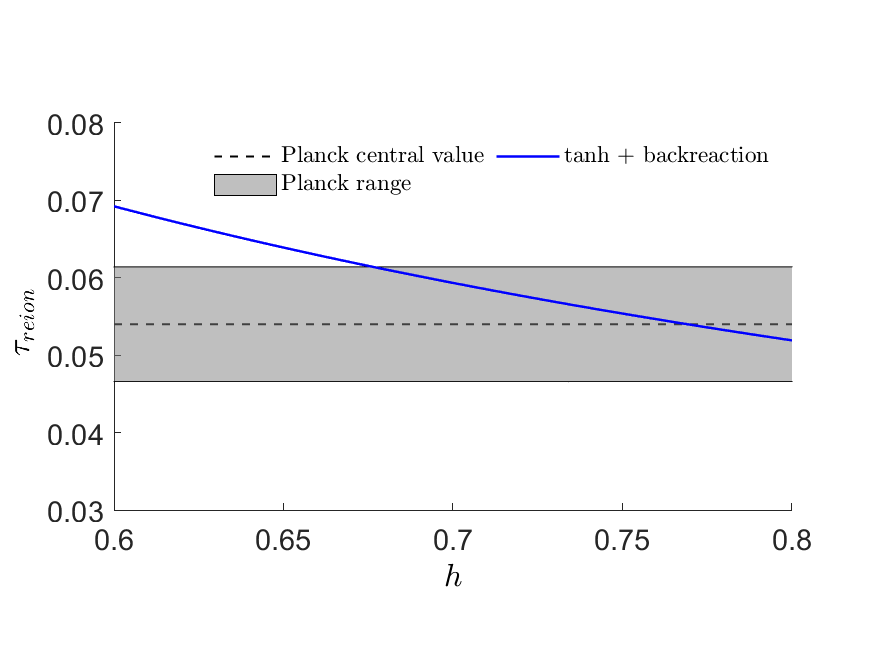}\\
	(a)&(b)
    \end{tabular}
\caption{\label{fig:tau_h0_vary} Subplot (a) is the plot of $\tau$ for varying values of $h$ for $\Lambda$CDM model. Subplot (b) is for the backreaction model. The grey dashed line shows the Planck reported value of $0.054$. In subplot (b), the value of the four backreaction parameters $(\mu_u,\sigma_u,\mu_o,\sigma_o)$ has been fixed at $(0.01,0.01,0.65,0.01)$. The dashed black line and the grey band show the Planck reported value of $0.054$ and the 68$\%$ confidence interval. We have used $z_{reion} = 7.68$ in this analysis.}
\end{figure*}

\begin{figure*}
    \centering
    \begin{tabular}{cccc}
	\includegraphics[trim={0 0 78 0},clip, width=0.29\textwidth]{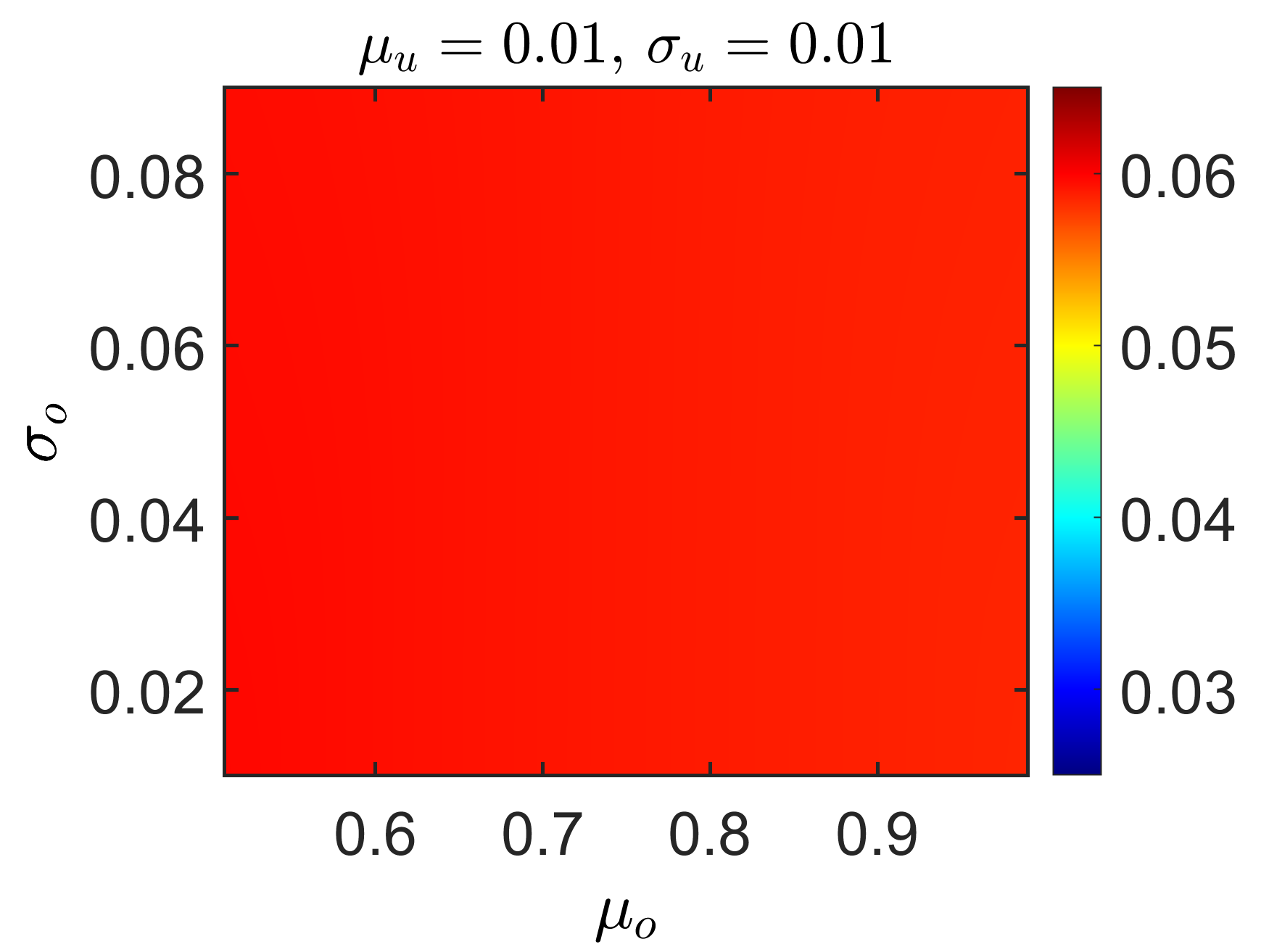}&
	\includegraphics[trim={70 0 78 0},clip, width=0.23\textwidth]{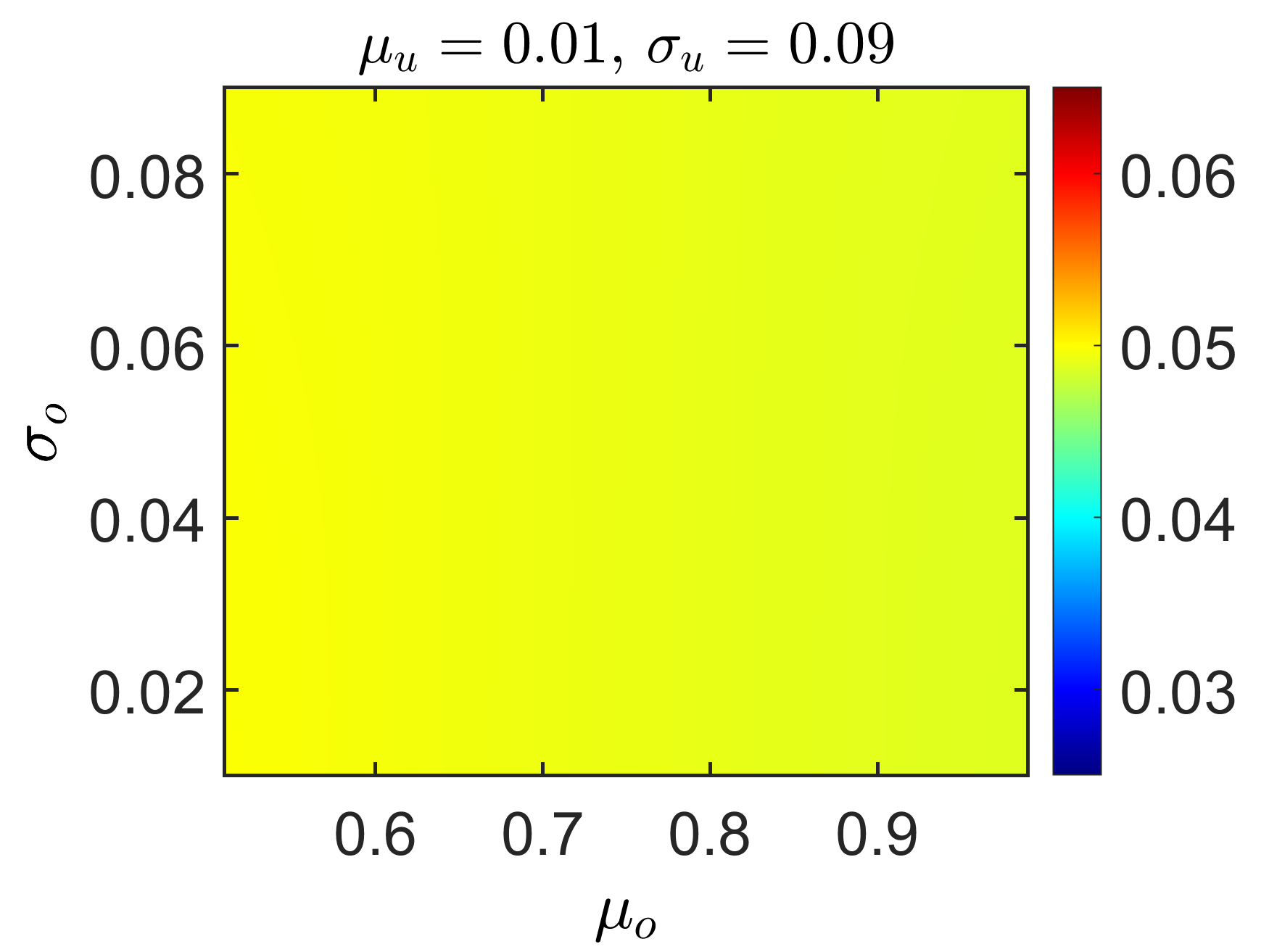}&
	\includegraphics[trim={70 0 78 0},clip, width=0.23\textwidth]{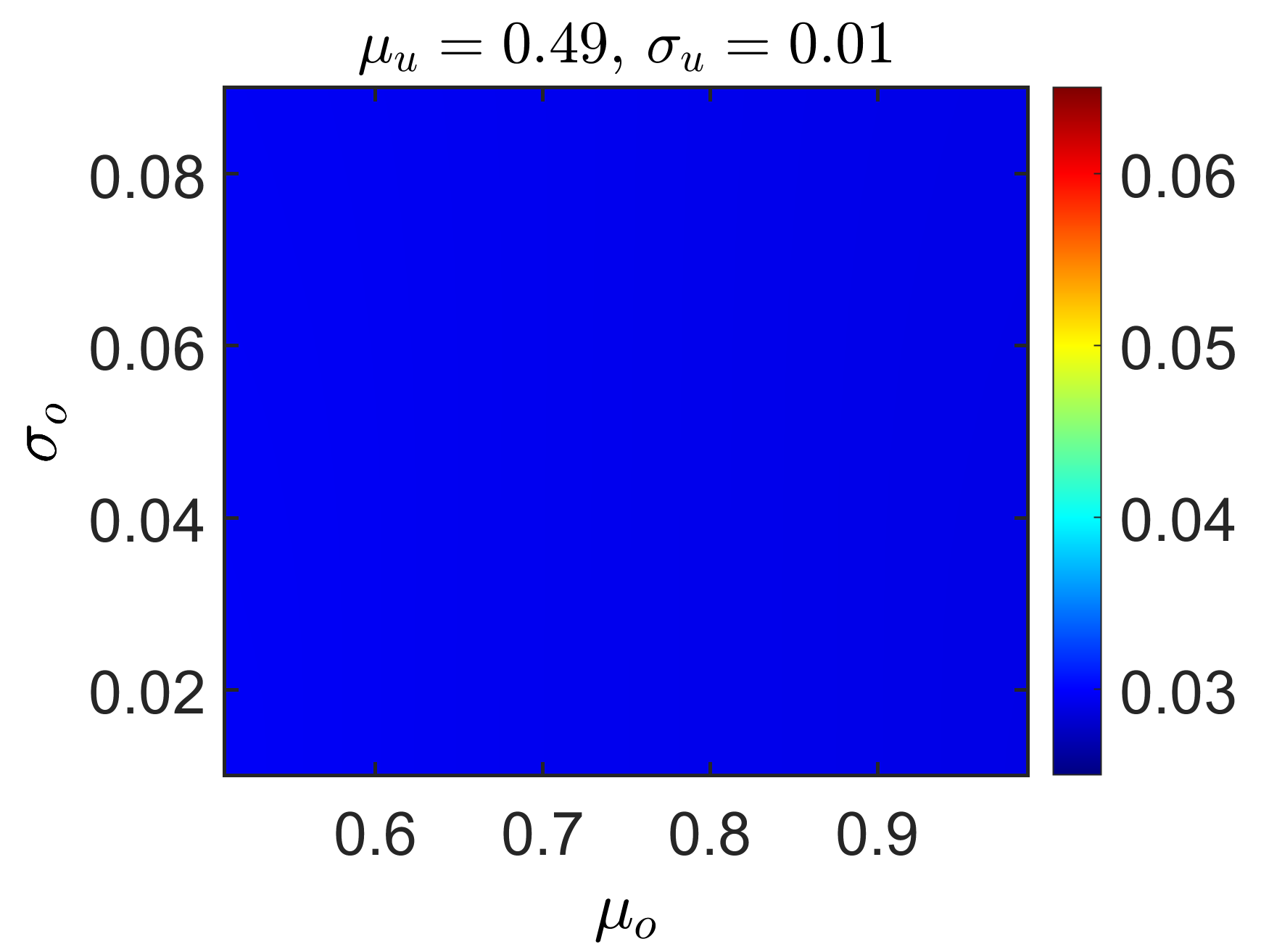}&
	\includegraphics[trim={70 0 78 0},clip, width=0.23\textwidth]{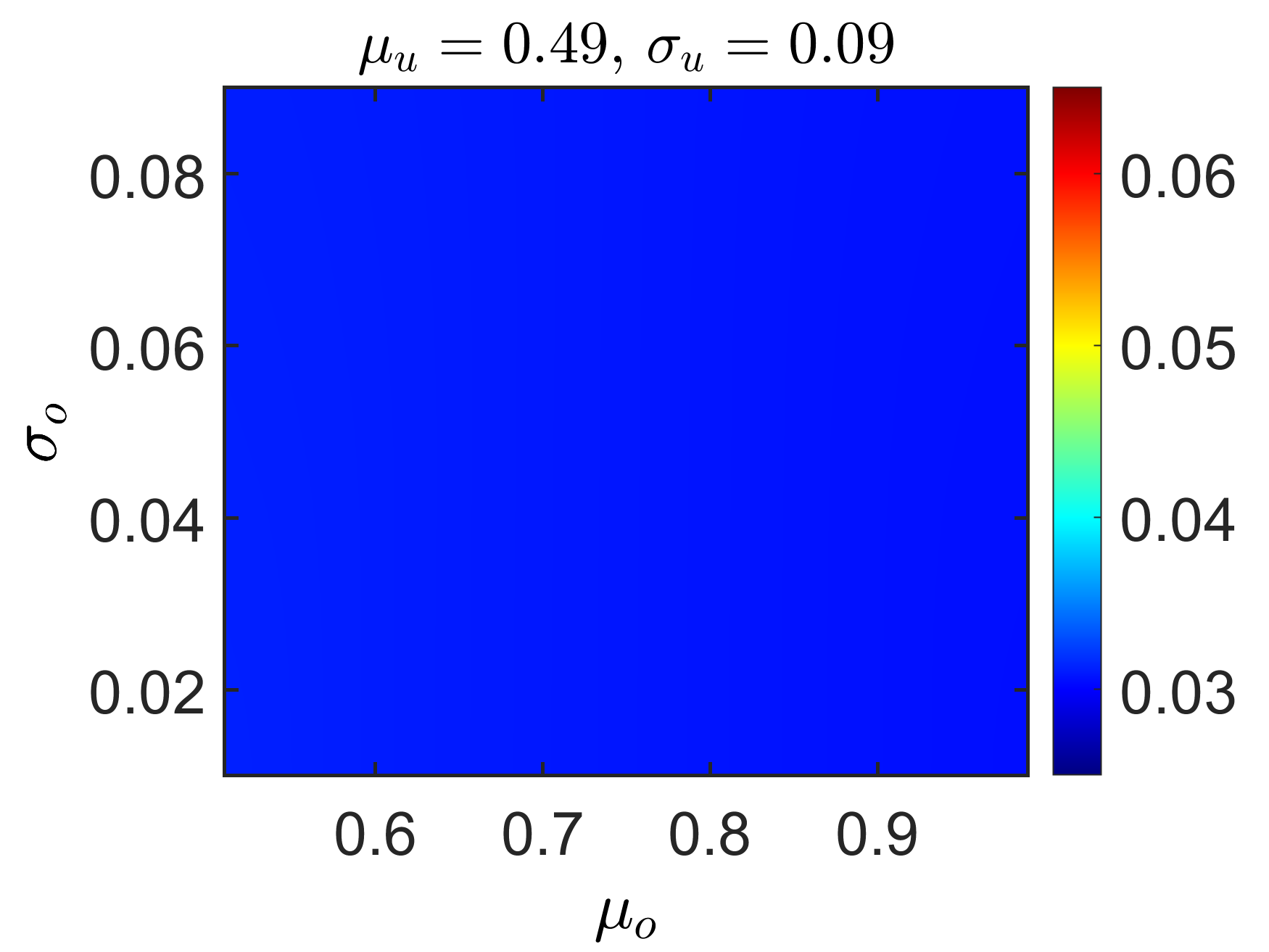}\\
	(a)&(b)&(c)&(d)\\
    \end{tabular}
    \begin{tabular}{c}
	\includegraphics[trim={25 35 20 240},clip, width=0.7\textwidth]{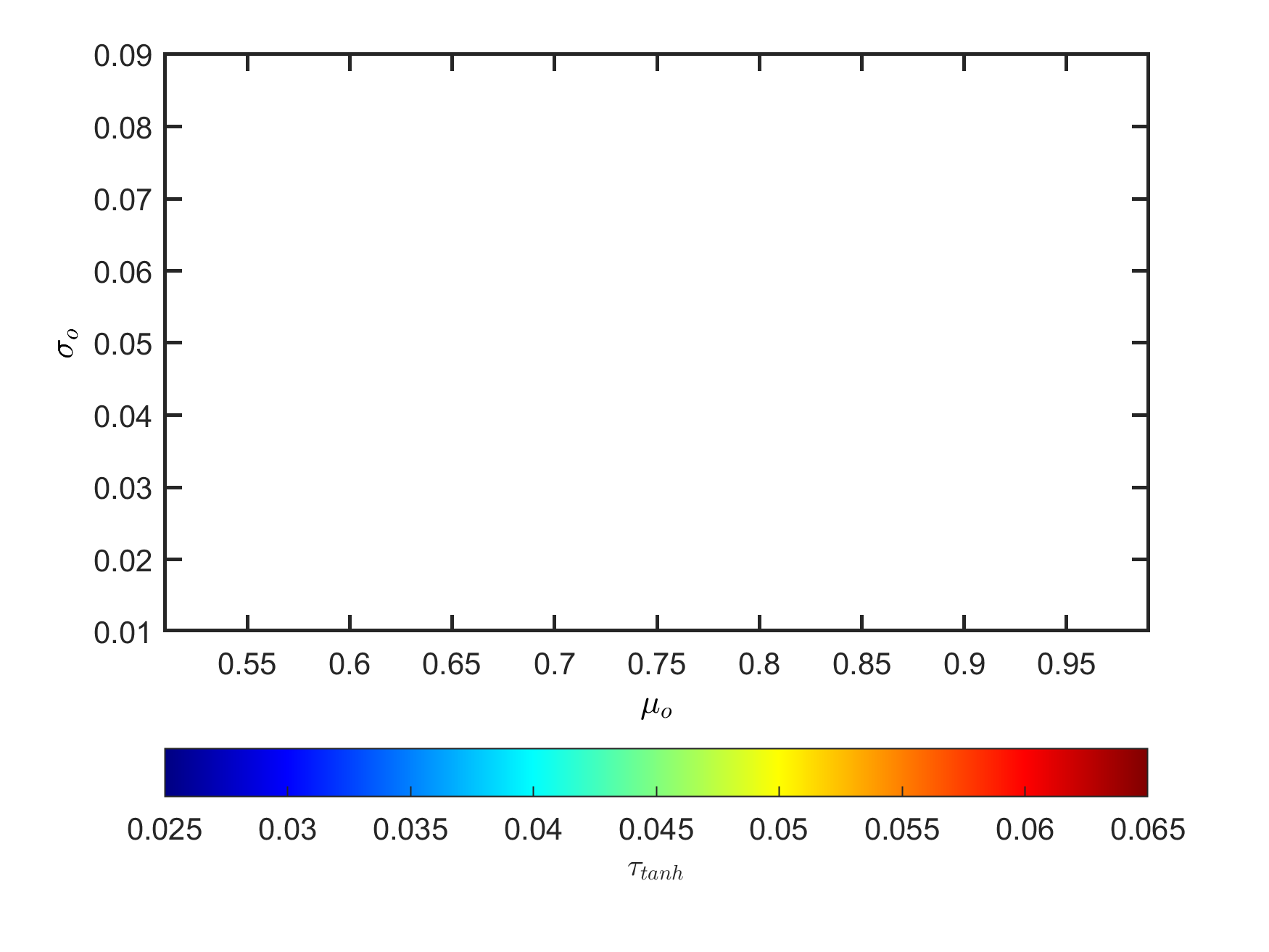}\\
       \large{$\tau_{reion}$}\\
    \end{tabular}
    \caption{\label{fig:tau_cont_o_combined} Color coded representation of $\tau_{reion}$ in the $\mu_o - \sigma_o$ plane for the tanh model of reionization. For subplots: (a) $\mu_u=0.01$, $\sigma_u=0.01$; (b) $\mu_u=0.01$, $\sigma_u=0.09$; (c) $\mu_u=0.49$, $\sigma_u=0.01$; (d) $\mu_u=0.49$, $\sigma_u=0.09$.}
\end{figure*}

\begin{figure*}
    \centering
    \begin{tabular}{cccc}
	\includegraphics[trim={0 0 78 0},clip, width=0.29\textwidth]{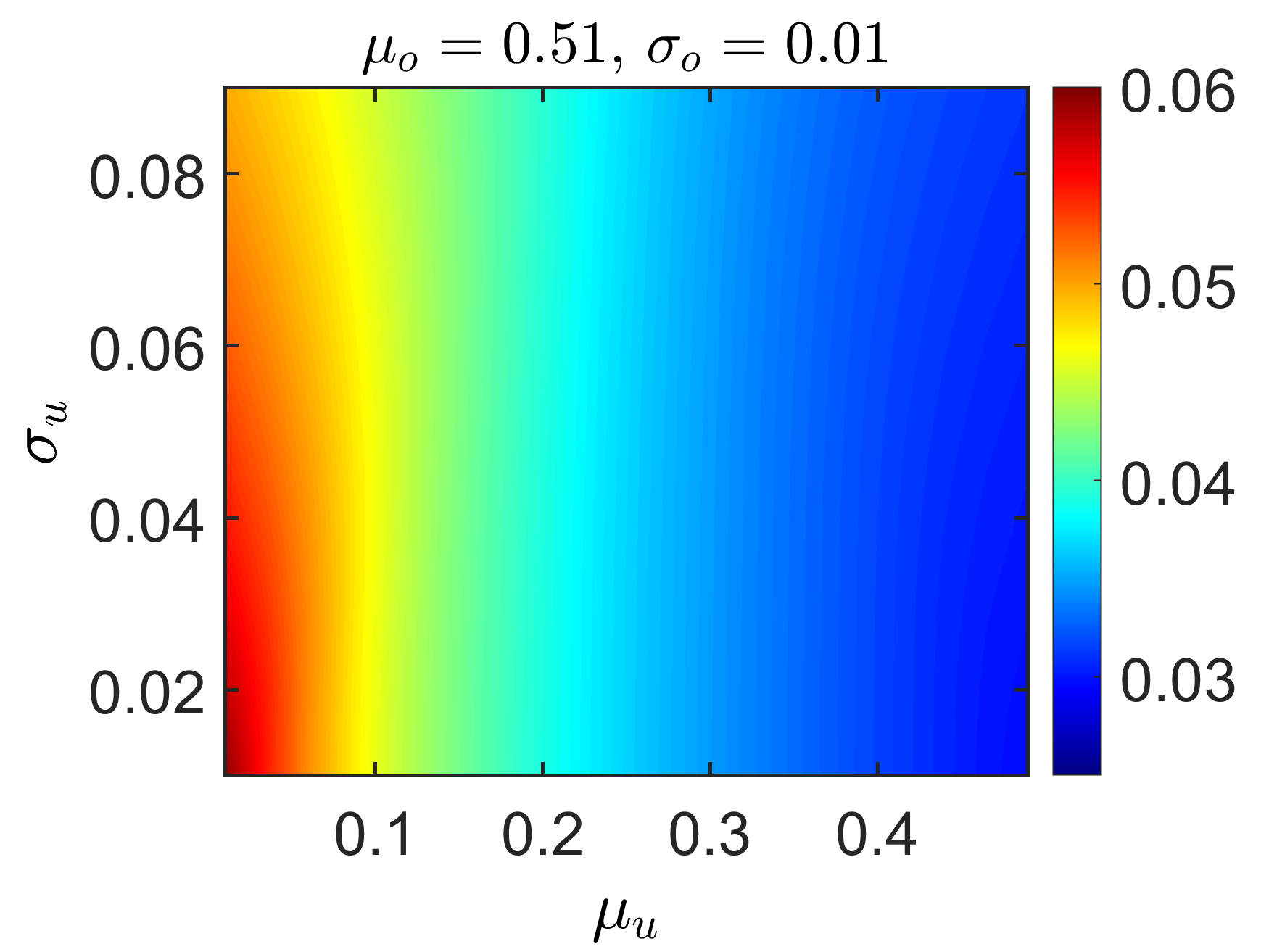}&
	\includegraphics[trim={70 0 78 0},clip, width=0.23\textwidth]{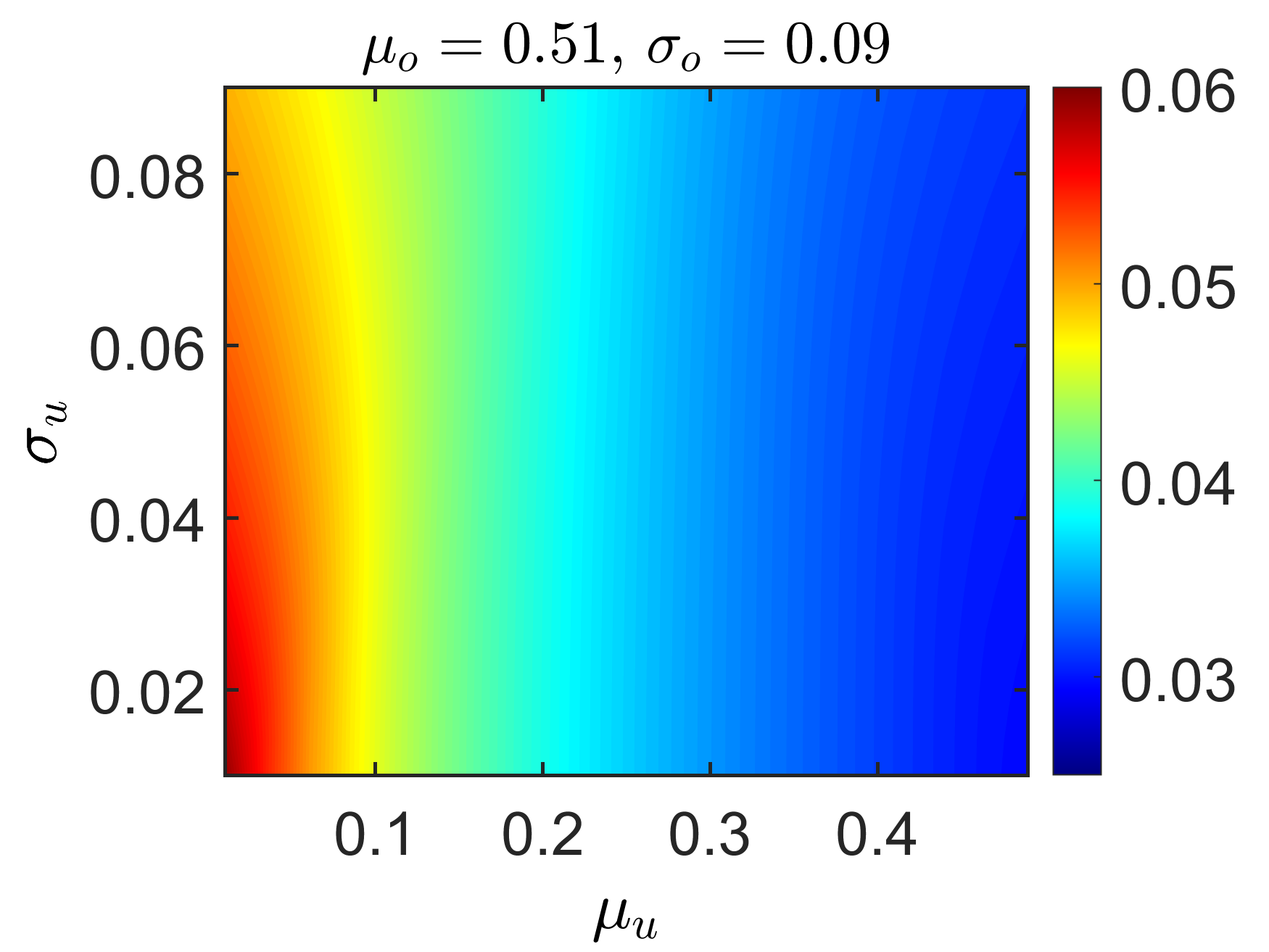}&
	\includegraphics[trim={70 0 78 0},clip, width=0.23\textwidth]{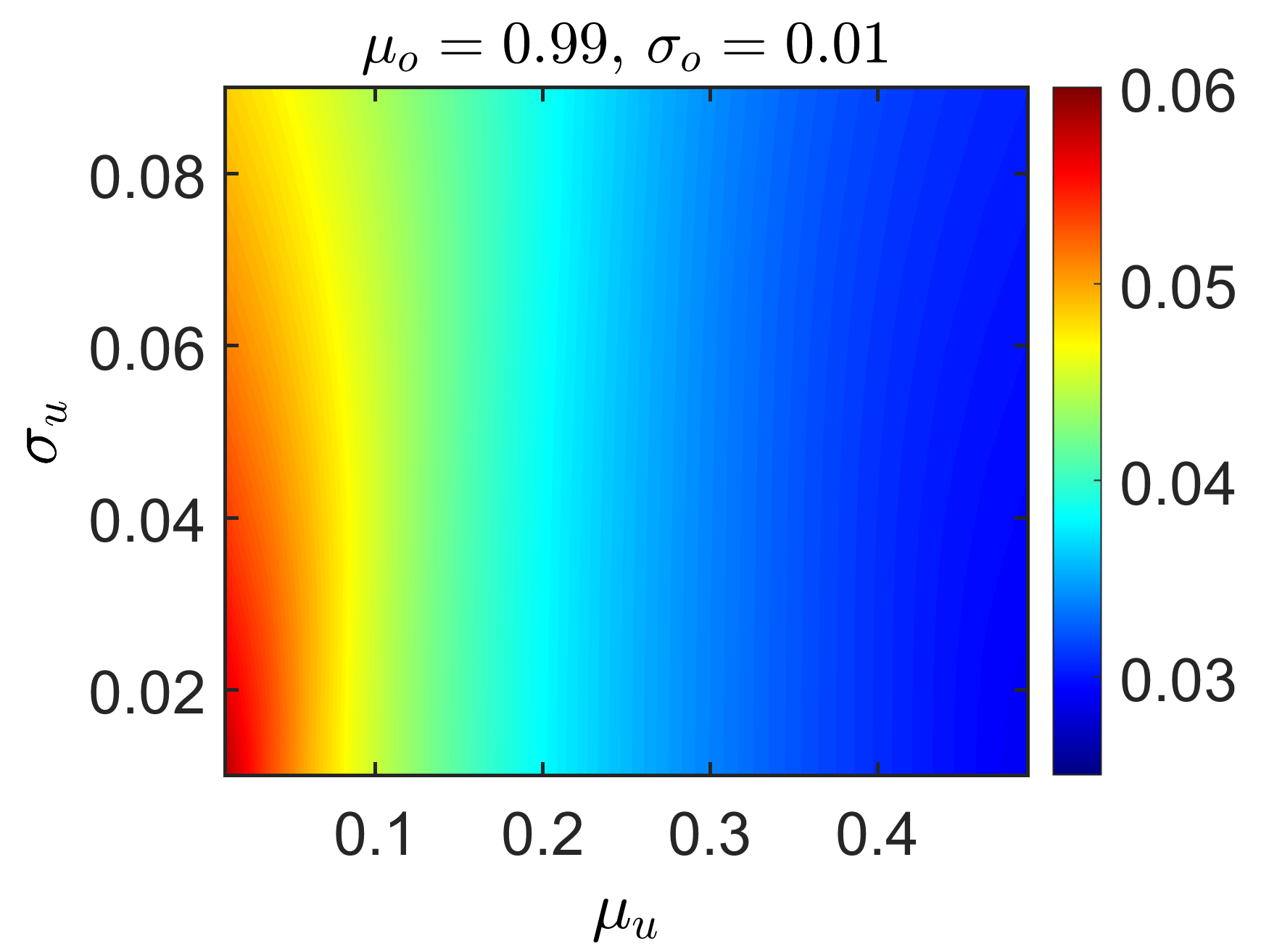}&
	\includegraphics[trim={70 0 78 0},clip, width=0.23\textwidth]{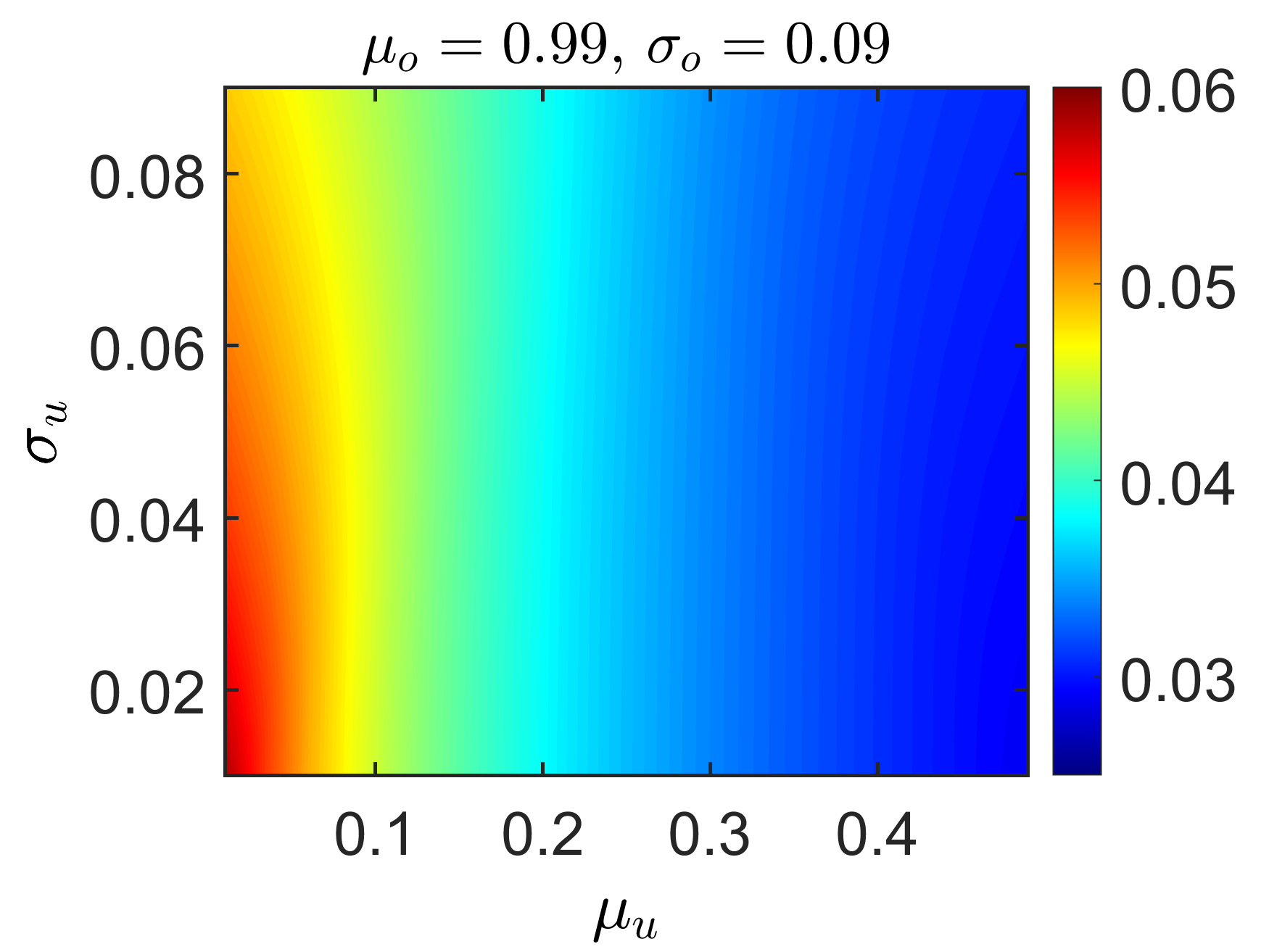}\\
	(a)&(b)&(c)&(d)\\
    \end{tabular}
    \begin{tabular}{c}
	\includegraphics[trim={25 35 20 240},clip, width=0.7\textwidth]{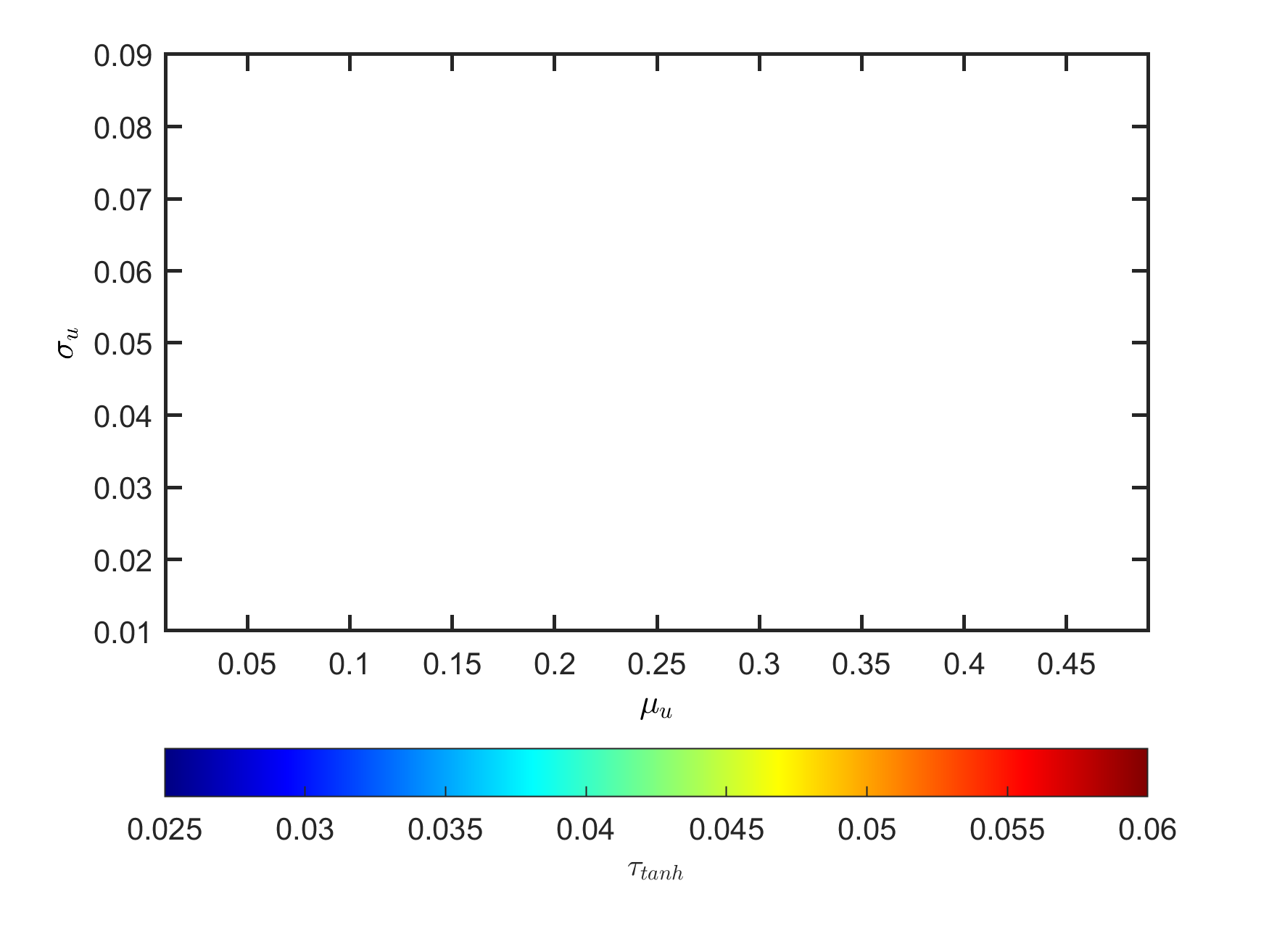}\\
    \large{$\tau_{reion}$}\\
    \end{tabular}
    \caption{\label{fig:tau_cont_u_combined} Color coded representation of $\tau_{reion}$ in the $\mu_u - \sigma_u$ plane for the tanh model of reionization. For subplots: (a) $\mu_o=0.51$, $\sigma_o=0.01$; (b) $\mu_o=0.51$, $\sigma_o=0.09$; (c) $\mu_o=0.99$, $\sigma_o=0.01$; (d) $\mu_o=0.99$, $\sigma_o=0.09$.}
\end{figure*}

 The backreaction model employed here is characterized by six parameters: $(\mu_u, \sigma_u, \mu_o, \sigma_o, h, n)$. Following earlier work \cite{Halder_future_2023, pandey_analyzing_2024}, we fix $n=100$ and adopt $h=0.7$ unless otherwise stated, leaving four free parameters $(\mu_u, \sigma_u, \mu_o, \sigma_o)$ that describe the matter distribution across underdense and overdense subregions. Figures~\ref{fig:tau_mu_sigma_vary}–\ref{fig:tau_cont_u_combined} present the dependence of $\tau_{reion}$ on these parameters, as well as on $h$ in both $\Lambda$CDM and the backreaction model. 

\par These plots are not intended as best-fit constraints, but rather as an illustration of how different physical ingredients of the model affect the reionization optical depth. The following section will confront the model predictions with observational data through a $\chi^2$ analysis. We have used $z_{reion} = 7.68$ in the analysis for this section.

\par \autoref{fig:tau_mu_sigma_vary} displays $\tau_{reion}$ for different values of the inhomogeneity parameters $\mu_o$, $\mu_u$, $\sigma_o$, and $\sigma_u$, while maintaining $n = 100$ and $h = 0.7$ as fixed values. All these plots contain the $\tanh$ model of reionization (blue line). Each subplot examines the impact of a single parameter on the reionization optical depth, varying it independently while holding the remaining three parameters constant. Specifically, panel (a) varies $\mu_o$ with $\sigma_o = \sigma_u = 0.01$ and $\mu_u = 0.09$ held fixed, panel (b) examines $\sigma_o$ variation with $\mu_o = 0.99$, $\sigma_u = 0.01$, and $\mu_u = 0.08$ kept constant, panel (c) shows $\mu_u$ variation with $\sigma_o = \sigma_u = 0.01$ and $\mu_o = 0.95$ fixed, and panel (d) investigates $\sigma_u$ variation with $\sigma_o = 0.01$, $\mu_o = 0.99$, and $\mu_u = 0.07$ held constant. These parameters ($\mu_o$, $\mu_u$, $\sigma_o$, $\sigma_u$) characterize the inhomogeneous matter distribution in the backreaction model, allowing us to assess how matter clustering affects the reionization process. The Hubble parameter is set to $H_0 = 100h$ km s$^{-1}$ Mpc$^{-1}$ with $h = 0.7$. The dashed black line indicates the Planck Collaboration VI 2020 measured value of $\tau_{reion} = 0.054$, while the grey band represents the corresponding $68\%$ confidence interval. The analysis reveals that variations in $\mu_u$ produce the most pronounced changes in $\tau_{reion}$ among the four inhomogeneity parameters examined. This arises from the fact that, at the present time in our model, underdense subregions occupy 91$\%$ of the total volume and therefore associated parameters play a significant role in determining the values of various cosmological quantities.

\par In \autoref{fig:tau_h0_vary}, $\tau_{reion}$ for varying values of $h$ has been plotted for the $\Lambda$CDM and the backreaction model. In subplot (a), the cosmological model is the $\Lambda$CDM model, while in subplot (b) it is the backreaction model. Both plots have been generated for the $\tanh$ model of reionization (blue line). For subplot (b), the value of the four backreaction model parameters $(\mu_u,\sigma_u,\mu_o,\sigma_o)$ has been fixed at $(0.01,0.01,0.65,0.01)$. These values were chosen by taking \autoref{fig:tau_mu_sigma_vary} into consideration.  It can be seen from the figure that, for the employed combination of backreaction model parameters, the Planck reported value of $\tau_{reion}$ can be obtained by the backreaction model (subplot b) for comparatively larger values of $h$ than in the $\Lambda$CDM model (subplot a). In comparison to the $\Lambda$CDM model, $\tau_{reion}$ for the backreaction model falls within the 68$\%$ confidence interval of the value reported by Planck for a narrower range of $h$.

\par \autoref{fig:tau_cont_o_combined} presents correlation plots illustrating the variation of $\tau_{reion}$ in the $\mu_o-\sigma_o$ parameter space for the tanh reionization model (subplots (a) - (d)). We have used $z_{reion} = 7.68$ and $h = 0.7$ in this analysis. The parameter $\mu_o$ spans the range $0.5-1.0$ along the horizontal axis, while $\sigma_o$ varies between $0.01-0.09$ along the vertical axis. The color scheme represents the magnitude of $\tau_{reion}$ according to the scale shown in the color bar. It varies from $0.025$ to $0.065$ for the range of backreaction model parameters employed here. Four different combinations of $(\mu_u,\sigma_u)$ values are examined across the subplots to assess their influence on the optical depth calculation.

\par Subplots (a) and (b) correspond to $(\mu_u,\sigma_u) = (0.01,0.01)$ and $(0.01,0.09)$, respectively. The results demonstrate that subplot (a) yields systematically higher $\tau_{reion}$ values compared to subplot (b), while both subplots exhibit minimal internal variation across the $\mu_o-\sigma_o$ parameter space. This behavior indicates that variations in the overdense region parameters $(\mu_o,\sigma_o)$ exert negligible influence on $\tau_{reion}$ when $(\mu_u,\sigma_u)$ are held constant. Conversely, the substantial difference between subplots (a) and (b) highlights the significant impact of $\sigma_u$ on the optical depth calculation for smaller values of $\mu_u$. Subplots (c) and (d), corresponding to $(\mu_u,\sigma_u) = (0.49,0.01)$ and $(0.49,0.09)$ respectively, exhibit markedly lower $\tau_{reion}$ values compared to subplots (a) and (b). The transition from $\sigma_u = 0.01$ to $\sigma_u = 0.09$ while maintaining $\mu_u = 0.49$ produces minimal variation in $\tau_{reion}$. This observation shows that at large values of $\mu_u$, the effect of variation of $\sigma_u$ on the value of $\tau_{reion}$ is negligible. Notably, the Planck Collaboration VI 2020 reported value of $\tau_{reion}$ is achieved within the parameter range explored in subplot (a). From \autoref{fig:tau_mu_sigma_vary}, it can be seen that to achieve Planck's reported value of $\tau_{reion}$, $\mu_u$ and $\sigma_u$ both need to be on the lower end of their allowed range. In subplot (a), both $\mu_u$ and $\sigma_u$ are at the lower end of their allowed range; hence, Planck's reported value is achieved in this subplot.

\par The comparison between subplots (a) and (c), where $\sigma_u$ remains constant while $\mu_u$ varies, reveals that changes in $\mu_u$ produce more substantial variations in $\tau_{reion}$ than equivalent changes in $\sigma_u$ (e.g., subplots (a) and (b)). These findings collectively demonstrate that underdense region parameters, particularly $\mu_u$, dominate in governing the global reionization optical depth. In contrast, the overdense region parameters exhibit limited influence on the overall dynamics.

\par In \autoref{fig:tau_cont_u_combined}, the variation of $\tau_{reion}$ for the tanh model of reionization (subplots (a) - (d)) in the $\mu_u-\sigma_u$ plane is shown for different sets of values of $(\mu_o,\sigma_o)$ using a contour plot. We have used $z_{reion} = 7.68$ and $h = 0.7$ in this analysis. The value of $\mu_u$ varies in the range of $0 - 0.5$ along the x-axis while $\sigma_u$ is varied along the y-axis in the range of $0.01 - 0.09$. The color scheme represents the magnitude of $\tau_{reion}$ according to the scale shown in the color bar. It varies from $0.025$ to $0.06$ for the range of backreaction model parameters employed here. In subplots (a) and (b) of the figures, $\mu_o$ is fixed at $0.51$, and $\sigma_o$ has the values of $0.01$ and $0.09$, respectively. In subplots (c) and (d) of the figures, $\mu_o$ is fixed at $0.99$ while $\sigma_u$ has the values of $0.01$ and $0.09$, respectively. These figures also highlight the insignificance of overdense parameters: all four subplots are similar, so changing them does not significantly affect the output. The only change in the plots is due to variation in the parameters $(\mu_u,\sigma_u)$ associated with the underdense subregions. These observations agree with the inferences from \autoref{fig:tau_mu_sigma_vary}.

\section{Observational data and Chi-Sq Analysis}\label{sec:obs_analysis}

We now examine the backreaction model against observational data to determine the optimal parameter values. A Bayesian analysis is conducted to compare our theoretical predictions with the PantheonPlus+SH0ES Type Ia supernova distance modulus versus redshift data \cite{Pantheon_likelihood, Panth_full_dataset_Scolnic_2022}. To facilitate this comparison between the backreaction model and observational data, we require a framework that relates the theoretical quantities to the observational ones. We employ the covariant scheme described specifically in \autoref{eq:covariant_sch_1} and \autoref{eq:covariant_sch_2}. The first equation of the covariant scheme \autoref{eq:covariant_sch_1} establishes the relationship between the theoretically calculated scale factor $a_{\mathcal{D}}$ from the backreaction model and the cosmological redshift. The second equation \autoref{eq:covariant_sch_2} connects the theoretically calculated average density $\langle\rho\rangle_{\mathcal{D}}$ from the backreaction model to the observational quantity, specifically the angular diameter distance $D_A$. Using standard cosmological distance relations, we can subsequently calculate the distance modulus from this angular diameter distance. 


\par To proceed with the supernova analysis, we utilize the standard distance modulus formulation for SN Ia \cite{Tripp1998, Panth_full_dataset_Scolnic_2022, Pantheon_likelihood}

\begin{equation}\label{eq:mu_panth}
    \mu = m_B + \alpha x_1 - \beta c - M - \delta_{\mu-bias}
\end{equation}

where $m_B \equiv -2.5\log_{10}(x_0)$, where $x_0$ is the overall amplitude of the light curve, $x_1$ is the stretch parameter corresponding to the width of the light curve, $c$ is the color of the light curve and $\alpha$ and $\beta$ are global nuisance parameters that associate stretch and color, respectively, with luminosity. We have used $\alpha = 0.148$ and $\beta = 3.112$ \cite{Panth_full_dataset_Scolnic_2022} in this analysis. M is the fiducial absolute magnitude of an SN Ia. $\delta_{\mu-bias}$ is the bias correction derived from simulations needed to account for selection effects and other issues in discovery. In this analysis, we have also included $M$ (\autoref{eq:mu_panth}), the fiducial magnitude of SN Ia, as one of the parameters. The parameters $M$ and $H_0$ are degenerate when analyzing SNe alone. However, in this analysis, we are using the full PantheonPlus+SH0ES dataset \cite{Pantheon_likelihood}, which includes the SH0ES Cepheid host distance anchors \cite{Riess_2022_fidu} in the likelihood that facilitates constraints on both $M$ and $H_0$.

\par Here, we use $\chi^2$ analysis and combined statistical and systematic covariance matrices ($C_{stat+syst} = C_{stat} + C_{syst}$) to constrain the backreaction model. We follow the formalism of \cite{Pantheon_likelihood} where cosmological parameters are constrained by minimizing a $\chi^2$ likelihood. We incorporate SH0ES Cepheid host galaxy distance measurements into this analysis, resulting in the following supernova distance residuals:

\[ \Delta\mathbf{D_i} =
  \begin{cases}
    \mu_i - \mu_i^{Cepheid}       & \quad  i \in \text{Cepheid hosts}\\
    \mu_i - \mu_{model}(z_i)  & \quad \text{otherwise },
  \end{cases}
\]

where $\mu_i^{Cepheid}$ is the Cepheid-calibrated host-galaxy distance provided by SH0ES. The model distances are defined as

\begin{equation}\label{eq:model_dL}
    \mu_{model}(z_i) = 5\log (d_L(z_i)/10 pc)
\end{equation}

\par For the backreaction model, $d_L$ can be calculated using the covariant scheme and standard cosmological distance relations relating angular diameter distance, $D_A$, to luminosity distance, $d_L$, as mentioned before.

\par In this analysis, we include the covariance matrix of PantheonPlus+SH0ES to calculate the likelihood;

\begin{equation}\label{eq:likelihod}
    -2\ln{\mathcal{(L)}} = \chi^2 = \Delta\mathbf{D}^T (C_{stat+syst}^{SN} + C_{stat+syst}^{Cepheid})^{-1}\Delta\mathbf{D},
\end{equation}

where $C_{stat+syst}^{SN}$ denotes the SN covariance with statistical and systematic uncertainties. 

\begin{figure*}
    \centering
    \includegraphics[width=\textwidth]{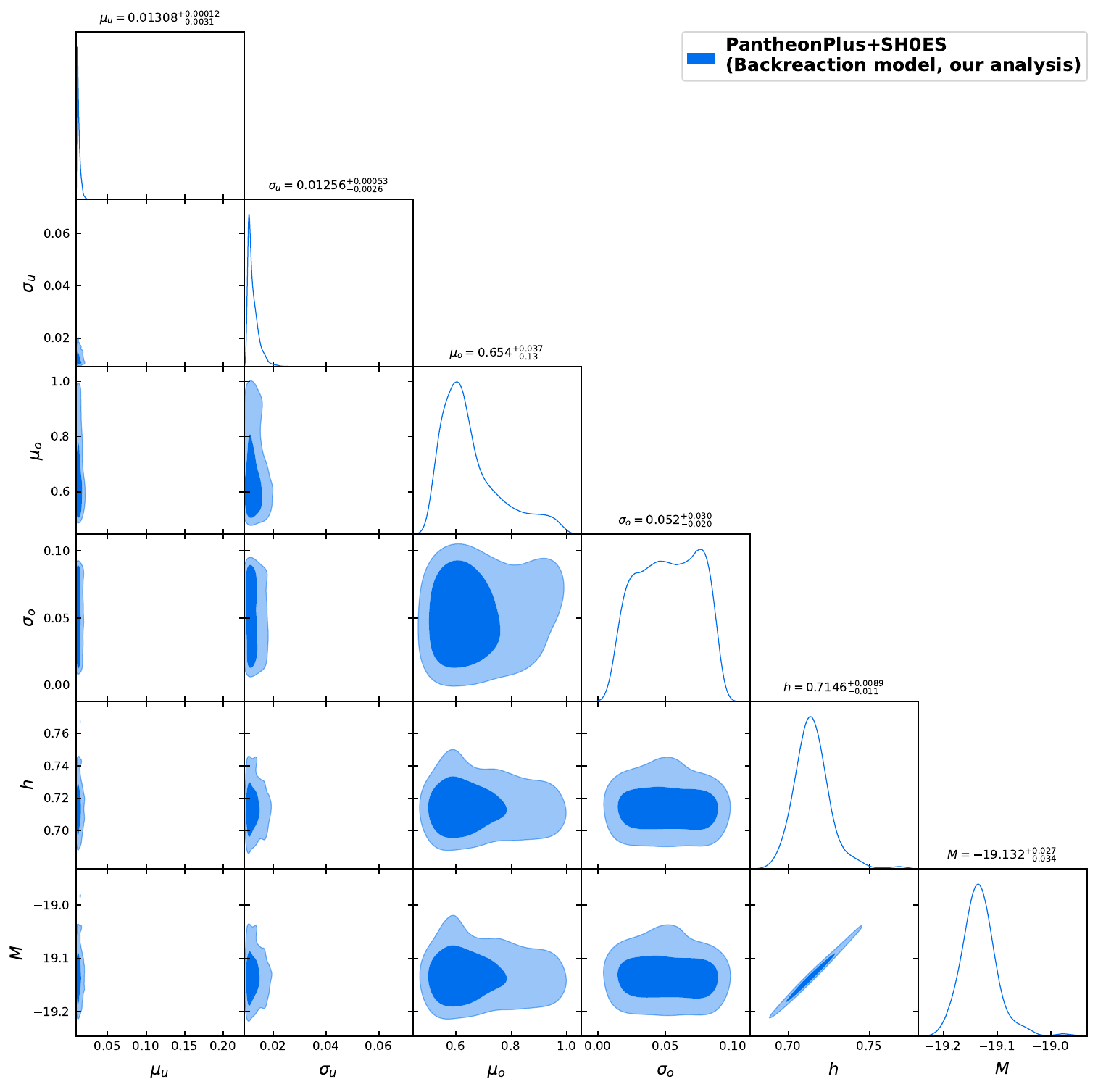}
    \caption{\label{fig:corner_MCMC_M} Corner plot showing the MCMC result for the backreaction model carried out using the observational results of the PantheonPlus+SH0ES supernova Ia data \cite{Pantheon_likelihood, Panth_full_dataset_Scolnic_2022}. The diagonal plots show the marginalized posterior densities for each parameter. Here, we have also included $M$, the fiducial magnitude of SN Ia, as one of the parameters. }
\end{figure*}

\par In this analysis, the resulting posterior distributions of different parameters \autoref{fig:corner_MCMC_M} are obtained using the Markov Chain Monte Carlo (MCMC) iteration method using the \texttt{MCMCSTAT} package \cite{mcmc1,mcmc2}. We use a total of $6\times 10^4$ number of events spread across six runs with the adaptation interval of $1000$ for a single run, within the parameter range given in \autoref{tab:prior}.

\begin{table}[h!]
\centering
\renewcommand{\arraystretch}{1.5} 
\setlength{\tabcolsep}{12pt} 
\begin{tabular}{|c|c|}
\hline
\textbf{Parameter} & \textbf{Range} \\ \hline
$\mu_u$ & $[0.01, 0.49]$ \\ \hline
$\sigma_u$ & $[0.01, 0.09]$ \\ \hline
$\mu_o$ & $[0.51, 0.99]$ \\ \hline
$\sigma_o$ & $[0.01, 0.09]$ \\ \hline
$h$ & $[0.65, 0.80]$ \\ \hline
$M$ & $[-18, -20]$ \\ \hline
\end{tabular}
\caption{\label{tab:prior} Parameter prior ranges used in the analysis.}
\end{table}

\par The topmost plots of each column of \autoref{fig:corner_MCMC_M} represent the marginalized posterior distribution for the parameters $\mu_u$, $\sigma_u$, $\mu_o$, $\sigma_o$, $h$ and $M$, respectively, obtained by marginalizing the other parameters, while the other plots of \autoref{fig:corner_MCMC_M} show the contour representation of the posterior distribution in different sets of two-parameter space. The darker and lighter regions denote the $1\sigma$ and $2\sigma$ confidence intervals in these contour plots. The diagonal panels display the 1-D histograms of the posterior distributions for each backreaction model parameter, obtained by marginalizing over the other parameters. The off-diagonal panels display 2D projections of the posterior probability distributions for each parameter pair, along with parameter correlations and contours. The optimum values in the plot are reported with 68$\%$ confidence limits.

\begin{table}[ht]
\centering
\begin{tabular}{lcc}
\toprule
\textbf{Parameter} & \textbf{68\% limits} & \textbf{95\% limits} \\
\hline
$\mu_u$ & $0.01308^{+0.00012}_{-0.0031}$ & $0.0131^{+0.0042}_{-0.0036}$ \\
$\sigma_u$ & $0.01256^{+0.00053}_{-0.0026}$ & $0.0126^{+0.0046}_{-0.0031}$ \\
$\mu_o$ & $0.654^{+0.037}_{-0.13}$ & $0.65^{+0.26}_{-0.16}$ \\
$\sigma_o$ & $0.052^{+0.030}_{-0.020}$ & $0.052^{+0.037}_{-0.039}$ \\
$h$ & $0.7146^{+0.0089}_{-0.011}$ & $0.715 \pm 0.023$ \\
$M$ & $-19.132^{+0.027}_{-0.034}$ & $-19.132 \pm 0.070$ \\
\hline
\end{tabular}
\caption{Backreaction model parameters showing 68\% (1$\sigma$) and 95\% (2$\sigma$) confidence limits obtained using PantheonPlus+SH0ES data.}
\label{tab:mod_params_M}
\end{table}

\par
\begin{figure*}
    \centering
    \includegraphics[width=\textwidth]{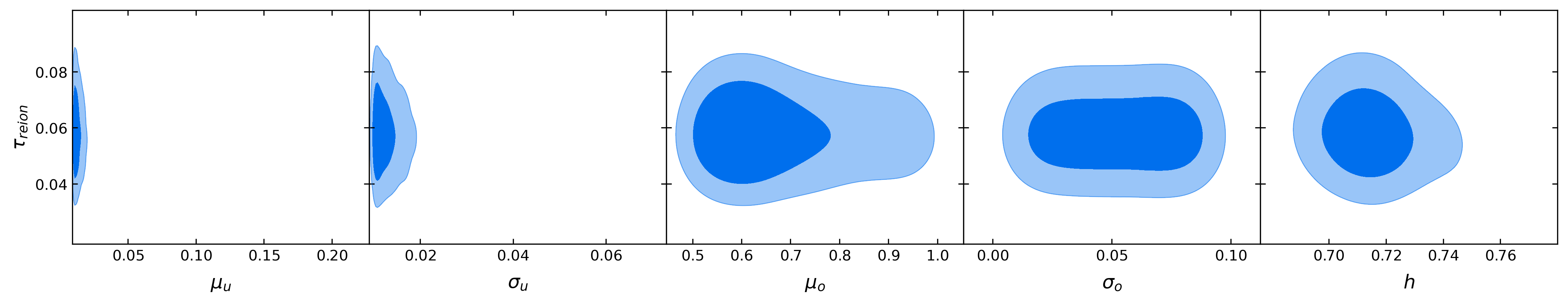}
    \caption{\label{fig:tau_tanh_vs_others} Contour plots showing the correlation between the backreaction model parameters and derived parameter - $\tau_{reion}$.}
\end{figure*}

 \autoref{tab:mod_params_M} shows the 68\% (1$\sigma$) and 95\% (2$\sigma$) confidence limits for the backreaction model parameters obtained from marginalized posterior distributions from the MCMC analysis. We have 60,000 values for the backreaction model parameters from the MCMC chains. We calculated the value of $\tau_{reion}$ for the tanh reionization model for these 60,000 sets of the backreaction model parameters' values using \autoref{eq:tau_z} and \autoref{eq:tanh_model} (see \autoref{subsec:tau_re}). We used the complete range of $z_{reion} = 7.68\pm0.79$ for this analysis. We then used these calculated values of $\tau_{reion}$ with the MCMC chains to convert this quantity into a derived parameter. This operation helps us in analyzing the correlation between the backreaction model parameters and $\tau_{reion}$ in \autoref{fig:tau_tanh_vs_others}. A negative correlation between $\tau_{reion}$ and $\mu_u$ can be seen from \autoref{fig:tau_mu_sigma_vary}. Increasing the value of $\mu_u$ results in a decrease in the value of $\tau_{reion}$. This negative correlation is not visible in \autoref{fig:tau_tanh_vs_others} as the range of $\mu_u$ constrained by the observational dataset is very narrow; hence, no correlation is visible.

\begin{table}[ht]
\centering
\begin{tabular}{lcccc}
\toprule
\multicolumn{5}{c}{\textbf{$\tau_{reion}$ Measurements}} \\
\hline
\textbf{S.No.}& \textbf{Dataset} & \textbf{Method} & \textbf{68\% limits} & \textbf{95\% limits} \\
\hline
(i) & Planck PR3 & TT,TE,EE+lowE & $0.0544^{+0.0070}_{-0.0081}$ & $0.054^{+0.017}_{-0.015}$ \\
(ii) & Planck PR3 & tanh+$\Lambda$CDM & $0.0535^{+0.0079}_{-0.0076}$ & $0.054^{+0.017}_{-0.015}$ \\
(iii) & PantheonPlus+SH0ES & tanh+$\Lambda$CDM & $0.0476 ^{+0.0069}_{-0.0065}$ & $0.048^{+0.013}_{-0.015}$ \\
(iv) & PantheonPlus+SH0ES & tanh+backreaction & $0.0581^{+0.0105}_{-0.0096}$ & $0.058^{+0.022}_{-0.019}$ \\
\hline
\end{tabular}
\caption{Optical depth to reionization measurements showing 68\% (1$\sigma$) and 95\% (2$\sigma$) confidence limits for $\tau_{reion}$ (i) from Planck PR3 TT,TE,EE+lowE (reported in \cite{Planck}); (ii) from tanh+$\Lambda$CDM model, calculated with $H_0$ constrained by Planck PR3 dataset; (iii) from tanh+$\Lambda$CDM model calculated with $H_0$ constrained by PantheonPlus+SH0ES dataset and (iv) from tanh+backreaction model calculated with $H_0$ constrained by PantheonPlus+SH0ES dataset. The distributions for the bottom three calculated cases are shown in \autoref{fig:optical_depth_final}.} 
\label{tab:tau_all_M}
\end{table}

\begin{table}[h!]
\centering
\begin{tabular}{lcccc}
\toprule
\multicolumn{5}{c}{\textbf{$H_0$ Measurements}} \\
\hline
\textbf{S.No.} & \textbf{Dataset} & \textbf{Cosmological Model} & \textbf{68\% limits} & \textbf{95\% limits} \\
\hline
(i) & Planck PR3 (TT,TE,EE+lowE) & $\Lambda\text{CDM}$ & $67.27 \pm 0.60$ & $67.3 \pm 1.2$ \\
(ii) & PantheonPlus+SH0ES & $\Lambda\text{CDM}$ & $73.45^{+1.04}_{-1.01}$ & $73.45^{+2.08}_{-1.99}$ \\
(iii) & PantheonPlus+SH0ES & Backreaction & $71.46^{+0.89}_{-1.1}$ & $71.5 \pm 2.3$ \\
\hline
\end{tabular}
\caption{Hubble constant ($H_0$) measurements with 68\% (1$\sigma$) and 95\% (2$\sigma$) confidence limits. All values in km/s/Mpc. (i) is the reported value from \cite{Planck}. (ii) is reported in \cite{Pantheon_likelihood}. (iii) is calculated here in this study.}
\label{tab:H0s_M}
\end{table}

\par
\begin{figure*}
    \centering
    \includegraphics[width=\textwidth]{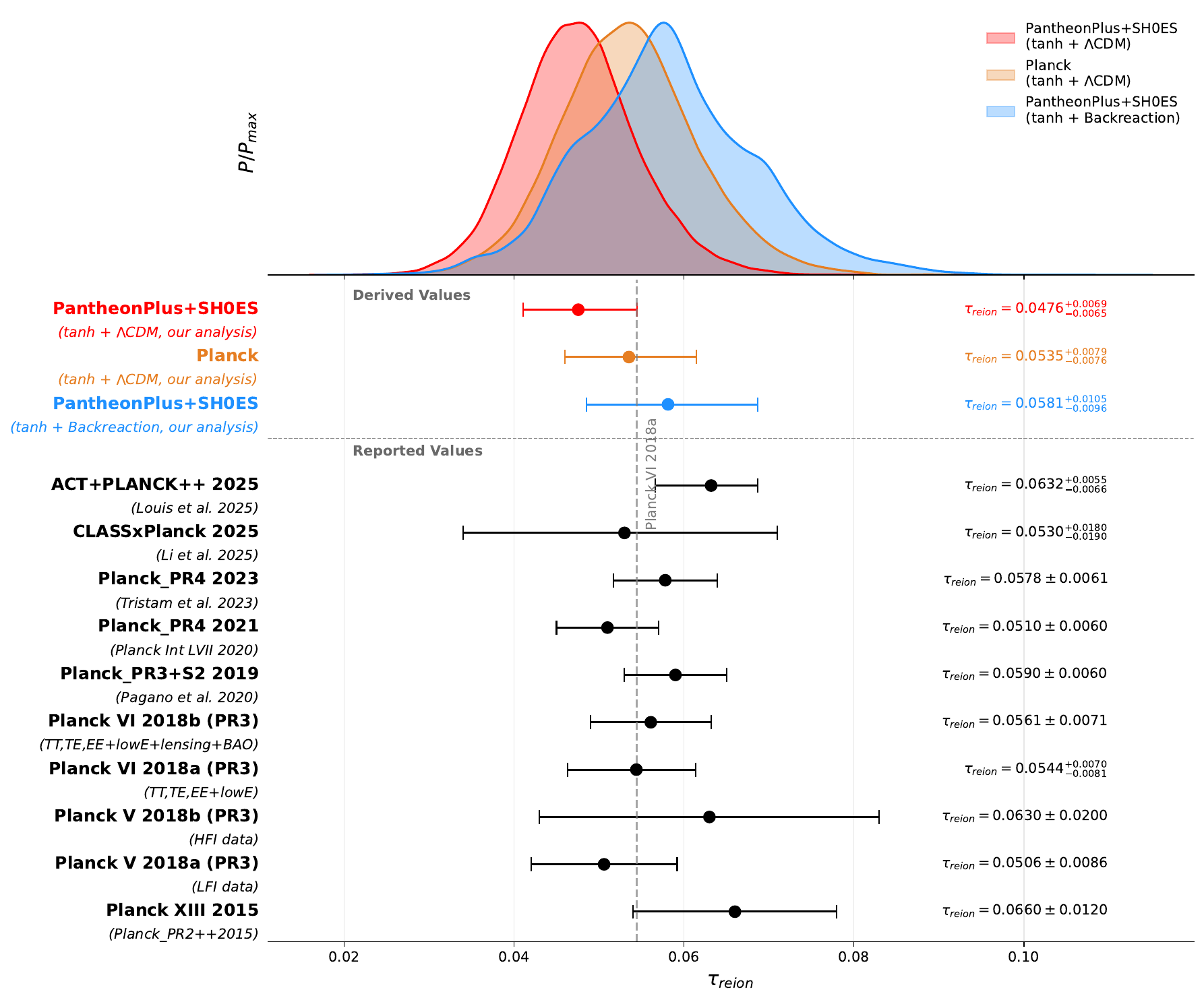}
\caption{\label{fig:optical_depth_final} The upper panel displays the marginalized posterior distributions of  \(\tau_{\mathrm{reion}}\) for the tanh reionization model, shown for both the  \(\Lambda\)CDM cosmology and the backreaction model. For the \(\Lambda\)CDM case, two distributions are presented: one with 
\(H_0\) constrained using the PantheonPlus+SH0ES dataset, and the other with 
\(H_0\) constrained by the Planck dataset. The backreaction model results 
are shown for parameters constrained by the PantheonPlus+SH0ES dataset. 
In all cases, the full allowed range of \(z_{\mathrm{re}}\) has been taken 
into account. 68 $\%$ confidence intervals of these distributions have been shown in the lower panel. The lower panel also summarizes the ranges of \(\tau_{\mathrm{reion}}\) along with their 68$\%$ confidence interval, reported in various studies based on different datasets within the context of the \(\Lambda\)CDM model.}
\end{figure*}

 \autoref{tab:tau_all_M} shows the 68\% (1$\sigma$) and 95\% (2$\sigma$) confidence limits for $\tau_{reion}$ for four cases. The calculation of $\tau_{reion}$ requires both the model of reionization and the cosmological model (see \autoref{subsec:tau_re}). The first case (i) is the direct calculation of $\tau_{reion}$ from the Planck PR3 dataset for the base-$\Lambda$CDM model using the TT,TE,EE+lowE spectra reported by \cite{Planck}. The remaining three cases (ii)–(iv) have been evaluated by us in this analysis. The second case is also for the Planck PR3 dataset (used here to constrain $H_0$), but this time using the tanh model of reionization and the $\Lambda$CDM cosmological model. The third and fourth cases employ the PantheonPlus+SH0ES dataset to constrain \(H_0\), in combination with the tanh parameterization of reionization, 
within the frameworks of the \(\Lambda\)CDM and backreaction cosmological 
models, respectively. The last three cases have been plotted in the upper portion of \autoref{fig:optical_depth_final}. In the lower panel of \autoref{fig:optical_depth_final}, we display previously reported $\tau_{reion}$ values from several observational datasets and studies, together with the $\tau_{reion}$ values obtained in our own analysis. 

\par The values listed for the first two cases in \autoref{tab:tau_all_M} are derived from the same dataset and assume the same cosmological model; only the methods used to obtain them differ. Consequently, they are expected to be equivalent, which is indeed what we observe. For the PantheonPlus+SH0ES dataset, the backreaction model yields a value of $\tau_{reion}$ that better aligns with the Planck PR3 results than the corresponding value from the $\Lambda$CDM model.

\par
\begin{figure*}
    \centering
	\includegraphics[width=0.75\textwidth]{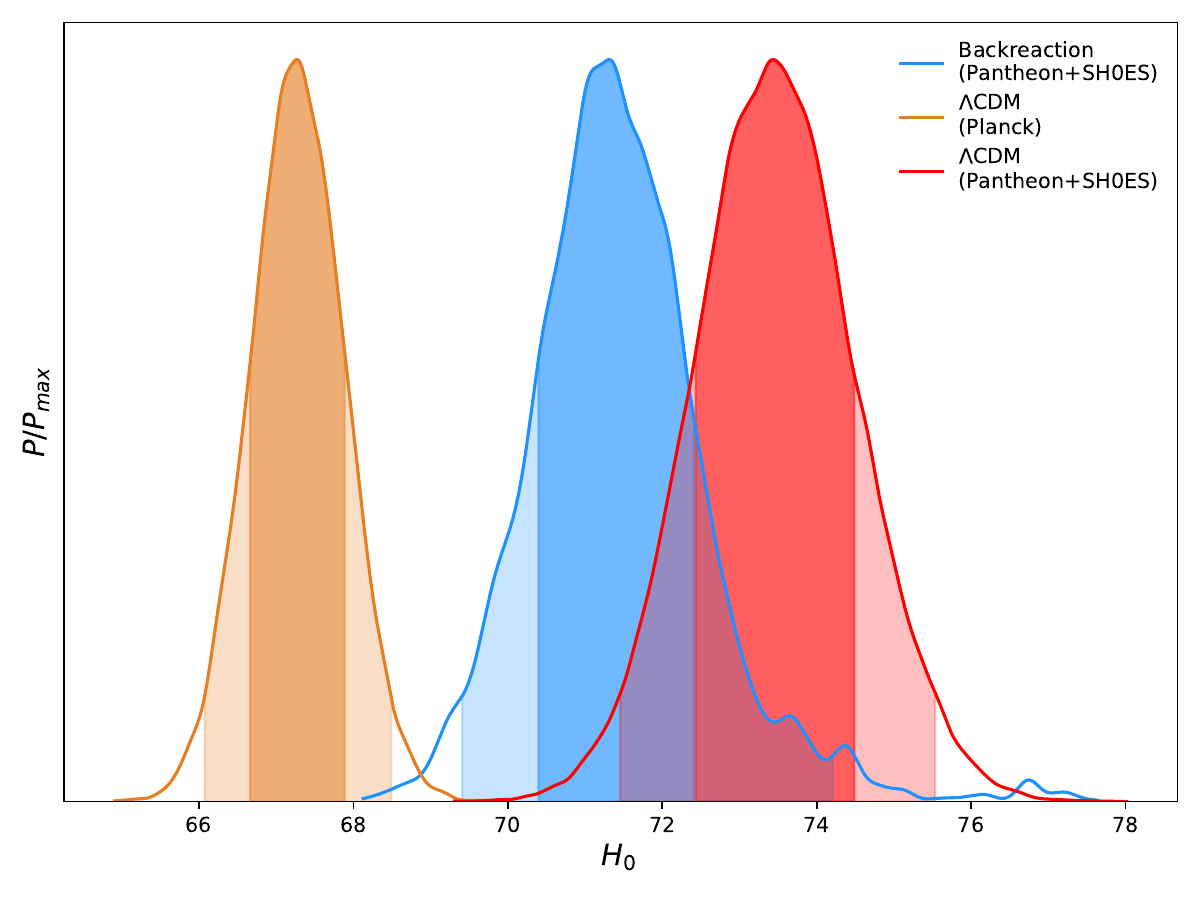}
\caption{\label{fig:H0s_all} This figure shows the distribution of $H_{0}$ obtained from Planck TT,TE,EE+lowE dataset (obtained by incorporating the $\Lambda$CDM model) \cite{Planck}, and for the PantheonPlus+SH0ES dataset using $\Lambda$CDM model and the backreaction model along with their 68 $\%$ and 95 $\%$ confidence intervals in different shades.}
\end{figure*}

 \autoref{tab:H0s_M} shows the 68\% (1$\sigma$) and 95\% (2$\sigma$) confidence limits for $H_{0}$ for three cases. The first case (i) is reported in \cite{Planck} for the Planck PR3 dataset and the $\Lambda$CDM cosmological model, where $H_0$ is calculated using the TT, TE,EE+lowE method. (ii) is reported in \cite{Pantheon_likelihood} for $\Lambda$CDM model using PantheonPlus+SH0ES dataset. (iii) is calculated here in this study for our backreaction model and PantheonPlus+SH0ES dataset. These quantities were then plotted in \autoref{fig:H0s_all}.

\par In \autoref{fig:H0s_all}, the tension between the $H_0$ estimates derived from the Planck PR3 data and the PantheonPlus+SH0ES data within the $\Lambda$CDM framework is illustrated. For the PantheonPlus+SH0ES dataset, the \(H_0\) distribution inferred in the backreaction framework shows significantly reduced tension with the Planck PR3 value, compared to the corresponding estimate derived within the standard cosmological model.

\par In summary, our methodology proceeds as follows. We first constrain the backreaction model parameters, including \(h\) and \(M\), using the low-redshift PantheonPlus+SH0ES dataset. With these best-fit parameter values, we then compute \(\tau_{\mathrm{reion}}\) within the framework of the tanh parameterization of 
reionization. This yields the corresponding prediction for 
\(\tau_{\mathrm{reion}}\) in the combined context of the tanh reionization model 
and the backreaction cosmology. \\
For comparison, we also evaluate \(\tau_{\mathrm{reion}}\) for the tanh 
reionization model within the \(\Lambda\)CDM framework, considering two cases: 
one in which \(H_0\) is constrained by the PantheonPlus+SH0ES dataset and another 
in which it is constrained by the Planck PR3 dataset. These results are shown in \autoref{fig:optical_depth_final} alongside the value of \(\tau_{\mathrm{reion}}\) reported by Planck Collaboration VI (2020), which depends only weakly on the underlying cosmological model, as well as with estimates from other works that use different observational datasets. We find that the central value of \(\tau_{\mathrm{reion}}\) inferred from the backreaction model parameters constrained by the PantheonPlus+SH0ES dataset agrees more closely with the values reported in numerous other studies using a variety of datasets (bottom half of \autoref{fig:optical_depth_final}) than does the corresponding \(\Lambda\)CDM-based estimate of \(\tau_{\mathrm{reion}}\) obtained from the same dataset.

\par The value of $H_0$ for the backreaction model (from $h$ in \autoref{fig:corner_MCMC_M}) that we obtain is within 2$\sigma$ agreement with the value obtained by \cite{Pantheon_likelihood} for the $\Lambda$CDM model using the same PantheonPlus+SH0ES dataset in \autoref{fig:H0s_all}. The value of $H_0$ obtained from the PantheonPlus+SH0ES dataset is not completely independent of the cosmological model, although the dependence on the cosmological model is very minimal. We also consider the value of $H_{0}$ for the Planck PR3 dataset for the $\Lambda$CDM cosmological model. This value is cosmological model dependent. The same figure shows that $H_0$ for the backreaction model from the PantheonPlus+SH0ES dataset is closer to that of the $\Lambda$CDM model from the Planck PR3 dataset, implying that the Hubble tension in this case is somewhat alleviated.


\section{Conclusions}\label{sec:conclusions}

\par Recent observations indicate that the Universe contains an inhomogeneous matter distribution at considerably large scales \cite{Labini_2009, wiegand_scale, lopez}. The effects of these inhomogeneities on various cosmological phenomena warrant scrutiny. In the present study, we explore the optical depth to reionization and the associated Hubble tension under the impact of backreaction from matter inhomogeneities.

\par In this analysis, we use the widely used Buchert formalism \cite{Buchert, Buchert2001} of averaging over inhomogeneities to evaluate the backreaction effect. The Buchert framework facilitates the relation of theoretically evaluated quantities with observables such as redshift and angular diameter distance \cite{rasanen1, rasanen2, Koksbang_2019, Koksbang2, Koksbang3}. Within this framework, we construct a backreaction model of multiple subregions with a Gaussian parameter distribution to mimic the Universe containing multiple voids and structures at the present epoch. We employ the covariant scheme to relate the theoretically evaluated parameters from the backreaction model to observational quantities. 

\par Using this backreaction framework, we compute the reionization optical depth, $\tau_{reion}$, for a particular reionization scenario, the tanh reionization model, considering both the standard $\Lambda$CDM cosmology and the backreaction cosmological model. The backreaction model we employ modifies the Hubble evolution, making it desirable to constrain its parameters using observational results. To correlate the backreaction model with observation data, we obtain the marginalized posterior densities for each model parameter through Markov Chain Monte Carlo (MCMC) simulations using PantheonPlus+SH0ES supernova Ia data \cite{Pantheon_likelihood, Panth_full_dataset_Scolnic_2022}. 

\par The MCMC analysis results in the dimensionless Hubble constant, one of the backreaction model parameters, $h = 0.7146^{+0.0089}_{-0.011}$, corresponding to a 68$\%$ (1$\sigma$) confidence interval. We next calculate $\tau_{reion}$ by treating this parameter as a derived parameter in the MCMC analysis. We then compare the obtained values with $\tau_{reion}$ from the Planck PR3 dataset. We also calculate $\tau_{reion}$ for the $\Lambda$CDM and the tanh reionization model using two datasets: PantheonPlus+SH0ES and Planck PR3.

\par The analysis carried out in this study shows that the Planck results, 
\(\tau_{\mathrm{reion}} = 0.0544^{+0.0070}_{-0.0081}\) and 
\(z_{\mathrm{reion}} = 7.68 \pm 0.79\), are compatible within one to two sigma in the backreaction model. In particular, these values are achievable within the 68\% (1$\sigma$) confidence interval for \(h = 0.7146^{+0.0089}_{-0.011}\), as constrained by the PantheonPlus+SH0ES dataset. Furthermore, the analysis indicates a reduction in the Hubble tension. Notably, this result is obtained without invoking exotic physics or non-standard models of dark matter and dark energy. In contrast, several previous studies have explored such extensions or modifications in attempts to address the Hubble tension 
\cite{DE_Hubble_tension, DE_Hubble_tension_2, Hubble_tension_review1, DM_DE_Hbble_tension_2, DM_DE_Hubble_tension, IDE_hubble_tension, Hubble_tension_review1}.

\par Before concluding, it may be noted that in the present study, we have relied exclusively on PantheonPlus+SH0ES data to place constraints on our backreaction model. Incorporating additional datasets, such as DESI \cite{DESI}, BAO \cite{eBOSS}, H0LiCOW \cite{H0LiCOW}, and ACT \cite{ACT}, could yield significantly tighter bounds on the model parameters. Though a comprehensive analysis incorporating all available cosmological datasets would provide the most rigorous test of the backreaction model, we focus exclusively on supernova data in this work. The results obtained herein motivate extending our analysis to include a larger suite of cosmological observations in future work.

\section{Acknowledgments}
SSP and SM would like to thank the Council of Scientific and Industrial Research (CSIR), Govt of India, for funding through the CSIR-SRF-NET fellowship. The author R acknowledges financial support from Grant PID2024-158938NB-I00, funded by MICIU/AEI/10.13039/501100011033 and by ``ERDF/EU -- A way of making Europe'', as well as from Project SA097P24, funded by the Junta de Castilla y Le\'on. R also acknowledges support from  ``Theoretical Astroparticle Physics'' (TAsP) iniziativa specifica of INFN, where part of this work was carried out. R further acknowledges support from the S. N. Bose National Centre for Basic Sciences for her visit, where part of this work has been done. We also acknowledge the use of the HPC facility Pegasus at IUCAA, Pune, India.

\bibliographystyle{apsrev4-2-author-truncate}
\bibliography{Hubble_tension.bib}

@article{He_abundance,
  title = {Big-bang nucleosynthesis enters the precision era},
  author = {Schramm, David N. and Turner, Michael S.},
  journal = {Rev. Mod. Phys.},
  volume = {70},
  issue = {1},
  pages = {303--318},
  numpages = {0},
  year = {1998},
  month = {Jan},
  publisher = {American Physical Society},
  doi = {10.1103/RevModPhys.70.303},
  url = {https://link.aps.org/doi/10.1103/RevModPhys.70.303}
}

@article{lopez,
    author = {Lopez, Alexia M and Clowes, Roger G and Williger, Gerard M},
    title = "{A Giant Arc on the Sky}",
    journal = {Monthly Notices of the Royal Astronomical Society},
    volume = {516},
    number = {2},
    pages = {1557-1572},
    year = {2022},
    month = {08},
    issn = {0035-8711},
    doi = {10.1093/mnras/stac2204},
    url = {https://doi.org/10.1093/mnras/stac2204},
}

@article{Ali_2017,
	doi = {10.1088/1475-7516/2017/01/054},
	url = {https://doi.org/10.1088/1475-7516/2017/01/054},
	year = 2017,
	month = {jan},
	publisher = {{IOP} Publishing},
	volume = {2017},
	number = {01},
	pages = {054--054},
	author = {Amna Ali and A.S. Majumdar},
	title = {Future evolution in a backreaction model and the analogous scalar field cosmology},
	journal = {Journal of Cosmology and Astroparticle Physics}
}

@Article{mcmc1,
author={Haario, Heikki
and Laine, Marko
and Mira, Antonietta
and Saksman, Eero},
title={DRAM: Efficient adaptive MCMC},
journal={Statistics and Computing},
year={2006},
month={Dec},
day={01},
volume={16},
number={4},
pages={339-354},
issn={1573-1375},
doi={10.1007/s11222-006-9438-0},
url={https://doi.org/10.1007/s11222-006-9438-0}
}

@article{mcmc2,
 ISSN = {13507265},
 URL = {http://www.jstor.org/stable/3318737},
 doi = {10.2307/3318737},
 author = {Heikki Haario and Eero Saksman and Johanna Tamminen},
 journal = {Bernoulli},
 number = {2},
 pages = {223--242},
 publisher = {International Statistical Institute (ISI) and Bernoulli Society for Mathematical Statistics and Probability},
 title = {An Adaptive Metropolis Algorithm},
 urldate = {2023-06-06},
 volume = {7},
 year = {2001}
}

@article{union,
doi = {10.1088/0004-637X/746/1/85},
url = {https://dx.doi.org/10.1088/0004-637X/746/1/85},
year = {2012},
month = {jan},
publisher = {The American Astronomical Society},
volume = {746},
number = {1},
pages = {85},
author = {N. Suzuki and others},
title = {THE HUBBLE SPACE TELESCOPE CLUSTER SUPERNOVA SURVEY. V. IMPROVING THE DARK-ENERGY CONSTRAINTS ABOVE z $>$ 1 AND BUILDING AN EARLY-TYPE-HOSTED SUPERNOVA SAMPLE$^{\star}$},
journal = {The Astrophysical Journal}
}

@article{wiegand_scale,
    author = {Wiegand, Alexander and Buchert, Thomas and Ostermann, Matthias},
    title = "{Direct Minkowski Functional analysis of large redshift surveys: a new high-speed code tested on the luminous red galaxy Sloan Digital Sky Survey-DR7 catalogue}",
    journal = {Monthly Notices of the Royal Astronomical Society},
    volume = {443},
    number = {1},
    pages = {241-259},
    year = {2014},
    month = {07},
    issn = {0035-8711},
    doi = {10.1093/mnras/stu1118},
    url = {https://doi.org/10.1093/mnras/stu1118},
}

@Inbook{Ellis1984,
	author="Ellis, G. F. R.",
	editor="Bertotti, B. and de Felice, F. and Pascolini, A.",
	title="Relativistic Cosmology: Its Nature, Aims and Problems",
	bookTitle="General Relativity and Gravitation: Invited Papers and Discussion Reports of the 10th International Conference on General Relativity and Gravitation, Padua, July 3--8, 1983",
	year="1984",
	publisher="Springer Netherlands",
	address="Dordrecht",
	pages="215--288",
	isbn="978-94-009-6469-3",
	doi="10.1007/978-94-009-6469-3_14",
	url="https://doi.org/10.1007/978-94-009-6469-3_14"
}

@article{Futamase,
  title = {Approximation Scheme for Constructing a Clumpy Universe in General Relativity},
  author = {Futamase, Toshifumi},
  journal = {Phys. Rev. Lett.},
  volume = {61},
  issue = {19},
  pages = {2175--2178},
  numpages = {0},
  year = {1988},
  month = {Nov},
  publisher = {American Physical Society},
  doi = {10.1103/PhysRevLett.61.2175},
  url = {https://link.aps.org/doi/10.1103/PhysRevLett.61.2175}
}

@Article{Zalaletdinov1992,
author={Zalaletdinov, Roustam M.},
title={Averaging out the Einstein equations},
journal={General Relativity and Gravitation},
year={1992},
month={Oct},
day={01},
volume={24},
number={10},
pages={1015-1031},
issn={1572-9532},
doi={10.1007/BF00756944},
url={https://doi.org/10.1007/BF00756944}
}

@Article{Zalaletdinov1993,
author={Zalaletdinov, Roustam M.},
title={Towards a theory of macroscopic gravity},
journal={General Relativity and Gravitation},
year={1993},
month={Jul},
day={01},
volume={25},
number={7},
pages={673-695},
issn={1572-9532},
doi={10.1007/BF00756937},
url={https://doi.org/10.1007/BF00756937}
}

@Article{Buchert,
author={Buchert, Thomas},
title={On Average Properties of Inhomogeneous Fluids in General Relativity: Dust Cosmologies},
journal={General Relativity and Gravitation},
year={2000},
month={Jan},
day={01},
volume={32},
number={1},
pages={105-125},
issn={1572-9532},
doi={10.1023/A:1001800617177},
url={https://doi.org/10.1023/A:1001800617177}
}

@article{Weigand_et_al,
  title = {Multiscale cosmology and structure-emerging dark energy: A plausibility analysis},
  author = {Wiegand, Alexander and Buchert, Thomas},
  journal = {Phys. Rev. D},
  volume = {82},
  issue = {2},
  pages = {023523},
  numpages = {24},
  year = {2010},
  month = {Jul},
  publisher = {American Physical Society},
  doi = {10.1103/PhysRevD.82.023523},
  url = {https://link.aps.org/doi/10.1103/PhysRevD.82.023523}
}

@article{Coley,
  title = {Cosmological Solutions in Macroscopic Gravity},
  author = {Coley, A. A. and Pelavas, N. and Zalaletdinov, R. M.},
  journal = {Phys. Rev. Lett.},
  volume = {95},
  issue = {15},
  pages = {151102},
  numpages = {4},
  year = {2005},
  month = {Oct},
  publisher = {American Physical Society},
  doi = {10.1103/PhysRevLett.95.151102},
  url = {https://link.aps.org/doi/10.1103/PhysRevLett.95.151102}
}

@Article{Buchert2001,
author={Buchert, Thomas},
title={On Average Properties of Inhomogeneous Fluids in General Relativity: Perfect Fluid Cosmologies},
journal={General Relativity and Gravitation},
year={2001},
month={Aug},
day={01},
volume={33},
number={8},
pages={1381-1405},
issn={1572-9532},
doi={10.1023/A:1012061725841},
url={https://doi.org/10.1023/A:1012061725841}
}

@article{Korzynski_2010,
doi = {10.1088/0264-9381/27/10/105015},
url = {https://dx.doi.org/10.1088/0264-9381/27/10/105015},
year = {2010},
month = {apr},
publisher = {},
volume = {27},
number = {10},
pages = {105015},
author = {Mikołaj Korzyński},
title = {Covariant coarse graining of inhomogeneous dust flow in general relativity},
journal = {Classical and Quantum Gravity}
}

@article{Clifton,
  title = {An exact quantification of backreaction in relativistic cosmology},
  author = {Clifton, Timothy and Rosquist, Kjell and Tavakol, Reza},
  journal = {Phys. Rev. D},
  volume = {86},
  issue = {4},
  pages = {043506},
  numpages = {12},
  year = {2012},
  month = {Aug},
  publisher = {American Physical Society},
  doi = {10.1103/PhysRevD.86.043506},
  url = {https://link.aps.org/doi/10.1103/PhysRevD.86.043506}
}

@article{Skarke,
  title = {Inhomogeneity implies accelerated expansion},
  author = {Skarke, Harald},
  journal = {Phys. Rev. D},
  volume = {89},
  issue = {4},
  pages = {043506},
  numpages = {5},
  year = {2014},
  month = {Feb},
  publisher = {American Physical Society},
  doi = {10.1103/PhysRevD.89.043506},
  url = {https://link.aps.org/doi/10.1103/PhysRevD.89.043506}
}

@article{Buchert_2015,
doi = {10.1088/0264-9381/32/21/215021},
url = {https://dx.doi.org/10.1088/0264-9381/32/21/215021},
year = {2015},
month = {oct},
publisher = {IOP Publishing},
volume = {32},
number = {21},
pages = {215021},
author = {T Buchert and M Carfora and G F R Ellis and E W Kolb and M A H MacCallum and J J Ostrowski and S Räsänen and B F Roukema and L Andersson and A A Coley and D L Wiltshire},
title = {Is there proof that backreaction of inhomogeneities is irrelevant in cosmology?},
journal = {Classical and Quantum Gravity}
}

@article{Buchert4,
author = {Buchert, Thomas and Coley, Alan A. and Kleinert, Hagen and Roukema, Boudewijn F. and Wiltshire, David L.},
title = {Observational challenges for the standard FLRW model},
journal = {International Journal of Modern Physics D},
volume = {25},
number = {03},
pages = {1630007},
year = {2016},
doi = {10.1142/S021827181630007X},

URL = { 
    
        https://doi.org/10.1142/S021827181630007X
    
    

}
}

@Article{Bose2013,
author={Bose, Nilok
and Majumdar, A. S.},
title={Effect of cosmic backreaction on the future evolution of an accelerating universe},
journal={General Relativity and Gravitation},
year={2013},
month={Oct},
day={01},
volume={45},
number={10},
pages={1971-1987},
issn={1572-9532},
doi={10.1007/s10714-013-1572-3},
url={https://doi.org/10.1007/s10714-013-1572-3}
}

@article{bose,
    author = {Bose, Nilok and Majumdar, A. S.},
    title = "{Future deceleration due to cosmic backreaction in presence of the event horizon}",
    journal = {Monthly Notices of the Royal Astronomical Society: Letters},
    volume = {418},
    number = {1},
    pages = {L45-L48},
    year = {2011},
    month = {11},
    issn = {1745-3925},
    doi = {10.1111/j.1745-3933.2011.01140.x},
    url = {https://doi.org/10.1111/j.1745-3933.2011.01140.x},
}

@article{Labini_2009,
doi = {10.1209/0295-5075/86/49001},
url = {https://dx.doi.org/10.1209/0295-5075/86/49001},
year = {2009},
month = {may},
publisher = {},
volume = {86},
number = {4},
pages = {49001},
author = {F. Sylos Labini and N. L. Vasilyev and L. Pietronero and Y. V. Baryshev},
title = {Absence of self-averaging and of homogeneity in the large-scale galaxy distribution},
journal = {Europhysics Letters}
}

@article{Rasanen_2004,
doi = {10.1088/1475-7516/2004/02/003},
url = {https://dx.doi.org/10.1088/1475-7516/2004/02/003},
year = {2004},
month = {feb},
publisher = {},
volume = {2004},
number = {02},
pages = {003},
author = {Syksy Räsänen},
title = {Dark energy from back-reaction},
journal = {Journal of Cosmology and Astroparticle Physics}
}

@inbook{wiltshire,
author = { Davis L. Wiltshire},
title = {DARK ENERGY WITHOUT DARK ENERGY},
booktitle = {Dark Matter in Astroparticle and Particle Physics},
chapter = {},
pages = {565-596},
doi = {10.1142/9789812814357_0053},
URL = {https://www.worldscientific.com/doi/abs/10.1142/9789812814357_0053}
}

@article{Kolb_2006,
doi = {10.1088/1367-2630/8/12/322},
url = {https://dx.doi.org/10.1088/1367-2630/8/12/322},
year = {2006},
month = {dec},
publisher = {},
volume = {8},
number = {12},
pages = {322},
author = {Edward W Kolb and Sabino Matarrese and Antonio Riotto},
title = {On cosmic acceleration without dark energy},
journal = {New Journal of Physics}
}

@article{Koksbang_2019,
doi = {10.1088/1475-7516/2019/10/036},
url = {https://dx.doi.org/10.1088/1475-7516/2019/10/036},
year = {2019},
month = {oct},
publisher = {},
volume = {2019},
number = {10},
pages = {036},
author = {S.M. Koksbang},
title = {Another look at redshift drift and the backreaction conjecture},
journal = {Journal of Cosmology and Astroparticle Physics}
}

@article{buchert_rasanen,
author = {Buchert, Thomas and R\"{a}s\"{a}nen, Syksy},
title = {Backreaction in Late-Time Cosmology},
journal = {Annual Review of Nuclear and Particle Science},
volume = {62},
number = {1},
pages = {57-79},
year = {2012},
doi = {10.1146/annurev.nucl.012809.104435},

URL = { 
    
        https://doi.org/10.1146/annurev.nucl.012809.104435
    
    

}
}

@BOOK{weinberg,
	author = {{Weinberg}, Steven},
	title = "{Gravitation and Cosmology: Principles and Applications of the General Theory of Relativity}",
	year = 1972,
	adsurl = {https://ui.adsabs.harvard.edu/abs/1972gcpa.book.....W}
}

@article{Planck,
	author = {{Planck Collaboration} and {Aghanim, N.} and {Akrami, Y.} and {Ashdown, M.} and {Aumont, J.} and {Baccigalupi, C.} and {Ballardini, M.} and {Banday, A. J.} and {Barreiro, R. B.} and {Bartolo, N.} and {Basak, S.} and {Battye, R.} and {Benabed, K.} and {Bernard, J.-P.} and {Bersanelli, M.} and {Bielewicz, P.} and {Bock, J. J.} and {Bond, J. R.} and {Borrill, J.} and {Bouchet, F. R.} and {Boulanger, F.} and {Bucher, M.} and {Burigana, C.} and {Butler, R. C.} and {Calabrese, E.} and {Cardoso, J.-F.} and {Carron, J.} and {Challinor, A.} and {Chiang, H. C.} and {Chluba, J.} and {Colombo, L. P. L.} and {Combet, C.} and {Contreras, D.} and {Crill, B. P.} and {Cuttaia, F.} and {de Bernardis, P.} and {de Zotti, G.} and {Delabrouille, J.} and {Delouis, J.-M.} and {Di Valentino, E.} and {Diego, J. M.} and {Dor\'e, O.} and {Douspis, M.} and {Ducout, A.} and {Dupac, X.} and {Dusini, S.} and {Efstathiou, G.} and {Elsner, F.} and {En\ss{}lin, T. A.} and {Eriksen, H. K.} and {Fantaye, Y.} and {Farhang, M.} and {Fergusson, J.} and {Fernandez-Cobos, R.} and {Finelli, F.} and {Forastieri, F.} and {Frailis, M.} and {Fraisse, A. A.} and {Franceschi, E.} and {Frolov, A.} and {Galeotta, S.} and {Galli, S.} and {Ganga, K.} and {G\'enova-Santos, R. T.} and {Gerbino, M.} and {Ghosh, T.} and {Gonz\'alez-Nuevo, J.} and {G\'orski, K. M.} and {Gratton, S.} and {Gruppuso, A.} and {Gudmundsson, J. E.} and {Hamann, J.} and {Handley, W.} and {Hansen, F. K.} and {Herranz, D.} and {Hildebrandt, S. R.} and {Hivon, E.} and {Huang, Z.} and {Jaffe, A. H.} and {Jones, W. C.} and {Karakci, A.} and {Keih\"anen, E.} and {Keskitalo, R.} and {Kiiveri, K.} and {Kim, J.} and {Kisner, T. S.} and {Knox, L.} and {Krachmalnicoff, N.} and {Kunz, M.} and {Kurki-Suonio, H.} and {Lagache, G.} and {Lamarre, J.-M.} and {Lasenby, A.} and {Lattanzi, M.} and {Lawrence, C. R.} and {Le Jeune, M.} and {Lemos, P.} and {Lesgourgues, J.} and {Levrier, F.} and {Lewis, A.} and {Liguori, M.} and {Lilje, P. B.} and {Lilley, M.} and {Lindholm, V.} and {L\'opez-Caniego, M.} and {Lubin, P. M.} and {Ma, Y.-Z.} and {Mac\'{\i}as-P\'erez, J. F.} and {Maggio, G.} and {Maino, D.} and {Mandolesi, N.} and {Mangilli, A.} and {Marcos-Caballero, A.} and {Maris, M.} and {Martin, P. G.} and {Martinelli, M.} and {Mart\'{\i}nez-Gonz\'alez, E.} and {Matarrese, S.} and {Mauri, N.} and {McEwen, J. D.} and {Meinhold, P. R.} and {Melchiorri, A.} and {Mennella, A.} and {Migliaccio, M.} and {Millea, M.} and {Mitra, S.} and {Miville-Desch\^enes, M.-A.} and {Molinari, D.} and {Montier, L.} and {Morgante, G.} and {Moss, A.} and {Natoli, P.} and {N\o{}rgaard-Nielsen, H. U.} and {Pagano, L.} and {Paoletti, D.} and {Partridge, B.} and {Patanchon, G.} and {Peiris, H. V.} and {Perrotta, F.} and {Pettorino, V.} and {Piacentini, F.} and {Polastri, L.} and {Polenta, G.} and {Puget, J.-L.} and {Rachen, J. P.} and {Reinecke, M.} and {Remazeilles, M.} and {Renzi, A.} and {Rocha, G.} and {Rosset, C.} and {Roudier, G.} and {Rubi\~no-Mart\'{\i}n, J. A.} and {Ruiz-Granados, B.} and {Salvati, L.} and {Sandri, M.} and {Savelainen, M.} and {Scott, D.} and {Shellard, E. P. S.} and {Sirignano, C.} and {Sirri, G.} and {Spencer, L. D.} and {Sunyaev, R.} and {Suur-Uski, A.-S.} and {Tauber, J. A.} and {Tavagnacco, D.} and {Tenti, M.} and {Toffolatti, L.} and {Tomasi, M.} and {Trombetti, T.} and {Valenziano, L.} and {Valiviita, J.} and {Van Tent, B.} and {Vibert, L.} and {Vielva, P.} and {Villa, F.} and {Vittorio, N.} and {Wandelt, B. D.} and {Wehus, I. K.} and {White, M.} and {White, S. D. M.} and {Zacchei, A.} and {Zonca, A.}},
	title = {Planck 2018 results - VI. Cosmological parameters},
	DOI= "10.1051/0004-6361/201833910",
	url= "https://doi.org/10.1051/0004-6361/201833910",
	journal = {A\&A},
	year = 2020,
	volume = 641,
	pages = "A6",
}

@article{Koksbang2,
    author = {Koksbang, S M},
    title = "{Observations in statistically homogeneous, locally inhomogeneous cosmological toy models without FLRW backgrounds}",
    journal = {Monthly Notices of the Royal Astronomical Society: Letters},
    volume = {498},
    number = {1},
    pages = {L135-L139},
    year = {2020},
    month = {09},
    issn = {1745-3925},
    doi = {10.1093/mnrasl/slaa146},
    url = {https://doi.org/10.1093/mnrasl/slaa146}
}

@article{Koksbang_PRL,
  title = {Searching for Signals of Inhomogeneity Using Multiple Probes of the Cosmic Expansion Rate $H(z)$},
  author = {Koksbang, S. M.},
  journal = {Phys. Rev. Lett.},
  volume = {126},
  issue = {23},
  pages = {231101},
  numpages = {6},
  year = {2021},
  month = {Jun},
  publisher = {American Physical Society},
  doi = {10.1103/PhysRevLett.126.231101},
  url = {https://link.aps.org/doi/10.1103/PhysRevLett.126.231101}
}

@article{Rasanen_2008,
doi = {10.1088/1475-7516/2008/04/026},
url = {https://dx.doi.org/10.1088/1475-7516/2008/04/026},
year = {2008},
month = {apr},
publisher = {},
volume = {2008},
number = {04},
pages = {026},
author = {Syksy Räsänen},
title = {Evaluating backreaction with the peak model of structure formation},
journal = {Journal of Cosmology and Astroparticle Physics}
}

@article{Pandey_2022,
doi = {10.1088/1475-7516/2022/06/021},
url = {https://dx.doi.org/10.1088/1475-7516/2022/06/021},
year = {2022},
month = {jun},
publisher = {IOP Publishing},
volume = {2022},
number = {06},
pages = {021},
author = {Shashank Shekhar Pandey and Arnab Sarkar and Amna Ali and A.S. Majumdar},
title = {Effect of inhomogeneities on the propagation of gravitational waves from binaries of compact objects},
journal = {Journal of Cosmology and Astroparticle Physics}
}

@article{Gasperini_2011,
doi = {10.1088/1475-7516/2011/07/008},
url = {https://dx.doi.org/10.1088/1475-7516/2011/07/008},
year = {2011},
month = {jul},
publisher = {},
volume = {2011},
number = {07},
pages = {008},
author = {M. Gasperini and  G. Marozzi and  F. Nugier and  G. Veneziano},
title = {Light-cone averaging in cosmology: formalism and applications},
journal = {Journal of Cosmology and Astroparticle Physics}
}

@article{rasanen1,
	doi = {10.1088/1475-7516/2009/02/011},
	url = {https://doi.org/10.1088/1475-7516/2009/02/011},
	year = 2009,
	month = {feb},
	publisher = {{IOP} Publishing},
	volume = {2009},
	number = {02},
	pages = {011--011},
	author = {Syksy Räsänen},
	title = {Light propagation in statistically homogeneous and isotropic dust universes},
	journal = {Journal of Cosmology and Astroparticle Physics}
}

@Article{Pandey2023,
author={Pandey, Shashank Shekhar
and Sarkar, Arnab
and Ali, Amna
and Majumdar, Archan S.},
title={Viscous attenuation of gravitational waves propagating through an inhomogeneous background},
journal={The European Physical Journal C},
year={2023},
month={May},
day={24},
volume={83},
number={5},
pages={435},
issn={1434-6052},
doi={10.1140/epjc/s10052-023-11605-9},
url={https://doi.org/10.1140/epjc/s10052-023-11605-9}
}

@article{rasanen2,
	doi = {10.1088/1475-7516/2010/03/018},
	url = {https://doi.org/10.1088/1475-7516/2010/03/018},
	year = 2010,
	month = {mar},
	publisher = {{IOP} Publishing},
	volume = {2010},
	number = {03},
	pages = {018--018},
	author = {Syksy Räsänen},
	title = {Light propagation in statistically homogeneous and isotropic universes with general matter content},
	journal = {Journal of Cosmology and Astroparticle Physics}
}

@article{Koksbang3,
	doi = {10.1088/1361-6382/ab376c},
	url = {https://doi.org/10.1088/1361-6382/ab376c},
	year = 2019,
	month = {aug},
	publisher = {{IOP} Publishing},
	volume = {36},
	number = {18},
	pages = {185004},
	author = {S M Koksbang},
	title = {Towards statistically homogeneous and isotropic perfect fluid universes with cosmic backreaction},
	journal = {Classical and Quantum Gravity}
}

@article{Koksbang4,
doi = {10.1088/1475-7516/2020/11/061},
url = {https://dx.doi.org/10.1088/1475-7516/2020/11/061},
year = {2020},
month = {nov},
publisher = {},
volume = {2020},
number = {11},
pages = {061},
author = {S.M. Koksbang},
title = {On the relationship between mean observations, spatial averages and the Dyer-Roeder approximation in Einstein-Straus models},
journal = {Journal of Cosmology and Astroparticle Physics}
}

@article{Koksbang5,
	doi = {10.1088/1475-7516/2016/01/009},
	url = {https://doi.org/10.1088/1475-7516/2016/01/009},
	year = 2016,
	month = {jan},
	publisher = {{IOP} Publishing},
	volume = {2016},
	number = {01},
	pages = {009--009},
	author = {S.M. Koksbang and S. Hannestad},
	title = {Redshift drift in an inhomogeneous universe: averaging and the backreaction conjecture},
	journal = {Journal of Cosmology and Astroparticle Physics}
}

@article{Koksbang6,
  title = {Quantifying effects of inhomogeneities and curvature on gravitational wave standard siren measurements of $H(z)$},
  author = {Koksbang, S. M.},
  journal = {Phys. Rev. D},
  volume = {106},
  issue = {6},
  pages = {063514},
  numpages = {12},
  year = {2022},
  month = {Sep},
  publisher = {American Physical Society},
  doi = {10.1103/PhysRevD.106.063514},
  url = {https://link.aps.org/doi/10.1103/PhysRevD.106.063514}
}

@article{Koksbang7,
  title = {Cosmic backreaction and the mean redshift drift from symbolic regression},
  author = {Koksbang, S. M.},
  journal = {Phys. Rev. D},
  volume = {107},
  issue = {10},
  pages = {103522},
  numpages = {19},
  year = {2023},
  month = {May},
  publisher = {American Physical Society},
  doi = {10.1103/PhysRevD.107.103522},
  url = {https://link.aps.org/doi/10.1103/PhysRevD.107.103522}
}

@article{Koksbang8,
  title = {Machine Learning Cosmic Backreaction and Its Effects on Observations},
  author = {Koksbang, S. M.},
  journal = {Phys. Rev. Lett.},
  volume = {130},
  issue = {20},
  pages = {201003},
  numpages = {6},
  year = {2023},
  month = {May},
  publisher = {American Physical Society},
  doi = {10.1103/PhysRevLett.130.201003},
  url = {https://link.aps.org/doi/10.1103/PhysRevLett.130.201003}
}

@article{Rasanen_2006_accelerated,
doi = {10.1088/1475-7516/2006/11/003},
url = {https://dx.doi.org/10.1088/1475-7516/2006/11/003},
year = {2006},
month = {nov},
publisher = {},
volume = {2006},
number = {11},
pages = {003},
author = {Syksy Räsänen},
title = {Accelerated expansion from structure formation},
journal = {Journal of Cosmology and Astroparticle Physics}
}

@article{Koksbang2023,
  title = {Cosmic backreaction and the mean redshift drift from symbolic regression},
  author = {Koksbang, S. M.},
  journal = {Phys. Rev. D},
  volume = {107},
  issue = {10},
  pages = {103522},
  numpages = {19},
  year = {2023},
  month = {May},
  publisher = {American Physical Society},
  doi = {10.1103/PhysRevD.107.103522},
  url = {https://link.aps.org/doi/10.1103/PhysRevD.107.103522}
}

@article{griffiths_tau,
    author = {Griffiths, Louise M. and Barbosa, Domingos and Liddle, Andrew R.},
    title = {Cosmic microwave background constraints on the epoch of reionization},
    journal = {Monthly Notices of the Royal Astronomical Society},
    volume = {308},
    number = {3},
    pages = {854-862},
    year = {1999},
    month = {09},
    issn = {0035-8711},
    doi = {10.1046/j.1365-8711.1999.02777.x},
    url = {https://doi.org/10.1046/j.1365-8711.1999.02777.x},
    eprint = {https://academic.oup.com/mnras/article-pdf/308/3/854/2962118/308-3-854.pdf},
}

@article{Lewis2008,
  title = {Cosmological parameters from WMAP 5-year temperature maps},
  author = {Lewis, Antony},
  journal = {Phys. Rev. D},
  volume = {78},
  issue = {2},
  pages = {023002},
  numpages = {9},
  year = {2008},
  month = {Jul},
  publisher = {American Physical Society},
  doi = {10.1103/PhysRevD.78.023002},
  url = {https://link.aps.org/doi/10.1103/PhysRevD.78.023002}
}

@ARTICLE{Tripp1998,
       author = {{Tripp}, Robert},
        title = "{A two-parameter luminosity correction for Type IA supernovae}",
      journal = {\aap},
         year = 1998,
        month = mar,
       volume = {331},
        pages = {815-820},
       adsurl = {https://ui.adsabs.harvard.edu/abs/1998A&A...331..815T}
}

@article{Pantheon_likelihood,
doi = {10.3847/1538-4357/ac8e04},
url = {https://dx.doi.org/10.3847/1538-4357/ac8e04},
year = {2022},
month = {oct},
publisher = {The American Astronomical Society},
volume = {938},
number = {2},
pages = {110},
author = {Brout, Dillon and Scolnic, Dan and Popovic, Brodie and Riess, Adam G. and Carr, Anthony and Zuntz, Joe and Kessler, Rick and Davis, Tamara M. and Hinton, Samuel and Jones, David and Kenworthy, W. D’Arcy and Peterson, Erik R. and Said, Khaled and Taylor, Georgie and Ali, Noor and Armstrong, Patrick and Charvu, Pranav and Dwomoh, Arianna and Meldorf, Cole and Palmese, Antonella and Qu, Helen and Rose, Benjamin M. and Sanchez, Bruno and Stubbs, Christopher W. and Vincenzi, Maria and Wood, Charlotte M. and Brown, Peter J. and Chen, Rebecca and Chambers, Ken and Coulter, David A. and Dai, Mi and Dimitriadis, Georgios and Filippenko, Alexei V. and Foley, Ryan J. and Jha, Saurabh W. and Kelsey, Lisa and Kirshner, Robert P. and Möller, Anais and Muir, Jessie and Nadathur, Seshadri and Pan, Yen-Chen and Rest, Armin and Rojas-Bravo, Cesar and Sako, Masao and Siebert, Matthew R. and Smith, Mat and Stahl, Benjamin E. and Wiseman, Phil},
title = {The Pantheon+ Analysis: Cosmological Constraints},
journal = {The Astrophysical Journal}
}

@article{Riess_2022_fidu,
doi = {10.3847/2041-8213/ac5c5b},
url = {https://dx.doi.org/10.3847/2041-8213/ac5c5b},
year = {2022},
month = {jul},
publisher = {The American Astronomical Society},
volume = {934},
number = {1},
pages = {L7},
author = {Riess, Adam G. and Yuan, Wenlong and Macri, Lucas M. and Scolnic, Dan and Brout, Dillon and Casertano, Stefano and Jones, David O. and Murakami, Yukei and Anand, Gagandeep S. and Breuval, Louise and Brink, Thomas G. and Filippenko, Alexei V. and Hoffmann, Samantha and Jha, Saurabh W. and D’arcy Kenworthy, W. and Mackenty, John and Stahl, Benjamin E. and Zheng, WeiKang},
title = {A Comprehensive Measurement of the Local Value of the Hubble Constant with 1 km s−1 Mpc−1 Uncertainty from the Hubble Space Telescope and the SH0ES Team},
journal = {The Astrophysical Journal Letters}
}

@article{Panth_full_dataset_Scolnic_2022,
doi = {10.3847/1538-4357/ac8b7a},
url = {https://dx.doi.org/10.3847/1538-4357/ac8b7a},
year = {2022},
month = {oct},
publisher = {The American Astronomical Society},
volume = {938},
number = {2},
pages = {113},
author = {Scolnic, Dan and Brout, Dillon and Carr, Anthony and Riess, Adam G. and Davis, Tamara M. and Dwomoh, Arianna and Jones, David O. and Ali, Noor and Charvu, Pranav and Chen, Rebecca and Peterson, Erik R. and Popovic, Brodie and Rose, Benjamin M. and Wood, Charlotte M. and Brown, Peter J. and Chambers, Ken and Coulter, David A. and Dettman, Kyle G. and Dimitriadis, Georgios and Filippenko, Alexei V. and Foley, Ryan J. and Jha, Saurabh W. and Kilpatrick, Charles D. and Kirshner, Robert P. and Pan, Yen-Chen and Rest, Armin and Rojas-Bravo, Cesar and Siebert, Matthew R. and Stahl, Benjamin E. and Zheng, WeiKang},
title = {The Pantheon+ Analysis: The Full Data Set and Light-curve Release},
journal = {The Astrophysical Journal}
}

@article{pandey_analyzing_2024,
    title = {Analyzing the 21-cm signal brightness temperature in the Universe with inhomogeneities},
  author = {Pandey, Shashank Shekhar and Halder, Ashadul and Majumdar, A. S.},
  journal = {Phys. Rev. D},
  volume = {110},
  issue = {4},
  pages = {043531},
  numpages = {18},
  year = {2024},
  month = {Aug},
  publisher = {American Physical Society},
  doi = {10.1103/PhysRevD.110.043531},
  url = {https://link.aps.org/doi/10.1103/PhysRevD.110.043531}
}

@article{LCDM_challenges,
title = {Challenges for ΛCDM: An update},
journal = {New Astronomy Reviews},
volume = {95},
pages = {101659},
year = {2022},
issn = {1387-6473},
doi = {https://doi.org/10.1016/j.newar.2022.101659},
url = {https://www.sciencedirect.com/science/article/pii/S1387647322000185},
author = {L. Perivolaropoulos and F. Skara}
}

@Article{LCDM_small_scale_problems,
AUTHOR = {Del Popolo, Antonino and Le Delliou, Morgan},
TITLE = {Small Scale Problems of the ?CDM Model: A Short Review},
JOURNAL = {Galaxies},
VOLUME = {5},
YEAR = {2017},
NUMBER = {1},
ARTICLE-NUMBER = {17},
URL = {https://www.mdpi.com/2075-4434/5/1/17},
ISSN = {2075-4434},
DOI = {10.3390/galaxies5010017}
}

@article{Hubble_tension_review1,
doi = {10.1088/1361-6382/ac086d},
url = {https://dx.doi.org/10.1088/1361-6382/ac086d},
year = {2021},
month = {jul},
publisher = {IOP Publishing},
volume = {38},
number = {15},
pages = {153001},
author = {Di Valentino, Eleonora and Mena, Olga and Pan, Supriya and Visinelli, Luca and Yang, Weiqiang and Melchiorri, Alessandro and Mota, David F and Riess, Adam G and Silk, Joseph},
title = {In the realm of the Hubble tension—a review of solutions*},
journal = {Classical and Quantum Gravity}
}

@Article{Hubble_tension_review2,
AUTHOR = {Hu, Jian-Ping and Wang, Fa-Yin},
TITLE = {Hubble Tension: The Evidence of New Physics},
JOURNAL = {Universe},
VOLUME = {9},
YEAR = {2023},
NUMBER = {2},
ARTICLE-NUMBER = {94},
URL = {https://www.mdpi.com/2218-1997/9/2/94},
ISSN = {2218-1997},
DOI = {10.3390/universe9020094}
}

@Article{Hubble_tension3,
AUTHOR = {Cervantes-Cota, Jorge L. and Galindo-Uribarri, Salvador and Smoot, George F.},
TITLE = {The Unsettled Number: Hubble’s Tension},
JOURNAL = {Universe},
VOLUME = {9},
YEAR = {2023},
NUMBER = {12},
ARTICLE-NUMBER = {501},
URL = {https://www.mdpi.com/2218-1997/9/12/501},
ISSN = {2218-1997},
DOI = {10.3390/universe9120501}
}

@article{DE_Hubble_tension,
  title = {Dark energy at early times, the Hubble parameter, and the string axiverse},
  author = {Karwal, Tanvi and Kamionkowski, Marc},
  journal = {Phys. Rev. D},
  volume = {94},
  issue = {10},
  pages = {103523},
  numpages = {10},
  year = {2016},
  month = {Nov},
  publisher = {American Physical Society},
  doi = {10.1103/PhysRevD.94.103523},
  url = {https://link.aps.org/doi/10.1103/PhysRevD.94.103523}
}

@Article{DE_Hubble_tension_2,
author={Zhao, Gong-Bo
and Raveri, Marco
and Pogosian, Levon
and Wang, Yuting
and Crittenden, Robert G.
and Handley, Will J.
and Percival, Will J.
and Beutler, Florian
and Brinkmann, Jonathan
and Chuang, Chia-Hsun
and Cuesta, Antonio J.
and Eisenstein, Daniel J.
and Kitaura, Francisco-Shu
and Koyama, Kazuya
and L'Huillier, Benjamin
and Nichol, Robert C.
and Pieri, Matthew M.
and Rodriguez-Torres, Sergio
and Ross, Ashley J.
and Rossi, Graziano
and S{\'a}nchez, Ariel G.
and Shafieloo, Arman
and Tinker, Jeremy L.
and Tojeiro, Rita
and Vazquez, Jose A.
and Zhang, Hanyu},
title={Dynamical dark energy in light of the latest observations},
journal={Nature Astronomy},
year={2017},
month={Sep},
day={01},
volume={1},
number={9},
pages={627-632},
issn={2397-3366},
doi={10.1038/s41550-017-0216-z},
url={https://doi.org/10.1038/s41550-017-0216-z}
}

@article{DM_DE_Hubble_tension,
  title = {Probing the interaction between dark matter and dark energy in the presence of massive neutrinos},
  author = {Kumar, Suresh and Nunes, Rafael C.},
  journal = {Phys. Rev. D},
  volume = {94},
  issue = {12},
  pages = {123511},
  numpages = {7},
  year = {2016},
  month = {Dec},
  publisher = {American Physical Society},
  doi = {10.1103/PhysRevD.94.123511},
  url = {https://link.aps.org/doi/10.1103/PhysRevD.94.123511}
}

@article{IDE_hubble_tension,
doi = {10.1088/1475-7516/2022/01/024},
url = {https://dx.doi.org/10.1088/1475-7516/2022/01/024},
year = {2022},
month = {jan},
publisher = {IOP Publishing},
volume = {2022},
number = {01},
pages = {024},
author = {Johnson, Joseph P. and Sangwan, Archana and Shankaranarayanan, S.},
title = {Observational constraints and predictions of the interacting dark sector with field-fluid mapping},
journal = {Journal of Cosmology and Astroparticle Physics}
}

@article{DM_DE_Hbble_tension_2,
doi = {10.1088/1475-7516/2019/11/044},
url = {https://dx.doi.org/10.1088/1475-7516/2019/11/044},
year = {2019},
month = {nov},
publisher = {},
volume = {2019},
number = {11},
pages = {044},
author = {Yang, Weiqiang and Pan, Supriya and Vagnozzi, Sunny and Valentino, Eleonora Di and Mota, David F. and Capozziello, Salvatore},
title = {Dawn of the dark: unified dark sectors and the EDGES Cosmic Dawn 21-cm signal},
journal = {Journal of Cosmology and Astroparticle Physics}
}

@article{Hubble_tension_inhomogeneous1,
    author = {Kasai, Masumi and Futamase, Toshifumi},
    title = {A possible solution to the Hubble constant discrepancy: Cosmology where the local volume expansion is driven by the domain average density},
    journal = {Progress of Theoretical and Experimental Physics},
    volume = {2019},
    number = {7},
    pages = {073E01},
    year = {2019},
    month = {07},
    issn = {2050-3911},
    doi = {10.1093/ptep/ptz066},
    url = {https://doi.org/10.1093/ptep/ptz066},
    eprint = {https://academic.oup.com/ptep/article-pdf/2019/7/073E01/28914425/ptz066.pdf},
}

@article{Hubble_tension_inhomogeneous2,
doi = {10.3847/2041-8213/aadf8c},
url = {https://dx.doi.org/10.3847/2041-8213/aadf8c},
year = {2018},
month = {sep},
publisher = {The American Astronomical Society},
volume = {865},
number = {1},
pages = {L4},
author = {Macpherson, Hayley J. and Lasky, Paul D. and Price, Daniel J.},
title = {The Trouble with Hubble: Local versus Global Expansion Rates in Inhomogeneous Cosmological Simulations with Numerical Relativity},
journal = {The Astrophysical Journal Letters}
}

@article{Hubble_tension_inhomogeneous3,
doi = {10.1088/1475-7516/2024/05/126},
url = {https://dx.doi.org/10.1088/1475-7516/2024/05/126},
year = {2024},
month = {may},
publisher = {IOP Publishing},
volume = {2024},
number = {05},
pages = {126},
author = {Miura, Taishi and Tanaka, Takahiro},
title = {Remarks on overestimating the effects of inhomogeneities on the Hubble constant},
journal = {Journal of Cosmology and Astroparticle Physics}
}

@article{Hubble_tension_inhomogeneous4,
title = {Hubble tension and matter inhomogeneities: A theoretical perspective},
journal = {Annals of Physics},
volume = {458},
pages = {169444},
year = {2023},
issn = {0003-4916},
doi = {https://doi.org/10.1016/j.aop.2023.169444},
url = {https://www.sciencedirect.com/science/article/pii/S0003491623002464},
author = {Marco {San Martín} and Carlos Rubio}
}

@article{Hubble_tension_inhomogeneous5,
  title = {Emerging spatial curvature can resolve the tension between high-redshift CMB and low-redshift distance ladder measurements of the Hubble constant},
  author = {Bolejko, Krzysztof},
  journal = {Phys. Rev. D},
  volume = {97},
  issue = {10},
  pages = {103529},
  numpages = {9},
  year = {2018},
  month = {May},
  publisher = {American Physical Society},
  doi = {10.1103/PhysRevD.97.103529},
  url = {https://link.aps.org/doi/10.1103/PhysRevD.97.103529}
}

@article{Halder_future_2023,
doi = {10.1088/1475-7516/2023/08/064},
url = {https://dx.doi.org/10.1088/1475-7516/2023/08/064},
year = {2023},
month = {aug},
publisher = {IOP Publishing},
volume = {2023},
number = {08},
pages = {064},
author = {Ashadul Halder and Shashank Shekhar Pandey and A.S. Majumdar},
title = {Future deceleration due to backreaction in a Universe with multiple inhomogeneous domains},
journal = {Journal of Cosmology and Astroparticle Physics}
}

@article{Heinesen_Hubble_tension,
doi = {10.1088/1361-6382/ab954b},
url = {https://dx.doi.org/10.1088/1361-6382/ab954b},
year = {2020},
month = {jul},
publisher = {IOP Publishing},
volume = {37},
number = {16},
pages = {164001},
author = {Heinesen, Asta and Buchert, Thomas},
title = {Solving the curvature and Hubble parameter inconsistencies through structure formation-induced curvature},
journal = {Classical and Quantum Gravity}
}

@article{cosmology_intertwined,
title = {Cosmology intertwined: A review of the particle physics, astrophysics, and cosmology associated with the cosmological tensions and anomalies},
journal = {Journal of High Energy Astrophysics},
volume = {34},
pages = {49-211},
year = {2022},
issn = {2214-4048},
doi = {https://doi.org/10.1016/j.jheap.2022.04.002},
url = {https://www.sciencedirect.com/science/article/pii/S2214404822000179},
author = {Elcio Abdalla and Guillermo Franco Abellán and Amin Aboubrahim and Adriano Agnello and Özgür Akarsu and Yashar Akrami and George Alestas and Daniel Aloni and Luca Amendola and Luis A. Anchordoqui and Richard I. Anderson and Nikki Arendse and Marika Asgari and Mario Ballardini and Vernon Barger and Spyros Basilakos and Ronaldo C. Batista and Elia S. Battistelli and Richard Battye and Micol Benetti and David Benisty and Asher Berlin and Paolo {de Bernardis} and Emanuele Berti and Bohdan Bidenko and Simon Birrer and John P. Blakeslee and Kimberly K. Boddy and Clecio R. Bom and Alexander Bonilla and Nicola Borghi and François R. Bouchet and Matteo Braglia and Thomas Buchert and Elizabeth Buckley-Geer and Erminia Calabrese and Robert R. Caldwell and David Camarena and Salvatore Capozziello and Stefano Casertano and Geoff C.-F. Chen and Jens Chluba and Angela Chen and Hsin-Yu Chen and Anton Chudaykin and Michele Cicoli and Craig J. Copi and Fred Courbin and Francis-Yan Cyr-Racine and Bożena Czerny and Maria Dainotti and Guido D'Amico and Anne-Christine Davis and Javier {de Cruz Pérez} and Jaume {de Haro} and Jacques Delabrouille and Peter B. Denton and Suhail Dhawan and Keith R. Dienes and Eleonora {Di Valentino} and Pu Du and Dominique Eckert and Celia Escamilla-Rivera and Agnès Ferté and Fabio Finelli and Pablo Fosalba and Wendy L. Freedman and Noemi Frusciante and Enrique Gaztañaga and William Giarè and Elena Giusarma and Adrià Gómez-Valent and Will Handley and Ian Harrison and Luke Hart and Dhiraj Kumar Hazra and Alan Heavens and Asta Heinesen and Hendrik Hildebrandt and J. Colin Hill and Natalie B. Hogg and Daniel E. Holz and Deanna C. Hooper and Nikoo Hosseininejad and Dragan Huterer and Mustapha Ishak and Mikhail M. Ivanov and Andrew H. Jaffe and In Sung Jang and Karsten Jedamzik and Raul Jimenez and Melissa Joseph and Shahab Joudaki and Marc Kamionkowski and Tanvi Karwal and Lavrentios Kazantzidis and Ryan E. Keeley and Michael Klasen and Eiichiro Komatsu and Léon V.E. Koopmans and Suresh Kumar and Luca Lamagna and Ruth Lazkoz and Chung-Chi Lee and Julien Lesgourgues and Jackson {Levi Said} and Tiffany R. Lewis and Benjamin L'Huillier and Matteo Lucca and Roy Maartens and Lucas M. Macri and Danny Marfatia and Valerio Marra and Carlos J.A.P. Martins and Silvia Masi and Sabino Matarrese and Arindam Mazumdar and Alessandro Melchiorri and Olga Mena and Laura Mersini-Houghton and James Mertens and Dinko Milaković and Yuto Minami and Vivian Miranda and Cristian Moreno-Pulido and Michele Moresco and David F. Mota and Emil Mottola and Simone Mozzon and Jessica Muir and Ankan Mukherjee and Suvodip Mukherjee and Pavel Naselsky and Pran Nath and Savvas Nesseris and Florian Niedermann and Alessio Notari and Rafael C. Nunes and Eoin {Ó Colgáin} and Kayla A. Owens and Emre Özülker and Francesco Pace and Andronikos Paliathanasis and Antonella Palmese and Supriya Pan and Daniela Paoletti and Santiago E. {Perez Bergliaffa} and Leandros Perivolaropoulos and Dominic W. Pesce and Valeria Pettorino and Oliver H.E. Philcox and Levon Pogosian and Vivian Poulin and Gaspard Poulot and Marco Raveri and Mark J. Reid and Fabrizio Renzi and Adam G. Riess and Vivian I. Sabla and Paolo Salucci and Vincenzo Salzano and Emmanuel N. Saridakis and Bangalore S. Sathyaprakash and Martin Schmaltz and Nils Schöneberg and Dan Scolnic and Anjan A. Sen and Neelima Sehgal and Arman Shafieloo and M.M. Sheikh-Jabbari and Joseph Silk and Alessandra Silvestri and Foteini Skara and Martin S. Sloth and Marcelle Soares-Santos and Joan {Solà Peracaula} and Yu-Yang Songsheng and Jorge F. Soriano and Denitsa Staicova and Glenn D. Starkman and István Szapudi and Elsa M. Teixeira and Brooks Thomas and Tommaso Treu and Emery Trott and Carsten {van de Bruck} and J. Alberto Vazquez and Licia Verde and Luca Visinelli and Deng Wang and Jian-Min Wang and Shao-Jiang Wang and Richard Watkins and Scott Watson and John K. Webb and Neal Weiner and Amanda Weltman and Samuel J. Witte and Radosław Wojtak and Anil Kumar Yadav and Weiqiang Yang and Gong-Bo Zhao and Miguel Zumalacárregui}
}

@article{Leandros1,
title = {Challenges for ΛCDM: An update},
journal = {New Astronomy Reviews},
volume = {95},
pages = {101659},
year = {2022},
issn = {1387-6473},
doi = {https://doi.org/10.1016/j.newar.2022.101659},
url = {https://www.sciencedirect.com/science/article/pii/S1387647322000185},
author = {L. Perivolaropoulos and F. Skara}
}

@Article{eleonora1,
AUTHOR = {Di Valentino, Eleonora},
TITLE = {Challenges of the Standard Cosmological Model},
JOURNAL = {Universe},
VOLUME = {8},
YEAR = {2022},
NUMBER = {8},
ARTICLE-NUMBER = {399},
URL = {https://www.mdpi.com/2218-1997/8/8/399},
ISSN = {2218-1997},
DOI = {10.3390/universe8080399}
}

@article{Nils_review,
   author = "Verde, Licia and Schöneberg, Nils and Gil-Marín, Héctor",
   title = "A Tale of Many H0", 
   journal= "Annual Review of Astronomy and Astrophysics",
   year = "2024",
   volume = "62",
   number = "Volume 62, 2024",
   pages = "287-331",
   doi = "https://doi.org/10.1146/annurev-astro-052622-033813",
   url = "https://www.annualreviews.org/content/journals/10.1146/annurev-astro-052622-033813",
   publisher = "Annual Reviews",
   issn = "1545-4282",
   type = "Journal Article",
  }

@article{CosmoVerse,
title = {The CosmoVerse White Paper: Addressing observational tensions in cosmology with systematics and fundamental physics},
journal = {Physics of the Dark Universe},
volume = {49},
pages = {101965},
year = {2025},
issn = {2212-6864},
doi = {https://doi.org/10.1016/j.dark.2025.101965},
url = {https://www.sciencedirect.com/science/article/pii/S221268642500158X},
author = {Eleonora {Di Valentino} and Jackson Levi Said and Adam Riess and Agnieszka Pollo and Vivian Poulin and Adrià Gómez-Valent and Amanda Weltman and Antonella Palmese and Caroline D. Huang and Carsten van de Bruck and Chandra Shekhar Saraf and Cheng-Yu Kuo and Cora Uhlemann and Daniela Grandón and Dante Paz and Dominique Eckert and Elsa M. Teixeira and Emmanuel N. Saridakis and Eoin Ó Colgáin and Florian Beutler and Florian Niedermann and Francesco Bajardi and Gabriela Barenboim and Giulia Gubitosi and Ilaria Musella and Indranil Banik and Istvan Szapudi and Jack Singal and Jaume Haro Cases and Jens Chluba and Jesús Torrado and Jurgen Mifsud and Karsten Jedamzik and Khaled Said and Konstantinos Dialektopoulos and Laura Herold and Leandros Perivolaropoulos and Lei Zu and Lluís Galbany and Louise Breuval and Luca Visinelli and Luis A. Escamilla and Luis A. Anchordoqui and M.M. Sheikh-Jabbari and Margherita Lembo and Maria Giovanna Dainotti and Maria Vincenzi and Marika Asgari and Martina Gerbino and Matteo Forconi and Michele Cantiello and Michele Moresco and Micol Benetti and Nils Schöneberg and Özgür Akarsu and Rafael C. Nunes and Reginald Christian Bernardo and Ricardo Chávez and Richard I. Anderson and Richard Watkins and Salvatore Capozziello and Siyang Li and Sunny Vagnozzi and Supriya Pan and Tommaso Treu and Vid Irsic and Will Handley and William Giarè and Yukei Murakami and Abdolali Banihashemi and Adèle Poudou and Alan Heavens and Alan Kogut and Alba Domi and Aleksander Łukasz Lenart and Alessandro Melchiorri and Alessandro Vadalà and Alexandra Amon and Alexander Bonilla Rivera and Alexander Reeves and Alexander Zhuk and Alfio Bonanno and Ali Övgün and Alice Pisani and Alireza Talebian and Amare Abebe and Amin Aboubrahim and Ana Luisa González Morán and András Kovács and Andreas Lymperis and Andreas Papatriantafyllou and Andrew R. Liddle and Andronikos Paliathanasis and Andrzej Borowiec and Anil Kumar Yadav and Anita Yadav and Anjan Ananda Sen and Anjitha John William and Anne Christine Davis and Anowar J. Shajib and Anthony Walters and Anto Idicherian Lonappan and Anton Chudaykin and Antonio Capodagli and Antonio da Silva and Antonio De Felice and Antonio Racioppi and Araceli Soler Oficial and Ariadna Montiel and Arianna Favale and Armando Bernui and Arrianne Crystal Velasco and Asta Heinesen and Athanasios Bakopoulos and Athanasios Chatzistavrakidis and Bahman Khanpour and Bangalore S. Sathyaprakash and Bartek Zgirski and Benjamin L’Huillier and Benoit Famaey and Bhuvnesh Jain and Bing Zhang and Biswajit Karmakar and Branko Dragovich and Brooks Thomas and Carlos Correa and Carlos G. Boiza and Catarina Marques and Celia Escamilla-Rivera and Charalampos Tzerefos and Chi Zhang and Chiara De Leo and Christian Pfeifer and Christine Lee and Christo Venter and Cláudio Gomes and Clecio Roque De bom and Cristian Moreno-Pulido and Damianos Iosifidis and Dan Grin and Daniel Blixt and Dan Scolnic and Daniele Oriti and Daria Dobrycheva and Dario Bettoni and David Benisty and David Fernández-Arenas and David L. Wiltshire and David Sanchez Cid and David Tamayo and David Valls-Gabaud and Davide Pedrotti and Deng Wang and Denitsa Staicova and Despoina Totolou and Diego Rubiera-Garcia and Dinko Milaković and Dominic W. Pesce and Dominique Sluse and Duško Borka and Ebrahim Yusofi and Elena Giusarma and Elena Terlevich and Elena Tomasetti and Elias C. Vagenas and Elisa Fazzari and Elisa G.M. Ferreira and Elvis Barakovic and Emanuela Dimastrogiovanni and Emil Brinch Holm and Emil Mottola and Emre Özülker and Enrico Specogna and Enzo Brocato and Erik Jensko and Erika Antonette Enriquez and Esha Bhatia and Fabio Bresolin and Felipe Avila and Filippo Bouchè and Flavio Bombacigno and Fotios K. Anagnostopoulos and Francesco Pace and Francesco Sorrenti and Francisco S.N. Lobo and Frédéric Courbin and Frode K. Hansen and Greg Sloan and Gabriel Farrugia and Gabriel Lynch and Gabriela Garcia-Arroyo and Gabriella Raimondo and Gaetano Lambiase and Gagandeep S. Anand and Gaspard Poulot and Genly Leon and Gerasimos Kouniatalis and Germano Nardini and Géza Csörnyei and Giacomo Galloni and Giada Bargiacchi and Giannis Papagiannopoulos and Giovanni Montani and Giovanni Otalora and Giulia De Somma and Giuliana Fiorentino and Giuseppe Fanizza and Giuseppe Gaetano Luciano and Giuseppe Sarracino and Gonzalo J. Olmo and Goran S. Djordjević and Guadalupe Cañas-Herrera and Hanyu Cheng and Harry Desmond and Hassan Abdalla and Houzun Chen and Hsu-Wen Chiang and Hume A. Feldman and Hussain Gohar and Ido Ben-Dayan and Ignacio Sevilla-Noarbe and Ignatios Antoniadis and Ilim Cimdiker and Inês S. Albuquerque and Ioannis D. Gialamas and Ippocratis Saltas and Iryna Vavilova and Isidro Gómez-Vargas and Ismael Ayuso and Ismailov Nariman Zeynalabdi and Ivan De Martino and Ivonne Zavala and J. Alberto Vázquez and Jacobo Asorey and Janusz Gluza and Javier Rubio and Jenny G. Sorce and Jenny Wagner and Jeremy Sakstein and Jessica Santiago and Jim Braatz and Joan Solà Peracaula and John Blakeslee and John Webb and Jose A.R. Cembranos and José Pedro Mimoso and Joseph Jensen and Juan García-Bellido and Judit Prat and Kathleen Sammut and Kay Lehnert and Keith R. Dienes and Kishan Deka and Konrad Kuijken and Krishna Naidoo and László Árpád Gergely and Laur Järv and Laura Mersini-Houghton and Leila L. Graef and Léo Vacher and Levon Pogosian and Lilia Anguelova and Lindita Hamolli and Lu Yin and Luca Caloni and Luca Izzo and Lucas Macri and Luis E. Padilla and Luz Ángela García and Maciej Bilicki and Mahdi Najafi and Manolis Plionis and Manuel Gonzalez-Espinoza and Manuel Hohmann and Marcel A. van der Westhuizen and Marcella Marconi and Marcin Postolak and Marco de Cesare and Marco Regis and Marek Biesiada and Maret Einasto and Margus Saal and Maria Caruana and Maria Petronikolou and Mariam Bouhmadi-López and Mariana Melo and Mariaveronica De Angelis and Marie-Noëlle Célérier and Marina Cortês and Mark Reid and Markus Michael Rau and Martin S. Sloth and Martti Raidal and Masahiro Takada and Masoume Reyhani and Massimiliano Romanello and Massimo Marengo and Mathias Garny and Matías Leizerovich and Matteo Martinelli and Matteo Tagliazucchi and Mehmet Demirci and Miguel A.S. Pinto and Miguel A. Sabogal and Miguel A. García-Aspeitia and Milan Milošević and Mina Ghodsi and Mustapha Ishak and Nelson J. Nunes and Nick Samaras and Nico Hamaus and Nico Schuster and Nicola Borghi and Nicola Deiosso and Nicola Tamanini and Nicolao Fornengo and Nihan Katırcı and Nikolaos E. Mavromatos and Nicholas Petropoulos and Nikolina Šarčević and Nils A. Nilsson and Nima Khosravi and Noemi Frusciante and Octavian Postavaru and Oem Trivedi and Oleksii Sokoliuk and Olga Mena and Paloma Morilla and Paolo Campeti and Paolo Salucci and Paula Boubel and Paweł Bielewicz and Pekka Heinämäki and Petar Suman and Petros Asimakis and Pierros Ntelis and Pran Nath and Predrag Jovanović and Purba Mukherjee and Radosław Wojtak and Rafaela Gsponer and Rafid H. Dejrah and Rahul Shah and Rasmi Hajjar and Rebecca Briffa and Rebecca Habas and Reggie C. Pantig and Renier Mendoza and Riccardo Della Monica and Richard Stiskalek and Rishav Roshan and Rita B. Neves and Roberto Molinaro and Roberto Terlevich and Rocco D’Agostino and Rodrigo Sandoval-Orozco and Ronaldo C. Batista and  Ruchika and Ruth Lazkoz and Saeed Rastgoo and Sahar Mohammadi and Salvatore Samuele Sirletti and Sandeep Haridasu and Sanjay Mandal and Saurya Das and Sebastian Bahamonde and Sebastian Grandis and Sebastian Trojanowski and Sergei D. Odintsov and Sergij Mazurenko and Shahab Joudaki and Sherry H. Suyu and Shouvik Roy Choudhury and Shruti Bhatporia and Shun-Sheng Li and Simeon Bird and Simon Birrer and Simone Paradiso and Simony Santos da Costa and Sofia Contarini and Sophie Henrot-Versillé and Spyros Basilakos and Stefano Casertano and Stefano Gariazzo and Stylianos A. Tsilioukas and Surajit Kalita and Suresh Kumar and Susana J. Landau and Sveva Castello and Swayamtrupta Panda and Tanja Petrushevska and Thanasis Karakasis and Thejs Brinckmann and Tiago B. Gonçalves and Tiziano Schiavone and Tom Abel and Tomi Koivisto and Torsten Bringmann and Umut Demirbozan and Utkarsh Kumar and Valerio Marra and Maurice H.P.M. van Putten and Vasileios Kalaitzidis and Vasiliki A. Mitsou and Vasilios Zarikas and Vedad Pasic and Venus Keus and Verónica Motta and Vesna Borka Jovanović and Víctor H. Cárdenas and Vincenzo Ripepi and Vincenzo Salzano and Violetta Impellizzeri and Vitor da Fonseca and Vittorio Ghirardini and Vladas Vansevičius and Weiqiang Yang and Wojciech Hellwing and Xin Ren and Yu-Min Hu and Yuejia Zhai and Abdul Malik Sultan and Abdurakhmon Nosirov and Adrienn Pataki and Alessandro Santoni and Aliya Batool and Amlan Chakraborty and Aneta Wojnar and Arman Tursunov and Avik De and Ayush Hazarika and Baojiu Li and Benjamin Bose and Bivudutta Mishra and Bobomurat Ahmedov and Claudia Scóccola and Crescenzo Tortora and D’Arcy Kenworthy and Daniel E. Holz and David F. Mota and David S. Pereira and Devon M. Williams and Dillon Brout and Dong Ha Lee and Eduardo Guendelman and Edward Olex and Emanuelly Silva and Emre Onur Kahya and Eva-Maria Mueller and Felipe Andrade-Oliveira and Feven Markos Hunde and F.R. Joaquim and Florian Pacaud and Francis-Yan Cyr-Racine and F. Pozo Nuñez and Gábor Rácz and Gene Carlo Belinario and Geraint F. Lewis and Gergely Dálya and Giorgio Laverda and Guido Risaliti and Guillermo Franco-Abellán and Hayden Zammit and Hayley Camilleri and Helene M. Courtois and Hooman Moradpour and Igor de Oliveira Cardoso Pedreira and Ilídio Lopes and István Csabai and James W. Rohlf and Jana Bogdanoska and Javier de Cruz Pérez and Joan Bachs-Esteban and Joseph Sultana and Julien Lesgourgues and Jun-Qian Jiang and Karem Peñaló Castillo and Kimet Jusufi and Lavinia Heisenberg and Laxmipriya Pati and Lón V.E. Koopmans and Lokesh kumar Duchaniya and Lucas Lombriser and María Pérez Garrote and Mariano Domínguez and Marine Samsonyan and Mark Pace and Martin Krššák and Masroor C. Pookkillath and Matteo Peronaci and Matteo Piani and Matthildi Raftogianni and Meet J. Vyas and Melina Michalopoulou and Merab Gogberashvili and Michael Klasen and Michele Cicoli and Miguel Quartin and Miguel Zumalacárregui and Milan S. Dimitrijević and Milos Dordevic and Mindaugas Karčiauskas and Morgan {Le Delliou} and Nastassia Grimm and Nicolás Augusto Kozameh and Nicoleta Voicu and Nicolina Pop and Nikos Chatzifotis and Odil Yunusov and Oliver Fabio Piattella and Pedro da Silveira Ferreira and Péter Raffai and Peter Schupp and Pilar Ruiz-Lapuente and Pradyumn Kumar Sahoo and Roberto V. Maluf and Ruth Durrer and S.A. Kadam and Sabino Matarrese and Samuel Brieden and Santiago González-Gaitán and Santosh V. Lohakare and Scott Watson and Shao-Jiang Wang and Simão Marques Nunes and Soumya Chakrabarti and Subinoy Das and Suvodip Mukherjee and Tajron Jurić and Tessa Baker and Theodoros Nakas and Tiago Barreiro and Upala Mukhopadhyay and Veljko Vujčić and Violetta Sagun and Vladimir A. Srećković and Wangzheng Zhang and Yo Toda and Yun-Song Piao and Zahra Davari}
}

@article{Buchert_obs_challenges,
author = {Buchert, Thomas and Coley, Alan A. and Kleinert, Hagen and Roukema, Boudewijn F. and Wiltshire, David L.},
title = {Observational challenges for the standard FLRW model},
journal = {International Journal of Modern Physics D},
volume = {25},
number = {03},
pages = {1630007},
year = {2016},
doi = {10.1142/S021827181630007X},

URL = { 
    
        https://doi.org/10.1142/S021827181630007X
    
    

},
eprint = { 
    
        https://doi.org/10.1142/S021827181630007X
    
    

}
}

@article{Labini_obs_challenges,
doi = {10.1088/0264-9381/28/16/164003},
url = {https://dx.doi.org/10.1088/0264-9381/28/16/164003},
year = {2011},
month = {aug},
publisher = {},
volume = {28},
number = {16},
pages = {164003},
author = {Labini, Francesco Sylos},
title = {Inhomogeneities in the universe},
journal = {Classical and Quantum Gravity}
}

@Article{Verevkin2011,
author={Verevkin, A. O.
and Bukhmastova, Yu. L.
and Baryshev, Yu. V.},
title={The non-uniform distribution of galaxies from data of the SDSS DR7 survey},
journal={Astronomy Reports},
year={2011},
month={Apr},
day={01},
volume={55},
number={4},
pages={324-340},
issn={1562-6881},
doi={10.1134/S1063772911020089},
url={https://doi.org/10.1134/S1063772911020089}
}

@article{Labini_obs_theo_2010,
doi = {10.1088/1742-5468/2010/11/P11029},
url = {https://dx.doi.org/10.1088/1742-5468/2010/11/P11029},
year = {2010},
month = {nov},
publisher = {},
volume = {2010},
number = {11},
pages = {P11029},
author = {Sylos Labini, Francesco and Pietronero, Luciano},
title = {The complex universe: recent observations and theoretical challenges},
journal = {Journal of Statistical Mechanics: Theory and Experiment}
}

@article{Aluri_obs_2023,
doi = {10.1088/1361-6382/acbefc},
url = {https://dx.doi.org/10.1088/1361-6382/acbefc},
year = {2023},
month = {apr},
publisher = {IOP Publishing},
volume = {40},
number = {9},
pages = {094001},
author = {Kumar Aluri, Pavan and Cea, Paolo and Chingangbam, Pravabati and Chu, Ming-Chung and Clowes, Roger G and Hutsemékers, Damien and Kochappan, Joby P and Lopez, Alexia M and Liu, Lang and Martens, Niels C M and Martins, C J A P and Migkas, Konstantinos and Ó Colgáin, Eoin and Pranav, Pratyush and Shamir, Lior and Singal, Ashok K and Sheikh-Jabbari, M M and Wagner, Jenny and Wang, Shao-Jiang and Wiltshire, David L and Yeung, Shek and Yin, Lu and Zhao, Wen},
title = {Is the observable Universe consistent with the cosmological principle?},
journal = {Classical and Quantum Gravity}
}

@article{backreaction_PRL,
  title = {Novel Approach to Cosmological Nonlinearities as an Effective Fluid},
  author = {Giani, Leonardo and von Marttens, Rodrigo and Camilleri, Ryan},
  journal = {Phys. Rev. Lett.},
  volume = {135},
  issue = {7},
  pages = {071004},
  numpages = {9},
  year = {2025},
  month = {Aug},
  publisher = {American Physical Society},
  doi = {10.1103/zr92-m7py},
  url = {https://link.aps.org/doi/10.1103/zr92-m7py}
}

@misc{lopez2025,
      title={Gigaparsec structures are nowhere to be seen in $\Lambda$CDM: an enhanced analysis of LSS in FLAMINGO-10K simulations}, 
      author={A. M Lopez and R. G. Clowes},
      year={2025},
      eprint={2504.14940},
      archivePrefix={arXiv},
      primaryClass={astro-ph.CO},
      url={https://arxiv.org/abs/2504.14940}, 
}

@misc{sawala2025,
      title={The Emperor's New Arc: gigaparsec patterns abound in a $\Lambda$CDM universe}, 
      author={Till Sawala and Meri Teeriaho and Carlos S. Frenk and John Helly and Adrian Jenkins and Gabor Racz and Matthieu Schaller and Joop Schaye},
      year={2025},
      eprint={2502.03515},
      archivePrefix={arXiv},
      primaryClass={astro-ph.CO},
      url={https://arxiv.org/abs/2502.03515}, 
}

@misc{sawala2025_1,
      title={The Giant Arc -- Filament of Figment?}, 
      author={Till Sawala and Meri Teeriaho},
      year={2025},
      eprint={2505.11072},
      archivePrefix={arXiv},
      primaryClass={astro-ph.CO},
      url={https://arxiv.org/abs/2505.11072}, 
}

@article{Hubble_tension_PRL_schwarz,
  title = {Value of ${H}_{0}$ in the Inhomogeneous Universe},
  author = {Ben-Dayan, Ido and Durrer, Ruth and Marozzi, Giovanni and Schwarz, Dominik J.},
  journal = {Phys. Rev. Lett.},
  volume = {112},
  issue = {22},
  pages = {221301},
  numpages = {5},
  year = {2014},
  month = {Jun},
  publisher = {American Physical Society},
  doi = {10.1103/PhysRevLett.112.221301},
  url = {https://link.aps.org/doi/10.1103/PhysRevLett.112.221301}
}

@article{Mukherjee_2025,
  title = {Constraining the Hubble parameter with the 21-cm brightness temperature signal in a universe with inhomogeneities},
  author = {Mukherjee, Subhadeep and Pandey, Shashank Shekhar and Majumdar, A. S.},
  journal = {Phys. Rev. D},
  volume = {112},
  issue = {6},
  pages = {063520},
  numpages = {14},
  year = {2025},
  month = {Sep},
  publisher = {American Physical Society},
  doi = {10.1103/w1wp-tqz2},
  url = {https://link.aps.org/doi/10.1103/w1wp-tqz2}
}

@misc{lodha2025extendeddarkenergyanalysis,
      title={Extended Dark Energy analysis using DESI DR2 BAO measurements}, 
      author={K. Lodha and R. Calderon and W. L. Matthewson and A. Shafieloo and M. Ishak and J. Pan and C. Garcia-Quintero and D. Huterer and G. Valogiannis and L. A. Ureña-López and N. V. Kamble and D. Parkinson and A. G. Kim and G. B. Zhao and J. L. Cervantes-Cota and J. Rohlf and F. Lozano-Rodríguez and J. O. Román-Herrera and M. Abdul-Karim and J. Aguilar and S. Ahlen and O. Alves and U. Andrade and E. Armengaud and A. Aviles and S. BenZvi and D. Bianchi and A. Brodzeller and D. Brooks and E. Burtin and R. Canning and A. Carnero Rosell and L. Casas and F. J. Castander and M. Charles and E. Chaussidon and J. Chaves-Montero and D. Chebat and T. Claybaugh and S. Cole and A. Cuceu and K. S. Dawson and A. de la Macorra and A. de Mattia and N. Deiosso and R. Demina and Arjun Dey and Biprateep Dey and Z. Ding and P. Doel and D. J. Eisenstein and W. Elbers and S. Ferraro and A. Font-Ribera and J. E. Forero-Romero and Lehman H. Garrison and E. Gaztañaga and H. Gil-Marín and S. Gontcho A Gontcho and A. X. Gonzalez-Morales and G. Gutierrez and J. Guy and C. Hahn and M. Herbold and H. K. Herrera-Alcantar and K. Honscheid and C. Howlett and S. Juneau and R. Kehoe and D. Kirkby and T. Kisner and A. Kremin and O. Lahav and C. Lamman and M. Landriau and L. Le Guillou and A. Leauthaud and M. E. Levi and Q. Li and C. Magneville and M. Manera and P. Martini and A. Meisner and J. Mena-Fernández and R. Miquel and J. Moustakas and D. Muñoz Santos and A. Muñoz-Gutiérrez and A. D. Myers and S. Nadathur and G. Niz and H. E. Noriega and E. Paillas and N. Palanque-Delabrouille and W. J. Percival and Matthew M. Pieri and C. Poppett and F. Prada and A. Pérez-Fernández and I. Pérez-Ràfols and C. Ramírez-Pérez and M. Rashkovetskyi and C. Ravoux and A. J. Ross and G. Rossi and V. Ruhlmann-Kleider and L. Samushia and E. Sanchez and D. Schlegel and M. Schubnell and H. Seo and F. Sinigaglia and D. Sprayberry and T. Tan and G. Tarlé and P. Taylor and W. Turner and M. Vargas-Magaña and M. Walther and B. A. Weaver and M. Wolfson and C. Yèche and P. Zarrouk and R. Zhou and H. Zou},
      year={2025},
      eprint={2503.14743},
      archivePrefix={arXiv},
      primaryClass={astro-ph.CO},
      url={https://arxiv.org/abs/2503.14743}, 
}

@article{supernova_deceleration,
    author = {Son, Junhyuk and Lee, Young-Wook and Chung, Chul and Park, Seunghyun and Cho, Hyejeon},
    title = {Strong progenitor age bias in supernova cosmology – II. Alignment with DESI BAO and signs of a non-accelerating universe},
    journal = {Monthly Notices of the Royal Astronomical Society},
    volume = {544},
    number = {1},
    pages = {975-987},
    year = {2025},
    month = {11},
    issn = {0035-8711},
    doi = {10.1093/mnras/staf1685},
    url = {https://doi.org/10.1093/mnras/staf1685},
    eprint = {https://academic.oup.com/mnras/article-pdf/544/1/975/65175128/staf1685.pdf},
}

@article{tau_intro,
	author = {{Wolz, Kevin} and {Krachmalnicoff, Nicoletta} and {Pagano, Luca}},
	title = {Inference of the optical depth to reionization τ from Planck CMB maps with convolutional neural networks},
	DOI= "10.1051/0004-6361/202345982",
	url= "https://doi.org/10.1051/0004-6361/202345982",
	journal = {A\&A},
	year = 2023,
	volume = 676,
	pages = "A30",
}

@article{ tau_intro2,
	author = {{Pagano, L.} and {Delouis, J.-M.} and {Mottet, S.} and {Puget, J.-L.} and {Vibert, L.}},
	title = {Reionization optical depth determination from Planck HFI data with ten percent accuracy},
	DOI= "10.1051/0004-6361/201936630",
	url= "https://doi.org/10.1051/0004-6361/201936630",
	journal = {A\&A},
	year = 2020,
	volume = 635,
	pages = "A99",
}

@article{tau_intro3,
    author = {de Belsunce, Roger and Gratton, Steven and Coulton, William and Efstathiou, George},
    title = {Inference of the optical depth to reionization from low multipole temperature and polarization Planck data},
    journal = {Monthly Notices of the Royal Astronomical Society},
    volume = {507},
    number = {1},
    pages = {1072-1091},
    year = {2021},
    month = {08},
    issn = {0035-8711},
    doi = {10.1093/mnras/stab2215},
    url = {https://doi.org/10.1093/mnras/stab2215},
    eprint = {https://academic.oup.com/mnras/article-pdf/507/1/1072/39812162/stab2215.pdf},
}

@article{Planck_2015,
	author = {{Planck Collaboration} and {Ade, P. A. R.} and {Aghanim, N.} and {Arnaud, M.} and {Ashdown, M.} and {Aumont, J.} and {Baccigalupi, C.} and {Banday, A. J.} and {Barreiro, R. B.} and {Bartlett, J. G.} and {Bartolo, N.} and {Battaner, E.} and {Battye, R.} and {Benabed, K.} and {Benoît, A.} and {Benoit-Lévy, A.} and {Bernard, J.-P.} and {Bersanelli, M.} and {Bielewicz, P.} and {Bock, J. J.} and {Bonaldi, A.} and {Bonavera, L.} and {Bond, J. R.} and {Borrill, J.} and {Bouchet, F. R.} and {Boulanger, F.} and {Bucher, M.} and {Burigana, C.} and {Butler, R. C.} and {Calabrese, E.} and {Cardoso, J.-F.} and {Catalano, A.} and {Challinor, A.} and {Chamballu, A.} and {Chary, R.-R.} and {Chiang, H. C.} and {Chluba, J.} and {Christensen, P. R.} and {Church, S.} and {Clements, D. L.} and {Colombi, S.} and {Colombo, L. P. L.} and {Combet, C.} and {Coulais, A.} and {Crill, B. P.} and {Curto, A.} and {Cuttaia, F.} and {Danese, L.} and {Davies, R. D.} and {Davis, R. J.} and {de Bernardis, P.} and {de Rosa, A.} and {de Zotti, G.} and {Delabrouille, J.} and {Désert, F.-X.} and {Di Valentino, E.} and {Dickinson, C.} and {Diego, J. M.} and {Dolag, K.} and {Dole, H.} and {Donzelli, S.} and {Doré, O.} and {Douspis, M.} and {Ducout, A.} and {Dunkley, J.} and {Dupac, X.} and {Efstathiou, G.} and {Elsner, F.} and {Enßlin, T. A.} and {Eriksen, H. K.} and {Farhang, M.} and {Fergusson, J.} and {Finelli, F.} and {Forni, O.} and {Frailis, M.} and {Fraisse, A. A.} and {Franceschi, E.} and {Frejsel, A.} and {Galeotta, S.} and {Galli, S.} and {Ganga, K.} and {Gauthier, C.} and {Gerbino, M.} and {Ghosh, T.} and {Giard, M.} and {Giraud-Héraud, Y.} and {Giusarma, E.} and {Gjerløw, E.} and {González-Nuevo, J.} and {Górski, K. M.} and {Gratton, S.} and {Gregorio, A.} and {Gruppuso, A.} and {Gudmundsson, J. E.} and {Hamann, J.} and {Hansen, F. K.} and {Hanson, D.} and {Harrison, D. L.} and {Helou, G.} and {Henrot-Versillé, S.} and {Hernández-Monteagudo, C.} and {Herranz, D.} and {Hildebrandt, S. R.} and {Hivon, E.} and {Hobson, M.} and {Holmes, W. A.} and {Hornstrup, A.} and {Hovest, W.} and {Huang, Z.} and {Huffenberger, K. M.} and {Hurier, G.} and {Jaffe, A. H.} and {Jaffe, T. R.} and {Jones, W. C.} and {Juvela, M.} and {Keihänen, E.} and {Keskitalo, R.} and {Kisner, T. S.} and {Kneissl, R.} and {Knoche, J.} and {Knox, L.} and {Kunz, M.} and {Kurki-Suonio, H.} and {Lagache, G.} and {Lähteenmäki, A.} and {Lamarre, J.-M.} and {Lasenby, A.} and {Lattanzi, M.} and {Lawrence, C. R.} and {Leahy, J. P.} and {Leonardi, R.} and {Lesgourgues, J.} and {Levrier, F.} and {Lewis, A.} and {Liguori, M.} and {Lilje, P. B.} and {Linden-Vørnle, M.} and {López-Caniego, M.} and {Lubin, P. M.} and {Macías-Pérez, J. F.} and {Maggio, G.} and {Maino, D.} and {Mandolesi, N.} and {Mangilli, A.} and {Marchini, A.} and {Maris, M.} and {Martin, P. G.} and {Martinelli, M.} and {Martínez-González, E.} and {Masi, S.} and {Matarrese, S.} and {McGehee, P.} and {Meinhold, P. R.} and {Melchiorri, A.} and {Melin, J.-B.} and {Mendes, L.} and {Mennella, A.} and {Migliaccio, M.} and {Millea, M.} and {Mitra, S.} and {Miville-Deschênes, M.-A.} and {Moneti, A.} and {Montier, L.} and {Morgante, G.} and {Mortlock, D.} and {Moss, A.} and {Munshi, D.} and {Murphy, J. A.} and {Naselsky, P.} and {Nati, F.} and {Natoli, P.} and {Netterfield, C. B.} and {Nørgaard-Nielsen, H. U.} and {Noviello, F.} and {Novikov, D.} and {Novikov, I.} and {Oxborrow, C. A.} and {Paci, F.} and {Pagano, L.} and {Pajot, F.} and {Paladini, R.} and {Paoletti, D.} and {Partridge, B.} and {Pasian, F.} and {Patanchon, G.} and {Pearson, T. J.} and {Perdereau, O.} and {Perotto, L.} and {Perrotta, F.} and {Pettorino, V.} and {Piacentini, F.} and {Piat, M.} and {Pierpaoli, E.} and {Pietrobon, D.} and {Plaszczynski, S.} and {Pointecouteau, E.} and {Polenta, G.} and {Popa, L.} and {Pratt, G. W.} and {Prézeau, G.} and {Prunet, S.} and {Puget, J.-L.} and {Rachen, J. P.} and {Reach, W. T.} and {Rebolo, R.} and {Reinecke, M.} and {Remazeilles, M.} and {Renault, C.} and {Renzi, A.} and {Ristorcelli, I.} and {Rocha, G.} and {Rosset, C.} and {Rossetti, M.} and {Roudier, G.} and {Rouillé d’Orfeuil, B.} and {Rowan-Robinson, M.} and {Rubiño-Martín, J. A.} and {Rusholme, B.} and {Said, N.} and {Salvatelli, V.} and {Salvati, L.} and {Sandri, M.} and {Santos, D.} and {Savelainen, M.} and {Savini, G.} and {Scott, D.} and {Seiffert, M. D.} and {Serra, P.} and {Shellard, E. P. S.} and {Spencer, L. D.} and {Spinelli, M.} and {Stolyarov, V.} and {Stompor, R.} and {Sudiwala, R.} and {Sunyaev, R.} and {Sutton, D.} and {Suur-Uski, A.-S.} and {Sygnet, J.-F.} and {Tauber, J. A.} and {Terenzi, L.} and {Toffolatti, L.} and {Tomasi, M.} and {Tristram, M.} and {Trombetti, T.} and {Tucci, M.} and {Tuovinen, J.} and {Türler, M.} and {Umana, G.} and {Valenziano, L.} and {Valiviita, J.} and {Van Tent, F.} and {Vielva, P.} and {Villa, F.} and {Wade, L. A.} and {Wandelt, B. D.} and {Wehus, I. K.} and {White, M.} and {White, S. D. M.} and {Wilkinson, A.} and {Yvon, D.} and {Zacchei, A.} and {Zonca, A.}},
	title = {Planck 2015 results - XIII. Cosmological parameters},
	DOI= "10.1051/0004-6361/201525830",
	url= "https://doi.org/10.1051/0004-6361/201525830",
	journal = {A\&A},
	year = 2016,
	volume = 594,
	pages = "A13",
}

@article{Planck_V_2018,
	author = {{Planck Collaboration} and {Aghanim, N.} and {Akrami, Y.} and {Ashdown, M.} and {Aumont, J.} and {Baccigalupi, C.} and {Ballardini, M.} and {Banday, A. J.} and {Barreiro, R. B.} and {Bartolo, N.} and {Basak, S.} and {Benabed, K.} and {Bernard, J.-P.} and {Bersanelli, M.} and {Bielewicz, P.} and {Bock, J. J.} and {Bond, J. R.} and {Borrill, J.} and {Bouchet, F. R.} and {Boulanger, F.} and {Bucher, M.} and {Burigana, C.} and {Butler, R. C.} and {Calabrese, E.} and {Cardoso, J.-F.} and {Carron, J.} and {Casaponsa, B.} and {Challinor, A.} and {Chiang, H. C.} and {Colombo, L. P. L.} and {Combet, C.} and {Crill, B. P.} and {Cuttaia, F.} and {de Bernardis, P.} and {de Rosa, A.} and {de Zotti, G.} and {Delabrouille, J.} and {Delouis, J.-M.} and {Di Valentino, E.} and {Diego, J. M.} and {Doré, O.} and {Douspis, M.} and {Ducout, A.} and {Dupac, X.} and {Dusini, S.} and {Efstathiou, G.} and {Elsner, F.} and {Enßlin, T. A.} and {Eriksen, H. K.} and {Fantaye, Y.} and {Fernandez-Cobos, R.} and {Finelli, F.} and {Frailis, M.} and {Fraisse, A. A.} and {Franceschi, E.} and {Frolov, A.} and {Galeotta, S.} and {Galli, S.} and {Ganga, K.} and {Génova-Santos, R. T.} and {Gerbino, M.} and {Ghosh, T.} and {Giraud-Héraud, Y.} and {González-Nuevo, J.} and {Górski, K. M.} and {Gratton, S.} and {Gruppuso, A.} and {Gudmundsson, J. E.} and {Hamann, J.} and {Handley, W.} and {Hansen, F. K.} and {Herranz, D.} and {Hivon, E.} and {Huang, Z.} and {Jaffe, A. H.} and {Jones, W. C.} and {Keihänen, E.} and {Keskitalo, R.} and {Kiiveri, K.} and {Kim, J.} and {Kisner, T. S.} and {Krachmalnicoff, N.} and {Kunz, M.} and {Kurki-Suonio, H.} and {Lagache, G.} and {Lamarre, J.-M.} and {Lasenby, A.} and {Lattanzi, M.} and {Lawrence, C. R.} and {Le Jeune, M.} and {Levrier, F.} and {Lewis, A.} and {Liguori, M.} and {Lilje, P. B.} and {Lilley, M.} and {Lindholm, V.} and {López-Caniego, M.} and {Lubin, P. M.} and {Ma, Y.-Z.} and {Macías-Pérez, J. F.} and {Maggio, G.} and {Maino, D.} and {Mandolesi, N.} and {Mangilli, A.} and {Marcos-Caballero, A.} and {Maris, M.} and {Martin, P. G.} and {Martínez-González, E.} and {Matarrese, S.} and {Mauri, N.} and {McEwen, J. D.} and {Meinhold, P. R.} and {Melchiorri, A.} and {Mennella, A.} and {Migliaccio, M.} and {Millea, M.} and {Miville-Deschênes, M.-A.} and {Molinari, D.} and {Moneti, A.} and {Montier, L.} and {Morgante, G.} and {Moss, A.} and {Natoli, P.} and {Nørgaard-Nielsen, H. U.} and {Pagano, L.} and {Paoletti, D.} and {Partridge, B.} and {Patanchon, G.} and {Peiris, H. V.} and {Perrotta, F.} and {Pettorino, V.} and {Piacentini, F.} and {Polenta, G.} and {Puget, J.-L.} and {Rachen, J. P.} and {Reinecke, M.} and {Remazeilles, M.} and {Renzi, A.} and {Rocha, G.} and {Rosset, C.} and {Roudier, G.} and {Rubiño-Martín, J. A.} and {Ruiz-Granados, B.} and {Salvati, L.} and {Sandri, M.} and {Savelainen, M.} and {Scott, D.} and {Shellard, E. P. S.} and {Sirignano, C.} and {Sirri, G.} and {Spencer, L. D.} and {Sunyaev, R.} and {Suur-Uski, A.-S.} and {Tauber, J. A.} and {Tavagnacco, D.} and {Tenti, M.} and {Toffolatti, L.} and {Tomasi, M.} and {Trombetti, T.} and {Valiviita, J.} and {Van Tent, B.} and {Vielva, P.} and {Villa, F.} and {Vittorio, N.} and {Wandelt, B. D.} and {Wehus, I. K.} and {Zacchei, A.} and {Zonca, A.}},
	title = {Planck 2018 results - V. CMB power spectra and likelihoods},
	DOI= "10.1051/0004-6361/201936386",
	url= "https://doi.org/10.1051/0004-6361/201936386",
	journal = {A\&A},
	year = 2020,
	volume = 641,
	pages = "A5",
}

@article{Planck_Int_LVII,
	author = {{Planck Collaboration} and {Akrami, Y.} and {Andersen, K. J.} and {Ashdown, M.} and {Baccigalupi, C.} and {Ballardini, M.} and {Banday, A. J.} and {Barreiro, R. B.} and {Bartolo, N.} and {Basak, S.} and {Benabed, K.} and {Bernard, J.-P.} and {Bersanelli, M.} and {Bielewicz, P.} and {Bond, J. R.} and {Borrill, J.} and {Burigana, C.} and {Butler, R. C.} and {Calabrese, E.} and {Casaponsa, B.} and {Chiang, H. C.} and {Colombo, L. P. L.} and {Combet, C.} and {Crill, B. P.} and {Cuttaia, F.} and {de Bernardis, P.} and {de Rosa, A.} and {de Zotti, G.} and {Delabrouille, J.} and {Di Valentino, E.} and {Diego, J. M.} and {Doré, O.} and {Douspis, M.} and {Dupac, X.} and {Eriksen, H. K.} and {Fernandez-Cobos, R.} and {Finelli, F.} and {Frailis, M.} and {Fraisse, A. A.} and {Franceschi, E.} and {Frolov, A.} and {Galeotta, S.} and {Galli, S.} and {Ganga, K.} and {Gerbino, M.} and {Ghosh, T.} and {González-Nuevo, J.} and {Górski, K. M.} and {Gruppuso, A.} and {Gudmundsson, J. E.} and {Handley, W.} and {Helou, G.} and {Herranz, D.} and {Hildebrandt, S. R.} and {Hivon, E.} and {Huang, Z.} and {Jaffe, A. H.} and {Jones, W. C.} and {Keihänen, E.} and {Keskitalo, R.} and {Kiiveri, K.} and {Kim, J.} and {Kisner, T. S.} and {Krachmalnicoff, N.} and {Kunz, M.} and {Kurki-Suonio, H.} and {Lasenby, A.} and {Lattanzi, M.} and {Lawrence, C. R.} and {Le Jeune, M.} and {Levrier, F.} and {Liguori, M.} and {Lilje, P. B.} and {Lilley, M.} and {Lindholm, V.} and {López-Caniego, M.} and {Lubin, P. M.} and {Macías-Pérez, J. F.} and {Maino, D.} and {Mandolesi, N.} and {Marcos-Caballero, A.} and {Maris, M.} and {Martin, P. G.} and {Martínez-González, E.} and {Matarrese, S.} and {Mauri, N.} and {McEwen, J. D.} and {Meinhold, P. R.} and {Mennella, A.} and {Migliaccio, M.} and {Mitra, S.} and {Molinari, D.} and {Montier, L.} and {Morgante, G.} and {Moss, A.} and {Natoli, P.} and {Paoletti, D.} and {Partridge, B.} and {Patanchon, G.} and {Pearson, D.} and {Pearson, T. J.} and {Perrotta, F.} and {Piacentini, F.} and {Polenta, G.} and {Rachen, J. P.} and {Reinecke, M.} and {Remazeilles, M.} and {Renzi, A.} and {Rocha, G.} and {Rosset, C.} and {Roudier, G.} and {Rubiño-Martín, J. A.} and {Ruiz-Granados, B.} and {Salvati, L.} and {Savelainen, M.} and {Scott, D.} and {Sirignano, C.} and {Sirri, G.} and {Spencer, L. D.} and {Suur-Uski, A.-S.} and {Svalheim, L. T.} and {Tauber, J. A.} and {Tavagnacco, D.} and {Tenti, M.} and {Terenzi, L.} and {Thommesen, H.} and {Toffolatti, L.} and {Tomasi, M.} and {Tristram, M.} and {Trombetti, T.} and {Valiviita, J.} and {Van Tent, B.} and {Vielva, P.} and {Villa, F.} and {Vittorio, N.} and {Wandelt, B. D.} and {Wehus, I. K.} and {Zacchei, A.} and {Zonca, A.}},
	title = {Planck intermediate results - LVII. Joint Planck LFI and HFI data processing},
	DOI= "10.1051/0004-6361/202038073",
	url= "https://doi.org/10.1051/0004-6361/202038073",
	journal = {A\&A},
	year = 2020,
	volume = 643,
	pages = "A42",
}

@article{tristram_planck,
	author = {{Tristram, M.} and {Banday, A. J.} and {Douspis, M.} and {Garrido, X.} and {Górski, K. M.} and {Henrot-Versillé, S.} and {Hergt, L. T.} and {Ilić, S.} and {Keskitalo, R.} and {Lagache, G.} and {Lawrence, C. R.} and {Partridge, B.} and {Scott, D.}},
	title = {Cosmological parameters derived from the final Planck data release (PR4)},
	DOI= "10.1051/0004-6361/202348015",
	url= "https://doi.org/10.1051/0004-6361/202348015",
	journal = {A\&A},
	year = 2024,
	volume = 682,
	pages = "A37",
}

@article{Li_2025,
doi = {10.3847/1538-4357/adc723},
url = {https://doi.org/10.3847/1538-4357/adc723},
year = {2025},
month = {jun},
publisher = {The American Astronomical Society},
volume = {986},
number = {2},
pages = {111},
author = {Li, Yunyang and Eimer, Joseph R. and Appel, John W. and Bennett, Charles L. and Brewer, Michael K. and Bruno, Sarah Marie and Bustos, Ricardo and Chan, Carol Yan Yan and Chuss, David T. and Cleary, Joseph and Dahal, Sumit and Datta, Rahul and Denes Couto, Jullianna and Denis, Kevin L. and Dünner, Rolando and Essinger-Hileman, Thomas and Harrington, Kathleen and Helson, Kyle and Hubmayr, Johannes and Iuliano, Jeffrey and Karakla, John and Marriage, Tobias A. and Miller, Nathan J. and Morales Perez, Carolina and Parker, Lucas P. and Petroff, Matthew A. and Reeves, Rodrigo A. and Rostem, Karwan and Ryan, Caleigh and Shi, Rui and Shukawa, Koji and Valle, Deniz A. N. and Watts, Duncan J. and Weiland, Janet L. and Wollack, Edward J. and Xu, Zhilei and Zeng, Lingzhen},
title = {A Measurement of the Largest-scale CMB E-mode Polarization with CLASS},
journal = {The Astrophysical Journal}
}

@article{jhaveri_PRD_optical_depth,
  title = {Turning a negative neutrino mass into a positive optical depth},
  author = {Jhaveri, Tanisha and Karwal, Tanvi and Hu, Wayne},
  journal = {Phys. Rev. D},
  volume = {112},
  issue = {4},
  pages = {043541},
  numpages = {12},
  year = {2025},
  month = {Aug},
  publisher = {American Physical Society},
  doi = {10.1103/6vd2-rbfn},
  url = {https://link.aps.org/doi/10.1103/6vd2-rbfn}
}

@article{huang_BAO,
    author = {Huang, Zhiqi},
    title = {Reionization optical depth and CMB–BAO tension in punctuated inflation},
    journal = {Monthly Notices of the Royal Astronomical Society},
    volume = {544},
    number = {2},
    pages = {2193-2199},
    year = {2025},
    month = {11},
    issn = {0035-8711},
    doi = {10.1093/mnras/staf1892},
    url = {https://doi.org/10.1093/mnras/staf1892},
    eprint = {https://academic.oup.com/mnras/article-pdf/544/2/2193/65133242/staf1892.pdf},
}

@article{H0LiCOW,
    author = {Wong, Kenneth C and Suyu, Sherry H and Chen, Geoff C-F and Rusu, Cristian E and Millon, Martin and Sluse, Dominique and Bonvin, Vivien and Fassnacht, Christopher D and Taubenberger, Stefan and Auger, Matthew W and Birrer, Simon and Chan, James H H and Courbin, Frederic and Hilbert, Stefan and Tihhonova, Olga and Treu, Tommaso and Agnello, Adriano and Ding, Xuheng and Jee, Inh and Komatsu, Eiichiro and Shajib, Anowar J and Sonnenfeld, Alessandro and Blandford, Roger D and Koopmans, Léon V E and Marshall, Philip J and Meylan, Georges},
    title = {H0LiCOW – XIII. A 2.4 per cent measurement of H0 from lensed quasars: 5.3σ tension between early- and late-Universe probes},
    journal = {Monthly Notices of the Royal Astronomical Society},
    volume = {498},
    number = {1},
    pages = {1420-1439},
    year = {2020},
    month = {09},
    issn = {0035-8711},
    doi = {10.1093/mnras/stz3094},
    url = {https://doi.org/10.1093/mnras/stz3094},
    eprint = {https://academic.oup.com/mnras/article-pdf/498/1/1420/33755111/stz3094.pdf},
}

\end{document}